\documentclass[12pt]{article}
\usepackage{amssymb,epsf,color}
\def\lsim{\mathrel{\rlap {\raise.5ex\hbox{$ < $}}
{\lower.5ex\hbox{$\sim$}}}}
\def\gsim{\mathrel{\rlap {\raise.5ex\hbox{$ > $}}
{\lower.5ex\hbox{$\sim$}}}} 
\def\sqr#1#2{{\vcenter{\vbox{\hrule height.#2pt

        \hbox{\vrule width.#2pt height#1pt \kern#1pt

           \vrule width.#2pt}

        \hrule height.#2pt}}}}

\def\lsim{{\displaystyle
{{\raise-8pt\hbox{$ <$}}
\atop{\raise5pt\hbox{$\sim$}}}}}
\def\gsim{{\displaystyle
{{\raise-8pt\hbox{$ >$}}
\atop{\raise5pt\hbox{$\sim$}}}}}
\def\slsim{{\displaystyle
{{\raise-8pt\hbox{$\scriptstyle <$}}
\atop{\raise5pt\hbox{$\scriptstyle \sim$}}}}}
\def\sgsim{{\displaystyle
{{\raise-8pt\hbox{$\scriptstyle  >$}}

\atop{\raise5pt\hbox{$\scriptstyle \sim$}}}}}

\newskip\humongous \humongous=0pt plus 1000pt minus 1000pt

\newcommand{\sumpf}[0]{\sum_{(H^{\rm f},G^{\rm f})}\! \! \! \!
{\raise
4pt
\hbox{$'$}}\,}

\newcommand{\sump}[0]{\sum_{(H,G)}\! \! {\raise 4pt \hbox{$'$}}\,}

\def\bs{\begin{subequations}}
\def\es{\end{subequations}}

\catcode`\@=11
\newcount\hour
\newcount\minute
\newtoks\amorpm

\hour=\time\divide\hour by60
\minute=\time{\multiply\hour by60 \global\advance\minute by-\hour}
\edef\standardtime{{\ifnum\hour<12 \global\amorpm={am}%
        \else\global\amorpm={pm}\advance\hour by-12 \fi

        \ifnum\hour=0 \hour=12 \fi
        \number\hour:\ifnum\minute<10 0\fi\number\minute\the\amorpm}}
\edef\militarytime{\number\hour:\ifnum\minute<10 0\fi\number\minute}
\def\draftlabel#1{{\@bsphack\if@filesw {\let\thepage\relax
   \xdef\@gtempa{\write\@auxout{\string
      \newlabel{#1}{{\@currentlabel}{\thepage}}}}}\@gtempa
   \if@nobreak \ifvmode\nobreak\fi\fi\fi\@esphack}
        \gdef\@eqnlabel{#1}}
\def\@eqnlabel{}
\def\@vacuum{}
\def\draftmarginnote#1{\marginpar{\raggedright\scriptsize\tt#1}}
\def\draft{\oddsidemargin -.2truein
        \def\@oddfoot{\sl preliminary draft \hfil
        \rm\thepage\hfil\sl\today\quad\militarytime}
        \let\@evenfoot\@oddfoot \overfullrule 3pt
        \let\label=\draftlabel
        \let\marginnote=\draftmarginnote
   \def\@eqnnum{(\theequation)\rlap{\kern\marginparsep\tt\@eqnlabel}%
\global\let\@eqnlabel\@vacuum}  }
\def\subequations{\refstepcounter{equation}%
  \edef\@savedequation{\the\c@equation}%
  \@stequation=\expandafter{\theequation}
  \edef\@savedtheequation{\the\@stequation}
  \edef\oldtheequation{\theequation}%
  \setcounter{equation}{0}%
  \def\theequation{\oldtheequation\alph{equation}}}

\def\endsubequations{\setcounter{equation}{\@savedequation}%
  \@stequation=\expandafter{\@savedtheequation}%
  \edef\theequation{\the\@stequation}\global\@ignoretrue
  \vspace*{-12pt} \\}

\def\bs{\begin{subequations}}
\def\es{\end{subequations}}
\relax
\def\thefootnote{\fnsymbol{footnote}}
\def\be{\begin{equation}}
\def\ee{\end{equation}}
\def\ba{\begin{eqnarray}}
\def\ea{\end{eqnarray}}

\newcommand{\ar}[2]{{#1\atopwithdelims[]#2}}

\def\ee{\end{equation}}
\def\bea{\begin{eqnarray}}
\def\eea{\end{eqnarray}}
\def\nn{\nonumber}

\newcommand{\uarrw}[0]{\mathrel{
{\raise.5ex\vbox{\hrule width 1cm}\hskip-6pt\rightarrow}}}
\def\thebibliography#1{%
\vskip 0.5cm \centerline{\bf References}
\list{%
[\arabic{enumi}]}{\settowidth\labelwidth{[#1]}
\leftmargin\labelwidth
\advance\leftmargin\labelsep
\usecounter{enumi}}
\def\newblock{\hskip .11em plus .33em minus .07em}
\sloppy\clubpenalty4000\widowpenalty4000
\sfcode`\.=1000\relax}

\renewcommand{\theequation}{\arabic{section}.\arabic{equation}}

\renewcommand{\section}{\setcounter{equation}{0}\@startsection%
{section}{1}{0mm}{-\baselineskip}{0.5\baselineskip}%
{\normalfont\normalsize\bfseries}}

\renewcommand{\subsection}{\@startsection%
{subsection}{2}{0mm}{-\baselineskip}{0.5\baselineskip}%
{\normalfont\normalsize\slshape}}

\renewcommand{\subsubsection}{\@startsection%
{subsubsection}{2}{0mm}{-\baselineskip}{0.5\baselineskip}%
{\normalfont\normalsize\slshape}}

\begin{document}
%
%
\renewcommand{\theequation}{\arabic{section}.\arabic{equation}}
\begin{titlepage}
\begin{flushright}
\end{flushright}
\begin{centering}
\vspace{1.0in}
\boldmath

{ \large \bf About combinatorics, and observables  }

\unboldmath
\vspace{1.5 cm}

{\bf Andrea Gregori}$^{\dagger}$ \\
\medskip
\vspace{3.2cm}
{\bf Abstract} \\
\end{centering} 
\vspace{.2in}
We investigate the most general ``phase space'' of configurations, consisting 
of all possible ways of assigning elementary attributes, ``energies'',
to elementary positions, ``cells''. We discuss how this space possesses 
structures that can be approximated by a quantum physics
scenario. In particular, we discuss how the Heisenberg's 
``Uncertainty Principle'' and a ``universe'' with a three-dimensional space 
arise, and what kind of mechanics rules it.

\vspace{6cm}

\hrule width 6.7cm
\noindent
$^{\dagger}$e-mail: agregori@libero.it

\end{titlepage}
\newpage
\setcounter{footnote}{0}
\renewcommand{\thefootnote}{\arabic{footnote}}

\tableofcontents

\vspace{1.5cm}

\noindent

\section{Introduction}
\label{intro}

In a recent work \cite{spi} we have discussed how, considering the 
superposition of all possible string configurations, weighted with their
occupation in the string phase space, we recover the actual properties 
and physics of the Universe,
as they are observed at present time and along its history and evolution. 
At any time, the main contribution
to the mean values of observables comes from the string configurations 
of minimal entropy. We have seen how this leads to a 
viable phenomenological scenario, highly predictive
and compatible with current experiments.
In this work, we approach the problem from a different, more general 
perspective: we investigate the most general 
possible phase space of ``spaces'', 
of arbitrary volume and number of coordinates,
describing whatever kind of degrees of freedom, asking ourselves if in this
apparently indistinct ``night of everything'' it is possible to disentangle
particular structures that appear more frequently than other ones.

We are used to order our observations according to phenomena that
take place in what we call space-time. An experiment, or, better,
an observation (through an experiment), basically consists in realizing that
something has changed: our ``eyes'' have been affected by something, that
we call ``light'', that has changed their configuration (molecular, atomic 
configuration). This light may carry information about changes in our
environment, that we refer either to gravitational phenomena, or 
electromagnetic ones, and so on...
In order to explain them we introduce energies, momenta, ``forces'', i.e.
interactions, and therefore we speak in terms of masses, couplings etc...
However, all in all, what all these concepts refer to is a change in
the ``geometry'' of our environment, a change that ``propagates'' to us,
and eventually results in a change in our brain, the ``observer''.
String Theory marks a big step forward along this line of thoughts, in that
it introduces a geometric origin of particles, masses, fields and couplings, 
which basically turn out to be geometric degrees of freedom, 
of an ``internal'' space.
In short, it implements the degrees of freedom of the geometries of
the space, and their changes during time (in one word, the ``space-time'')
in a fibered space. In its basic formulation, base and fiber appear on the
same footing.
It remains however that the ``conjugates'' variables to space and time,
namely energy and momentum, are introduced ``by hand'', as
separate concepts, related to the modes of expansion of the string.
Indeed, in order to introduce them, besides the ``target space'' we 
precisely need the ``string'', the object that ``lives'' in this spaces.
Since the time it was realized that the various perturbative string
constructions belong to a net of slices of a unique theory, which in certain
limits (11d supergravity) is not based on the introduction of a string, and 
in other dual approaches it requires more extended objects in order
to express all its degrees of freedom (membranes), it seems
the more and more reasonable to think that these objects 
are in themselves not so fundamental,
being perhaps just convenient ``parametrizations'' of the real thing.
This appears to be the parametrization of a geometric problem.
On the other hand, since the time of General Relativity we know that
energy and geometry are related, and ``interchangeable''. Understanding
energy is therefore the same as understanding geometry.

But what is after
all geometry other than a way of saying that, by moving along
a path in space, we will encounter or not some modifications? 
Assigning a ``geometry''
is a way of parametrizing modifications, differentiations, ``unevenesses'',
through a map from a space of ``attributes'', 
whose coordinate we call ``curvature'',
to another space, that we call ``the'' space. 
When we speak of ``attributes'', we mean the very basic
possibility of assigning an ``elementary object'' belonging to a set,
a ``space'', to ``positions'', ``elementary cells'' of a target set/space.
An elementary cell can either be occupied or not, by being assigned
an elementary object or not. We deal therefore with a ``combinatoric
of the distribution of occupations of cells''. 
Indeed, had we to start with a generic ``phase space of everything'',
in order to recognize structures inside this space 
we should be able to measure, to say 
without ambiguity if something is larger, equal or smaller than something
else. In practice, we should get rid of infinities, introduce a regulator 
(this is by the way what one does any time, when it is a matter of
performing physical computations in spaces with infinitely many
degrees of freedom) \footnote{``Infinite'', ``infinitely extended'',
as well as also the specular concept of ``zero size'', point-like, are
concepts that perhaps belong more to our mental abstraction than  
to real life. The fact that we can think at them, and that up to a certain
extent they turn out to be useful in order to organize our way of mathematical
thinking, doesn't necessarily mean that they are also fundamental in the
physical world.}. 
This would be done by working with a finite, although arbitrary,
number $n$ of coordinates, in a finite, although arbitrary volume $V$.
In order to measure things, we should also introduce a ``length'' $\ell$, 
the size of the ``elementary cell'', 
such that volumes of submanifolds or subsets of this
space would be ``counted'' in terms of $v_{(p)} = (\ell)^p$, $1 \leq p \leq n$.
In order to ``count'' degrees of freedom, we must work with discrete
quantities, reduce the space to a lattice. In itself
an elementary cell is ``adimensional'', in the sense that,
expressing volumes of any space dimensionality in terms of number
of cells, we can compare any powers of length.
Only in this way we can really say if a ``segment'' may contain more or
less degrees of freedom than another one, and we can also compare
without ambiguity the size of a ``segment'' to the one of a discrete
set of ``points''. In this way, we can deal with any kind of geometry.

In this work, we assume that the
basic formulation of the problem is the discrete one, given in terms of
``unit cells'', and investigate the combinatorics of the applications
of ``cells into cells''. We consider therefore a space in which
everything is given in terms of number of ``cells'':
a point is one cell, a two-dimensional square is given by $N \times N$
cells etc...
There are no units a priori distinguishing the measure of space from the one
of its attributes (in other words, ``space'' and ``momentum/energy'' are 
measured in the same way, in terms of unit cells).
The problem of geometries becomes in this way a problem of combinatorics,
and of their interpretation.
We start our analysis in section~\ref{setup} by investigating the
combinatorics of the ``distribution of attributes'', namely, the applications
of ``cells'' into a space of ``positions of cells'', and discuss how,
and in which sense, certain structures dominate. This allows to see an
``order'' in this ``darkness''. We discuss how a ``geometry'' shows up,
and how geometric inhomogeneities, that we can interpret as the discrete 
version of ``wave packets'',
arise. We recover in this way, through a completely different approach,
all the known concepts of particles and masses. In the ``phase space''
constituted by all possible configurations
we introduce a ``time ordering'' based on the inclusion of sets 
of configurations.
At any time, what appears to be the ``Universe'' is the superpositions
of an infinite number of configurations, weighted according the their
``geometric'' occupation in the phase space, 
namely by the exponential of their entropy.
Evaluation of entropies enables
to see that three space dimensions are favoured; statistically,
the ``space-time'' looks therefore mostly ``four dimensional''.

Time evolution turns out to be neither deterministic, nor probabilistic.
On the other hand, under certain conditions, after the 
introduction of approximations and simplifications enabling one to
concretely solve the combinatoric problem with a viable effective theory,
one is led to the probabilistic interpretation usually associated to
quantum mechanics. Indeed, the Heisenberg's Uncertainty Principle shows up
as an inequality encoding the indeterminacy introduced by ignoring the
contribution to the mean values due to a full bunch of configurations,
for which there is no interpretation in terms of particles and fields,
interacting in a space-time of well defined dimensionality.
The Uncertainty Principle turns out to be not only a bound on our
possibility of measuring quantities, but a bound on the 
meaning in itself of these quantities. 
It arises not simply as a bound on
the precision with which we can know certain observables, but as the threshold
beyond which they cannot even be consistently defined. In a certain sense,
they make sense only as ``mean, average values'' we can introduce only 
together with a certain degree of ``fuzziness''.

We devote section~\ref{detprob} to a discussion
of the issues of causality and in what limit ``quantum mechanics'' arises
in this framework.  
We pass then (section \ref{stringT}) to discuss what is the role played
by string theory in this scenario:
in which sense and up to what extent it provides an approximation
to the description of the combinatoric/geometric scenario, of which 
Quantum String Theory constitutes an implementation in 
the framework of a continuum (differentiable) space.
Strictly speaking, String Theory deals only with a subset 
of configurations,
a subspace of the full phase space, but, through the implementation
of quantization, and therefore of the Heisenberg's Uncertainty Principle,
it considers also the neglected configurations of the phase space, relating
them to the uncertainty ``built in'' in its basic definition.
In other words, it comes already provided with a 
``fuzziness'' that incorporates in its range the contribution
of all the other possible configurations.

We briefly reconsider the basics of string theory
in this perspective, discussing the relation to the 
``combinatoric'' approach.
In particular, we reconsider (section \ref{mimacro})
the entropy weighted sum of Ref.~\cite{spi};
we point out how, at ``fixed time'', the entropy of a string vacuum, intended
as the entropy of the states a string configuration is built of, computed
according to their probability within the string vacuum, is ``dual''
to the entropy of the whole string configuration itself, viewed as an element
in the space of all the possible configurations. As a consequence,
the most often realized string configurations, 
those of maximal ``absolute'' entropy, correspond to
those of minimal ``string'' entropy, as  
expressed in the functional of Ref.~\cite{spi}. 
Indeed, string theory gives a 
``collective'' representation of configurations in phase space, centered
on effective ``mean values'' of geometries and inhomogeneities (parametrized
through particles, fields, and their masses and charges),
embedded in a space already 
provided with a ``time'' coordinate. However, in the perturbative
string constructions the latter appears to be in a 
``decompactification phase'',
and does not coincide  with the ``physical time'' parametrizing the
evolution in the full phase space. The identification works only
for the string configurations of minimal entropy, which represent somehow
the ``on shell'' description of the Universe.
Along the path of minimal string entropy (or maximal phase space entropy)
configurations, the ``average'' geometry of the universe is that of a 
three-dimensional sphere. Near-to-minimal
configurations contribute on the other hand for inhomogeneities that give 
rise to ``local concentrations'' of energy/curvature: 
particles, wave packets, galaxy clusters etc... 
These perturbations of the ``regular'' geometry
are of the order of the Heisenberg's Uncertainties. 

In section~\ref{dimspt} we discuss how the Universe, as it appears 
to an observer, builds up; 
in particular, we discuss what is the meaning of a boundary,
an horizon, in such a spheric geometry, and the non-trivial
relation between what one \emph{sees},
and what indeed \emph{is} inside this space.

This work is thought as complementary 
of \cite{spi}. Therefore, many
discussions are not repeated here, and the reader is invited to refer
to \cite{spi} in order to complete the information.

\section{The general set up}
\label{setup}

Consider the system constituted by the following two ''cells'':
\be
\epsfxsize=3cm
\epsfbox{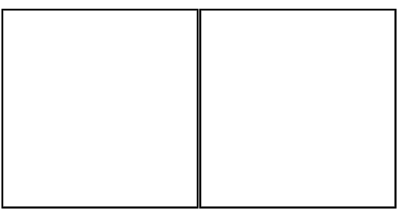}
\label{2cells1}
\ee
Let's assume that the only degrees of freedom this system possesses are that
each one of the two cells can independently be white or black.
We have the following possible configurations:
\be
\epsfxsize=3cm
\epsfbox{2cells1.eps}
\label{2cells2}
\ee
\be
\epsfxsize=3cm
\epsfbox{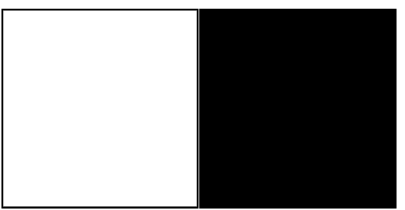}
\label{2cells3}
\ee
\be
\epsfxsize=3cm
\epsfbox{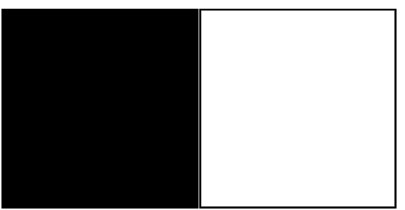}
\label{2cells4}
\ee
\be
\epsfxsize=3cm
\epsfbox{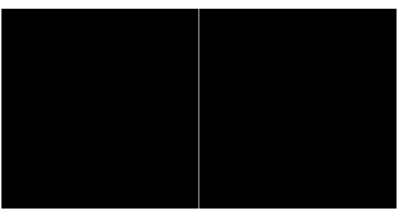}
\label{2cells5}
\ee
This is the ``phase space'' of our system.
The configuration ``one cell white, one cell black'' is realized two
times, while the configuration ``two cells white'' and ``two cells black''
are realized each one just once. 
Let's now abstract from the practical fact that these pictures 
appear inserted in a page, 
in which the presence of a written text clearly selects an orientation.
When considered as a ``universe'', something standing alone in its own,
configuration \ref{2cells3} and \ref{2cells4} are equivalent.
In the average, for an observer possessing the same ``symmetry'' of
this system (we will come back later to the subtleties of
the presence of an observer), the ``universe'' will appear 
something like the following:   
\be
\epsfxsize=3cm
\epsfbox{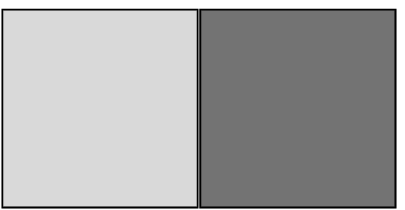}
\label{2cells6}
\ee
or, equivalently, the following: 
\be
\epsfxsize=3cm
\epsfbox{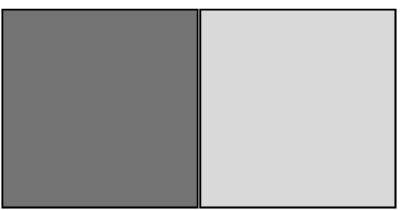}
\label{2cells7}
\ee
namely, the ``sum'':
\be
\epsfxsize=3cm
\epsfbox{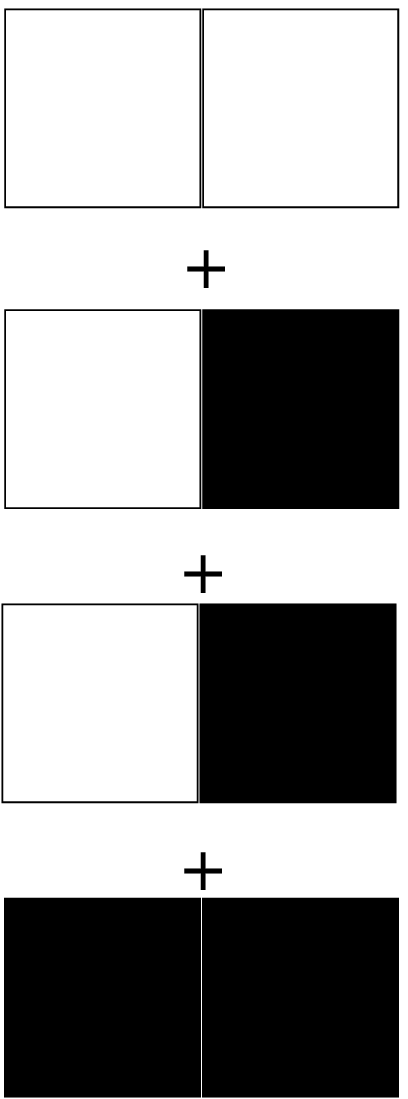}
\label{2cells8}
\ee
or equivalently the sum:
\be
\epsfxsize=3cm
\epsfbox{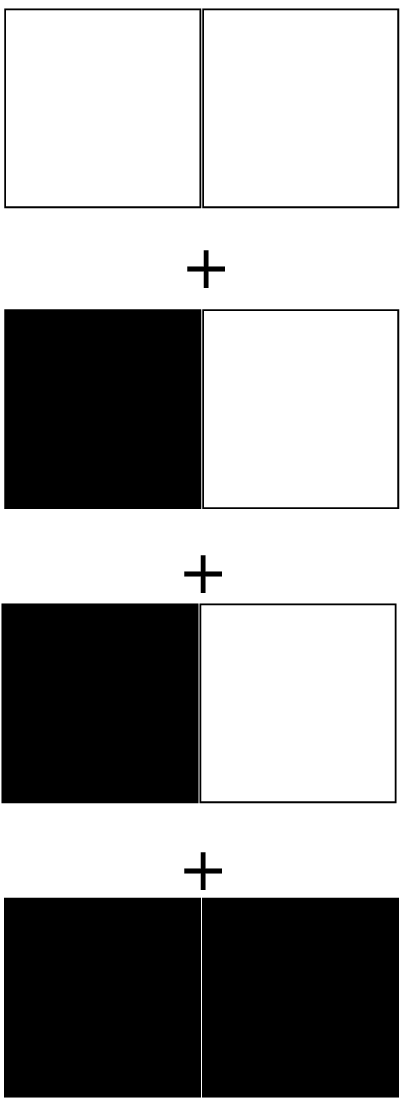}
\label{2cells9}
\ee
Notice that the observer ``doesn't know'' that we have rotated
the second and third term, because he possesses the same
symmetries of the system, and therefore is not able to distinguish
the two cases by comparing the orientation with, say, the
orientation of the characters of the text. What he sees, is
a universe consisting of two cells which appear slightly differentiated, 
one ``light grey'', the other ``dark grey''.

The system just described can be viewed as a two-dimensional space,
in which one coordinate specifies the position of a cell
along the ``space'',
and the other coordinate the attribute of each position, namely, the colour.
Our two-dimensional ``phase space'' is 
made by $2 ({\rm space}) \times 2 ({\rm colours})$ cells.
By definition the volume occupied in the phase
space by each configuration (two white; two black; one white one black)
is proportional to the logarithm of its entropy. 
The highest occupation corresponds
to the configuration with highest entropy. The effective appearance,
one light-grey one dark-grey, \ref{2cells6} or \ref{2cells7}, 
mostly resembles the highest entropy configuration.
  
Let's now consider in general cells and colours. The colours are attributes we 
can assign to the cells, which represent the positions in our space.
On the other hand, these ``degrees of freedom'' can themselves be viewed as 
coordinates. Indeed, if in our space with $m ({\rm space}) \times n 
({\rm colours})$ we have $n > m$, then we have more degrees of freedom
than places to allocate them. In this case, it is more appropriate
to invert the interpretation, and speak of $n$ places to which to assign
the $m$ cells. The colours become the space and the cells their ``attributes''.
Therefore, in the following we consider always $n \leq m$.

\subsection{Distributing degrees of freedom}
\label{ddf}

\begin{figure}
\centerline{
\epsfxsize=8cm
\epsfbox{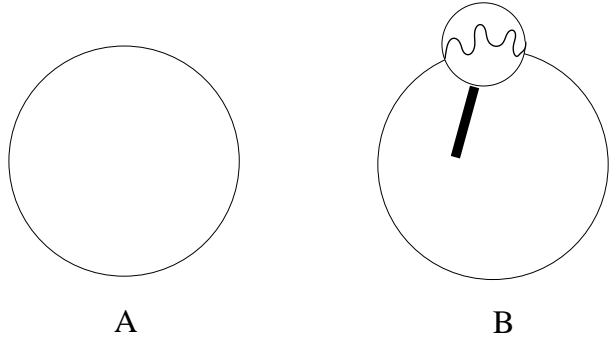}
}
\vspace{0.3cm}
\caption{On the left (A) is drawn a sphere (here for simplicity a 1-sphere, 
a circle, on the right (B) a surface with more energy (curvature) than a 
sphere, here showing out under a ``loupe'' as a wavy line.}     
\label{overcurve}
\end{figure}

Consider now a generic ``multi-dimensional'' space, consisting of 
$M_1^{p_1} \times \ldots \times M_i^{p_i} \ldots \times M_n^{p_n}$ 
``elementary cells''. Since an elementary, ``unit'' 
cell is basically a-dimensional, it makes sense to measure the volume
of this $p$-dimensional space, $p = \sum^n_i p_i$, in terms of unit cells:
$V = M_1^{p_i} \times \ldots \times M_n^{p_n} \stackrel{\rm def}{\equiv} 
M^p $. Although with the same volume, from the point of 
view of the combinatorics of cells and attributes this space is deeply
different from a one-dimensional space with $M^p$ cells. 
However, independently on the dimensionality, 
to such a space we can in any case assign, in the sense of ``distribute'',
$N \leq M^p$ ``elementary'' attributes. Indeed, in order to preserve the
basic interpretation of the ``$N$'' coordinate as ``attributes'' and
the ``$M$'' degrees of freedom as ``space'' coordinates, to which attributes
are assigned, it is necessary that $N \leq M_n$, $\forall \, n$ 
\footnote{In the case 
$N > M_n$ for some $n$, we must interchange the interpretation of 
the $N$ as attributes and instead consider them as a space coordinate, whereas
it is $M_n$ that are going to be seen as a coordinate of attributes.}.   
What are these attributes? Cells, simply cells:
our space doesn't know about ``colours'', it is simply a mathematical
structure of cells, and cells that we attribute in certain positions
to cells. By doing so, we are constructing a discrete 
``function'' $y = f (\vec{x})$, where 
$y$ runs in the ``attributes'' and $\vec{x} \in \{ M^{\otimes p} \}$
belongs to our $p$-dimensional space. 
We \underline{define} the phase space as the space of the assignments, the
``maps'':
\be
N \, \to \, \prod_i \otimes M_i^{\otimes p_i} \, , ~~~~~ M_i 
\, \geq \, N \, . 
\ee 
For large $M_i$ and $N$, we can approximate
the discrete degrees of freedom with continuous coordinates: $M_i \to r_i$,
$N \to R$.
We have therefore a $p$-dimensional space with volume $\prod r_i^{p_i}$, and
a continuous map $\vec{x} \in \{ \vec{r}^{\vec p} \} 
\stackrel{f}{\to} y \in \{ R \}$,
where $y$ spans the space up to $R = \prod r_i^{p_i} \equiv r^p$ and no more. 
In the following we will always consider $M_i \gg N$, while keeping $V = M^p$
finite. This has to considered as a regularization condition,
to be evetually relaxed by letting $V \to \infty$.
We ask now: what is the most realized configuration, namely, are there
special combinatorics in such a phase space that single out ``preferred''
structures, in the same sense as in our ``two-cells $\times$ two colours''   
example we found that the system in the average appears 
``light-grey--dark-grey''?
In order to come out of this complicated combinatoric problem, let's call
$y$ ``total energy''. The value $y$ is given by the total number of unit 
cells of this coordinate. Measured in units of the elementary cell, energy
goes from $1$ to $N \sim R$. Distributing, assigning cells from our 
``energy coordinate'' to our $p$-dimensional space corresponds then to 
assigning a curvature, a ``geometry'' to this space. Indeed, for us
assigning a geometry will be equivalent to assigning an energy distribution.
In this perspective, a curved surface in $p$ dimensions has not to be seen 
as a geometric set of points embedded in a $p+1$-dimensional flat space, 
but as a particular configuration due to a distribution of the energy 
cells along the coordinates of the $p$-dimensional space  $\prod M_i^{p_i}$,
$\sum p_i = p$, which we \emph{interpret} in geometric terms as a curved
space. This entails an implicit choice of units of energy as compared to
units measuring space, that we now consider as simply set to 1, as is done
in quantum-relativistic mechanics when choosing the so-called 
reduced Planck units: $c = \hbar = M_{pl} = 1$. The geometry of a 
$p$-sphere of radius $r$ is therefore characterized by an energy density 
scaling like:
\be
\rho(E) ~ \approx ~ K(p) \, {N \over r^p} \; \propto \; {1 \over r^2} \, ,
\label{rhoscal}
\ee 
where $K(p)$ is a typical symmetry factor depending on the dimensionality of
the sphere. 
At any given $N$ (and fixed volume $V \gg N$),
the most entropic configurations are the 
``maximally symmetric'' ones, i.e.
those that look like spheres in the above sense.
Equation \ref{rhoscal} sets also a bound on the \emph{mean}
amount of energy we can 
put within a space region of radius $r$ at any dimension $p$. 
If we try to force more energy to stay within such a region, 
we ``overcurve'' it, producing a less symmetric configuration 
(see figure~\ref{overcurve}).
For $p = 3$, equation \ref{rhoscal} reads:
\be
E_{p = 3}\, (\sim N) ~ \propto ~ r \, .   
\label{Np3}
\ee
In classical terms, this is basically the Schwarzschild bound
relating the radius of a Black Hole to its mass (= total rest energy).
In practice, we are saying that configurations over the Schwarzschild bound
(as well as configurations below this bound) 
are less entropic, unfavoured in the phase space as compared to the spheres. 
We will come back on this issue later in this work.

\subsection{Entropy of spheres}
\label{eSp}

\begin{figure}
\centerline{
\epsfxsize=8cm
\epsfbox{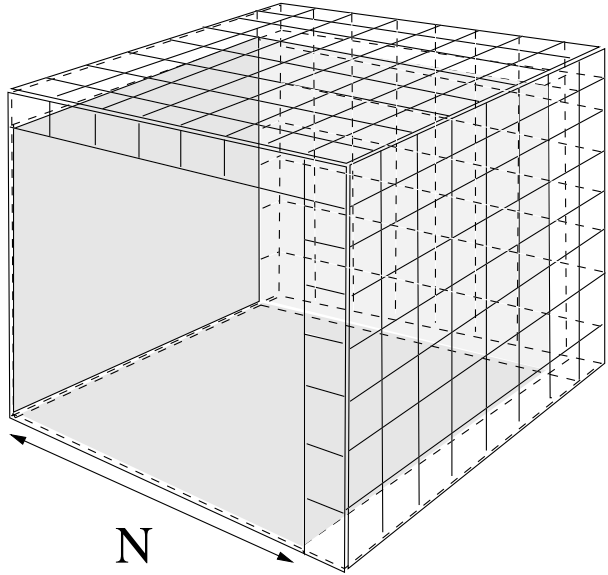}
}
\vspace{0.3cm}
\caption{The growth from $m \to m+1$ of a ``$p$ cube'', here schematically
represented as a 3-cube.}     
\label{cube-1}
\end{figure}

\begin{figure}
\centerline{
\epsfxsize=10cm
\epsfbox{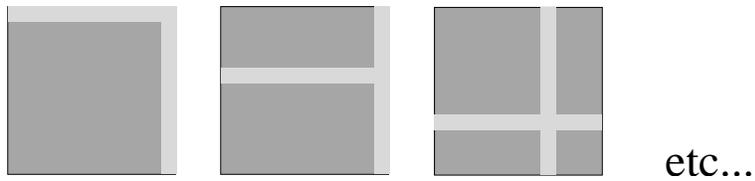}
}
\vspace{0.3cm}
\caption{The new cells, represented in white in figure~\ref{cube-1}, and
schematically represented here in light grey in a projection to two dimensions,
can be added in any position of the cube.}     
\label{distributions}
\end{figure}

We want to see now what is the entropy, or equivalently, the weight $W$
in the phase space, of a $p$-sphere of radius $m$. As it was in the 
previous section for the total energy $N$, here too 
we consider $m \ll M$. The weight of a sphere in the phase space
will be given by the 
number of times such a sphere can be formed by permuting its points, times
the number of choices of the position of, say, its center, in the whole space.
Since we eventually are going to take the limit $V \to \infty$,
we don't consider here this second contribution, which
is going to produce an infinite factor, equal for each kind of geometry, for 
any finite amount of total energy $N$. We will therefore concentrate here
to the first contribution, the one that distinguishes from sphere to sphere,
and from sphere to other geometries.
To this purpose, we solve
the ``differential equation'' (more properly, a finite difference equation)
of the increase in the combinatoric when passing from $m$ to $m + 1$.
In a $p$-dimensional space, illustrated as a ``$p$-cube'' in 
figure~\ref{cube-1},
when we pass from $m$ to $m + 1$ we increase the volume by an
amount of cells scaling as the boundary of the space. For instance,
in the case of a
three-cube we get some $3 m^2$ more cells. We can think to place the
one more degree of freedom in one of these cells, but we can also permute the 
position of the planes, while a subspace of volume $m^p$ has the same
distribution of cells as before (see figure~\ref{distributions}). 
We have therefore $\sim (m+1)^p$, and not only
$\sim p m^{p-1}$ possibilities. In order to preserve the
type of geometry, we must multiply the new gained volume times
the ``matter density'' of the sphere, $\sim 1 / m^2$. We recall in fact that
in this framework boundary conditions, such as for instance those allowing to
conclude that we have the curved geometry of a sphere, are not to be thought
as geometric conditions to be imposed on the coordinates, as if a geometry
could exist independently of the energy content of space. 
A sphere is by definition identified with the space with
an energy density corresponding to (i.e. generating)
the appropriate curvature. 
In this way one can see that:
\be
W(m+1) \, \sim \, W(m) \times   {(m+1)^p \over m^2 } \, . 
\ee 
The rate of increase of the number
of configurations scales therefore like:
\be
W(m+1)_{p-{\rm sphere}} \, \approx \, W(m)_{p-{\rm sphere}} 
\times  m^{p-2} \, , ~~~~ p \, \geq \, 2 \, ,
\label{Wmp-2}
\ee 
where on the second factor of the r.h.s. we have been a bit loose 
in the evaluation of the exact volumes, 
neglecting minorities like
a difference between $m$ and $m+1$ and the correct normalization
of the curvature/energy density of a $p$-sphere: what we are interested in is
understanding the main behaviour at 
large $m$. Translated into a difference equation, \ref{Wmp-2} means:
\be
{\Delta W(m)_p \over W(m)_p} \, \sim \,  \, m^{p-2} \, , ~~~~p \, \geq \, 2 
\, .
\ee  
Through a  passage to differential equations, $m \to x$, $x$ a continuous 
variable, we can integrate and obtain:
\be
S_{( p \geq 2)} \, \propto \, \ln W(m) \, \sim \, {1 \over p-1} \, m^{p-1} \, .
\label{Spm}
\ee
We obtain therefore the typical scaling law of the entropy of 
a $p$-dimensional black hole (see for instance \cite{Rabinowitz:2001ag}).

For $p = 1$, we cannot have a sphere. With an energy density $\rho(E)$
scaling as $1 / R^2$, the total energy in one dimension would be
$E \sim \rho(E) \times R \sim 1 /R$, but
we cannot have a total energy $E \sim 1 / R$,
lower than one. The only possible configurations
are ``less symmetric'' than a sphere. For a total energy $N = m = R$
there is no free space at all: $m$ cells can only occupy the $m$ free 
positions, and therefore $\Delta W = 0$. In general,
for any energy $n < m$ that anyway scales proportionally to $m$: $n /m = q$,
we have: 
\be
W(m+1)_1 \, \sim W(m)_1 \times q \, ,
\ee
and therefore:
\be
d W / W ~ \sim ~ 0 \, , ~~~~~~p = 1 \, , 
\ee
leading to:
\be
S_{(p = 1)} \, \propto \, \ln W ~ \sim ~ {\rm const.} 
\ee
\newpage
\noindent
This analysis allows us to conclude that:
\begin{itemize}
\item \emph{\underline{At any energy $N$, the 
most entropic configuration is the one corresponding to the geo-}}

\emph{\underline{metry of a 
three-sphere}}. 
\end{itemize}

\noindent
Higher dimensional spheres have an unfavoured ratio
entropy/energy. Three dimensions are then statistically
``selected out'' as the dominant space dimensionality. 
In higher dimension $p$, the condition for having a sphere of radius
$r \sim m$ reads in fact:
\be
{N \over m^p} \, \sim \, {1 \over m^2} \, ,
\ee
which implies that the radius must be shorter than the total energy:
\be
m \, \sim \,  N^{1 \over p-2} ~ < \, N \, , ~~~ p > 2 \, .
\ee
(We are measuring everything in terms
of number of cells, therefore we can freely play with dimensions.
Strictly speaking, what we are saying is that in a $p$-dimensional
sphere of radius $\sim m$, the total energy must scale like $\sim m^{p-2}$).
From expression~\ref{Spm} we derive:
\be
S_{(p > 2)} \vert_N \, \sim \, {1 \over p-1} \, 
m^{p-1} \, \sim \, {1 \over p-1} \, N^{p-1 \over p -2} \, .
\ee 
Therefore, as $p$ increases,
the weight $W = \exp S$ decreases exponentially as compared to
$p = 3$:
\be
{ W(N)_{p} \over W(N)_{3}} \, \approx \, {\rm e}^{N^{ -{p - 3 \over p-2}}}
\, = \, {\rm e}^{- \left({p - 3 \over p-2} \right) N }  \, , ~~~ p \geq 3 
\, . 
\ee
A second observation is that:
\begin{itemize}
\item
\emph{\underline{there do not exist two configurations with the same entropy}}:
if they have the same entropy, they are perceived as the same configuration.
\end{itemize} \label{noneq}
\noindent
The reason is that we have a combinatoric problem, and,
at fixed $N$, the volume of occupation
in the phase space is related to the symmetry group of the configuration.
In practice, we classify configurations through combinatorics: 
a configuration corresponds to a certain combinatoric group. Now,
discrete groups with the same volume, i.e. the same number of elements,
are homeomorphic. This means that they describe
the same configuration. Configurations and entropies are therefore
in bijection with discrete groups, and this removes the degeneracy.
Different entropy = different occupation volume = different volume of the
symmetry group, in practice  this means that we have a different configuration.
As a consequence, at fixed $N$, 
\emph{\underline{the correction to the mean value of the 
curvature}}, as due to the $p \neq 3$ configurations, is simply the sum over
their weights, counted with multiplicity 1:
\be
\Delta \left( {1 \over N^2}  \right) \, \approx \, 
\sum_p {1 \over N^{2 \over p - 2}} \, \exp { N^{p - 1 \over p-2}} \, ,
~~~ p > 2 \, ,
\ee
and \emph{\underline{is bounded by the three-dimensional term}}:
\be
\Delta \left( {1 \over N^2}  \right) \, <  \, 
\left( {1 \over N^2}  \right) \, . 
\label{bound3}
\ee
The lower-than-three dimensional spheres
provide a negligible contribution, because their contributions to
the total energy, $\sim 1$ ($p = 2$) and $\sim 1/N$ ($p = 1$), are 
exponentially suppressed by a factor
$W_{(2)} = {\rm e}^{S_{(p)}} \approx {\rm e}^{-N} W_3$ and
$W_{(1)} \approx {\rm e}^{-N^2} W_3 $ respectively.
The amount of uncertainty we introduce in neglecting them is 
therefore of an order smaller than the curvature itself, $\sim {1 \over N^2}$.
It remains to see what is the contribution of non-spheric configurations.
This will be considered in section~\ref{UncP}.

\subsection{How do inhomogeneities arise}
\label{inho}

We have seen that the dominant geometry, the spheric geometry, corresponds to a
homogeneous distribution of cells along the positions of the space, that 
we illustrate in figure~\ref{inhomo-1},
\be
\epsfxsize=4cm
\epsfbox{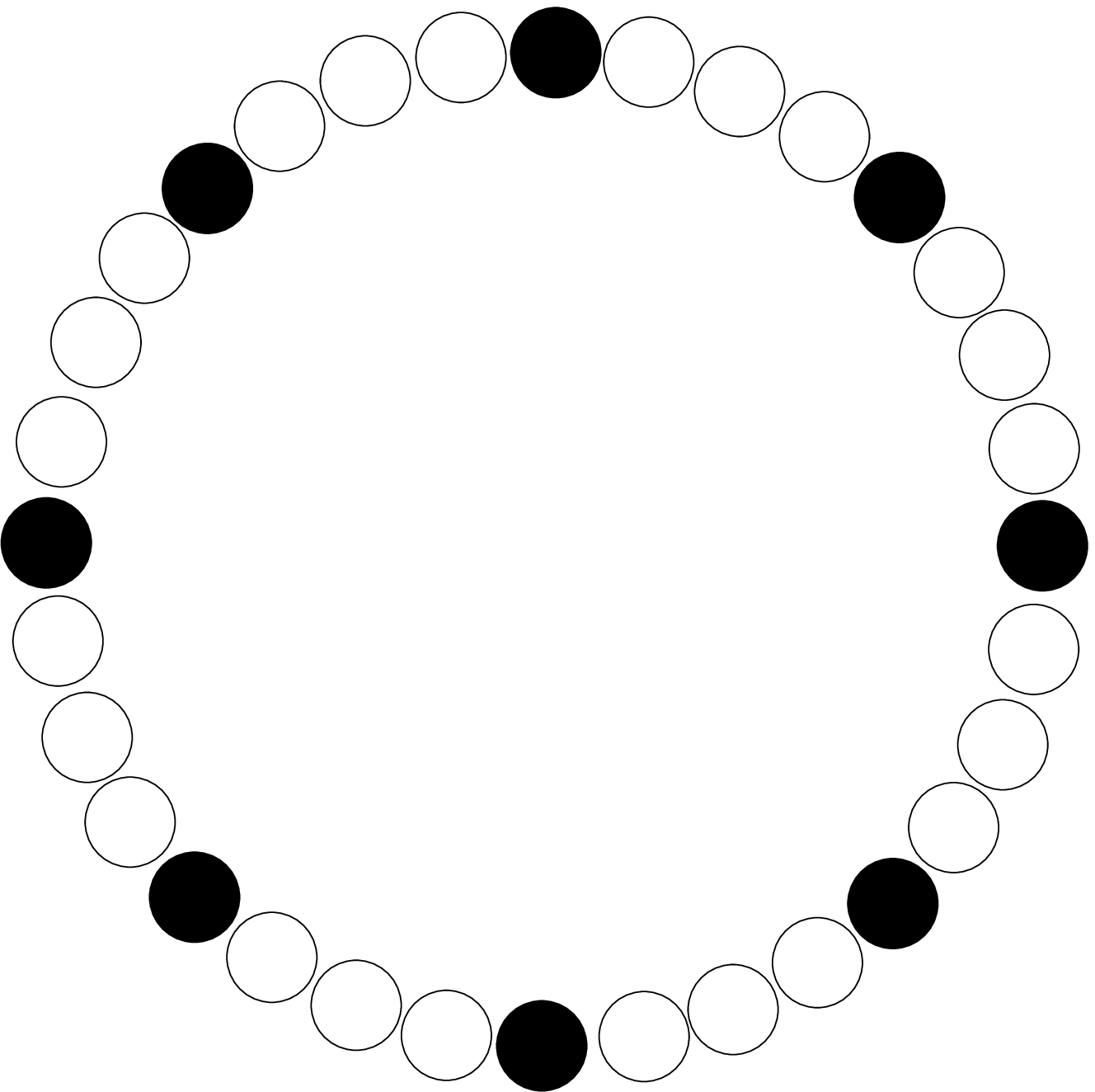}
\label{inhomo-1}
\ee
where we mark in black the occupied cells.
However, also the following configurations have spheric symmetry:
\be
\epsfxsize=3cm
\epsfbox{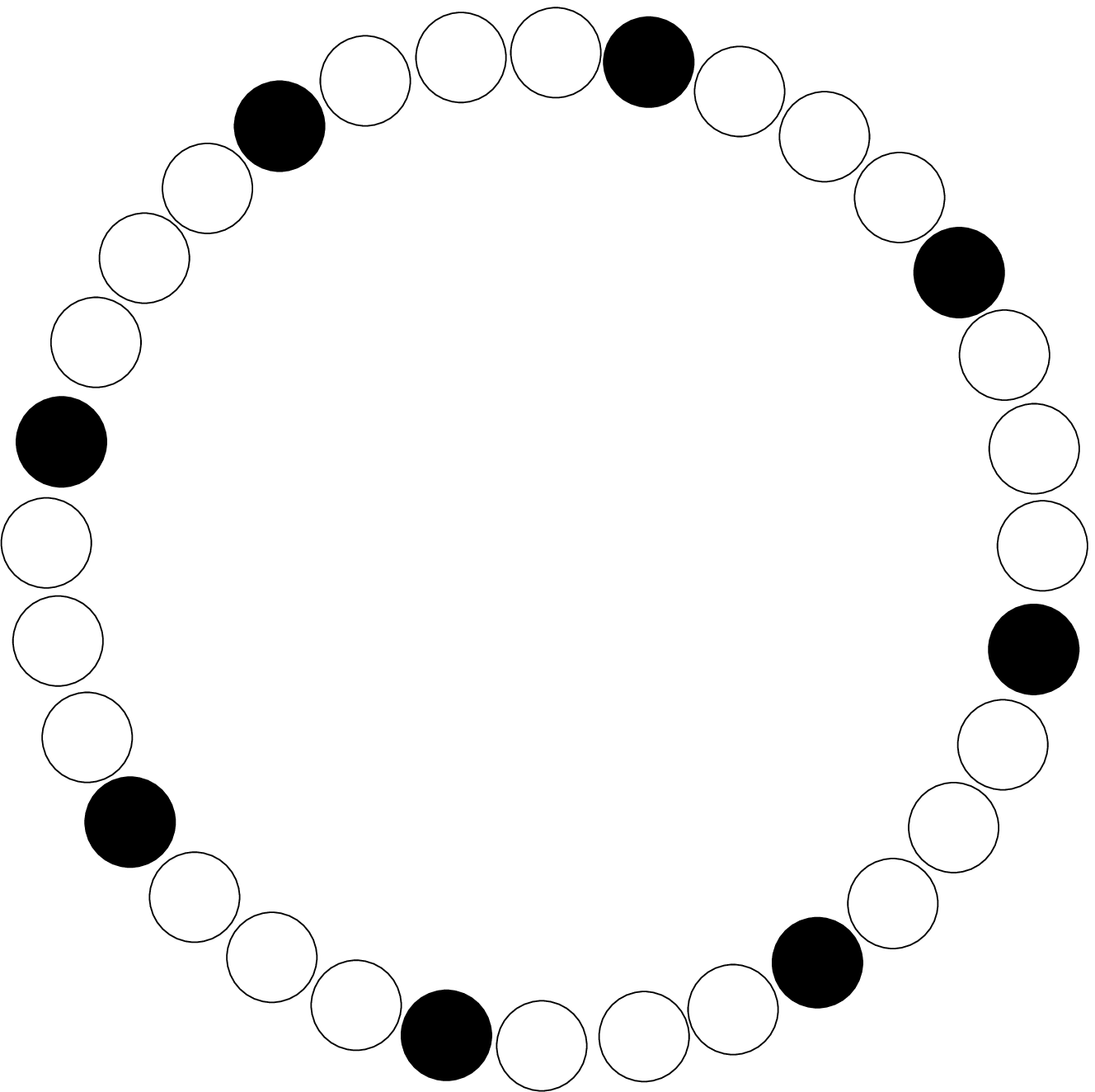}
~~~~~~~~~~~
\epsfxsize=3cm
\epsfbox{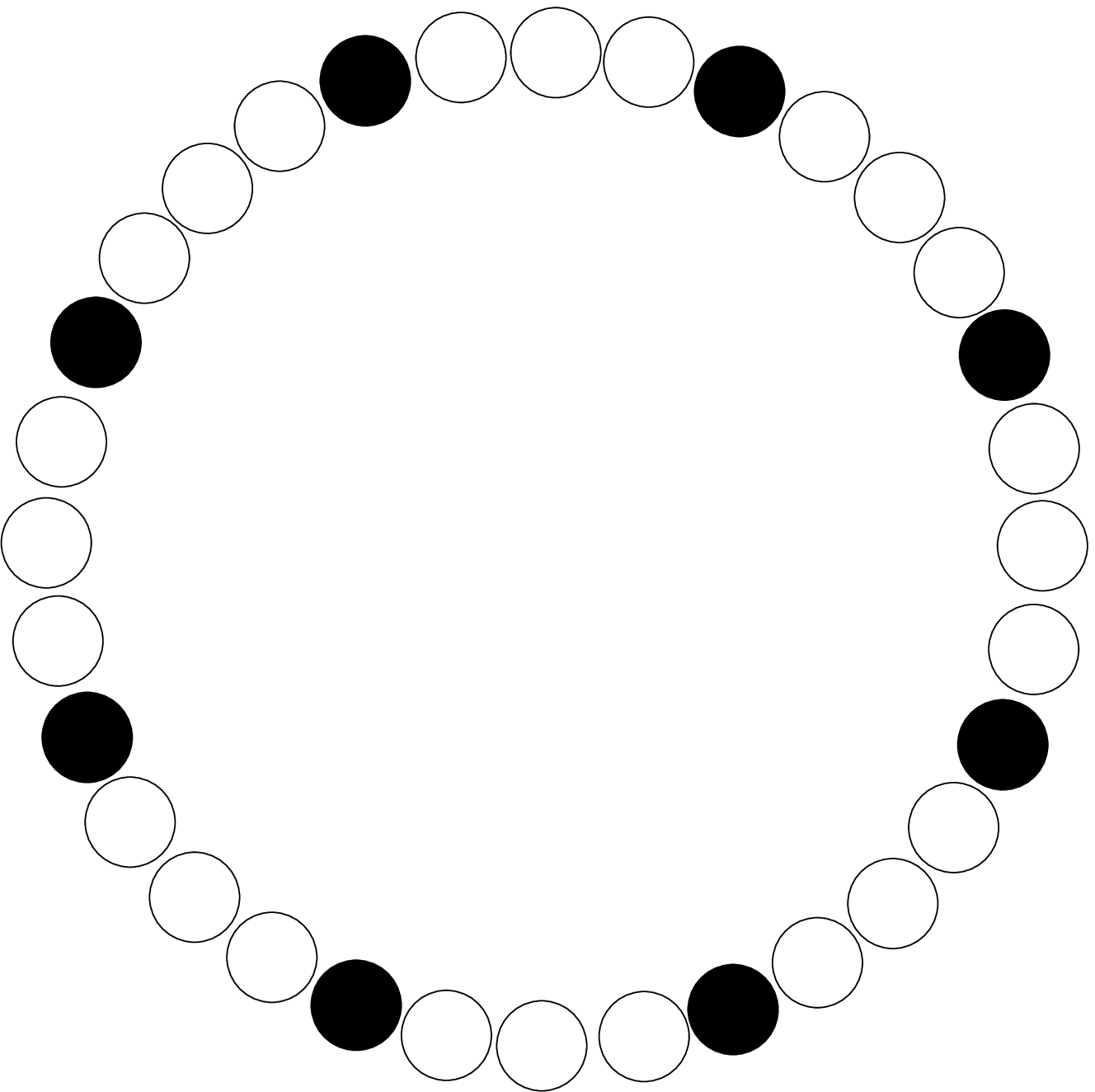}
~~~~~~~~~~~
\epsfxsize=3cm
\epsfbox{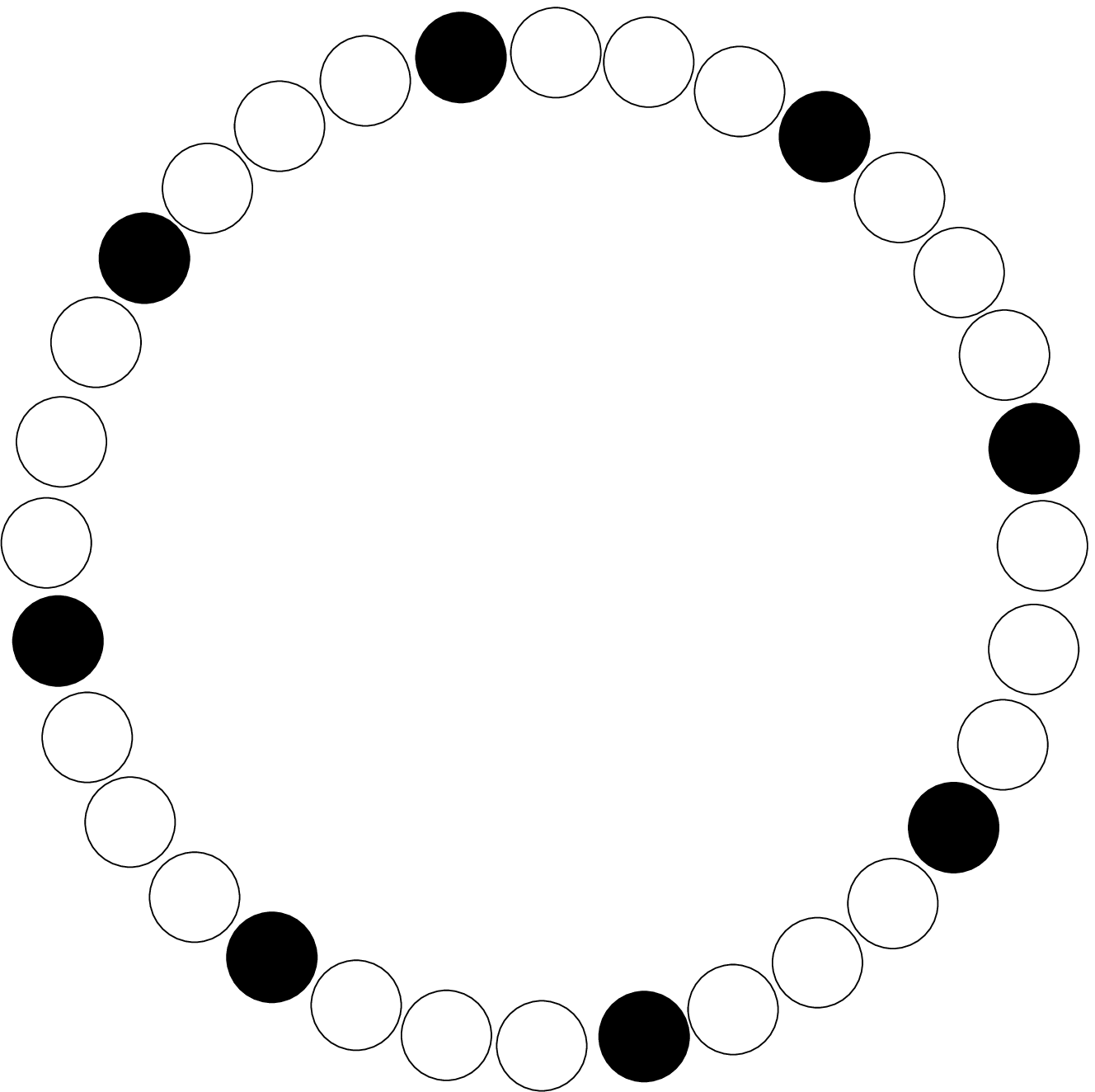}
\label{inhomo-4}
\ee
They are obtained from the previous one by shifting clockwise by one position 
the occupied cell.
One would think that they should sum up to an apparent averaged
distribution like the following:
\be
\epsfxsize=4cm
\epsfbox{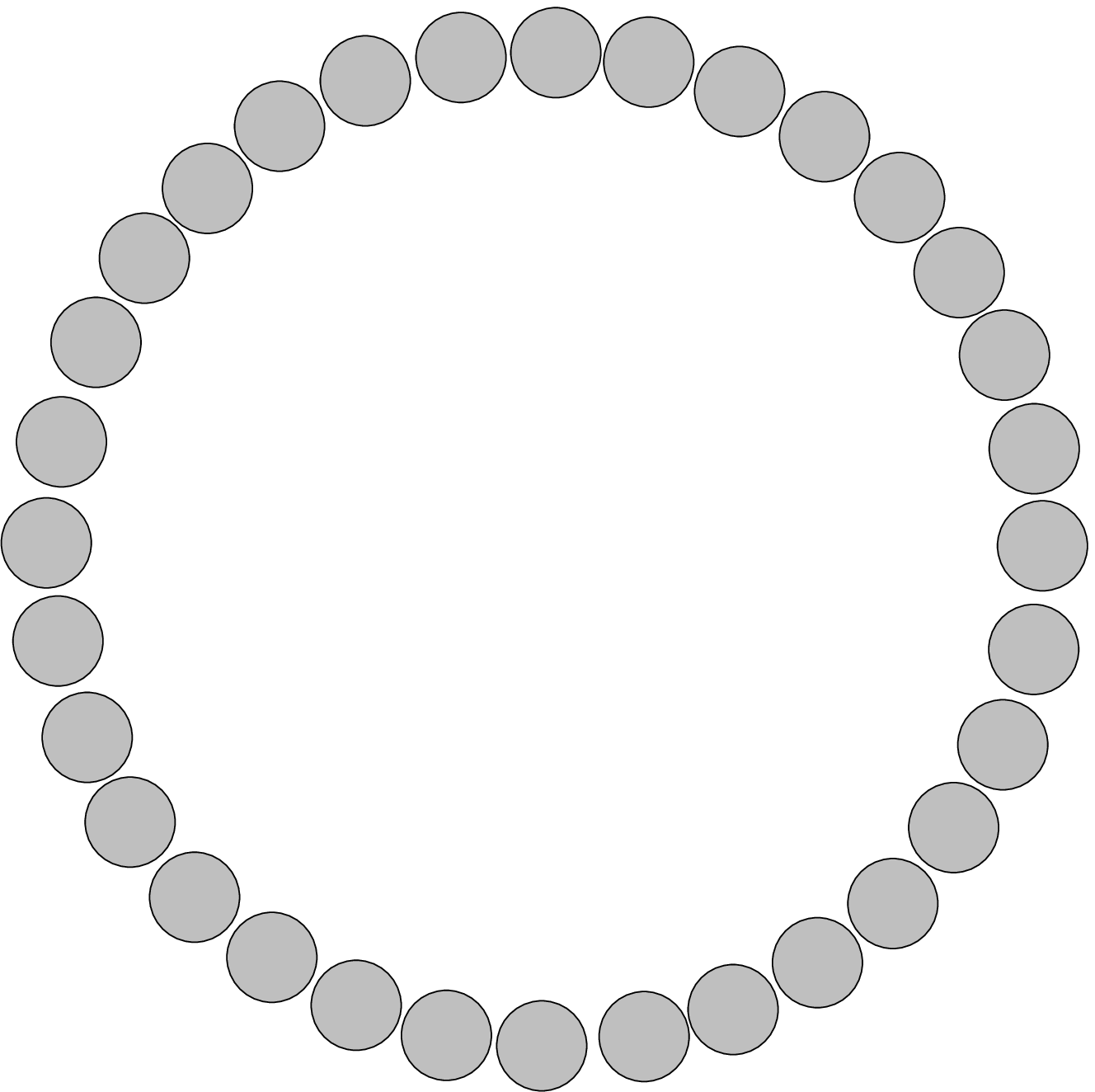}
\label{inhomo-5}
\ee 
This is not true: the Universe will indeed look like in 
figure~\ref{inhomo-5}, however this will be the ``smeared out'' result
of the configuration~\ref{inhomo-1}. As long as there are no reference points
in the space, which is an absolute space, all the above configurations
are indeed the same configuration, because nobody can tell in which sense
a configuration differs from the other one: ``shifted clockwise''
or ``counterclockwise'' with respect to what?
We will discuss later how the presence of an
observer by definition breaks some symmetries. Let's see here how
inhomogeneities (and therefore also configurations that we call
``observers'') do arise. Configurations with almost maximal, 
although non-maximal entropy, correspond to a slight breaking of the 
homogeneity of space. 
For instance, the following configuration, in which only one cell is shifted
in position, while all the other ones remain as in \ref{inhomo-1}:
\be
\epsfxsize=4cm
\epsfbox{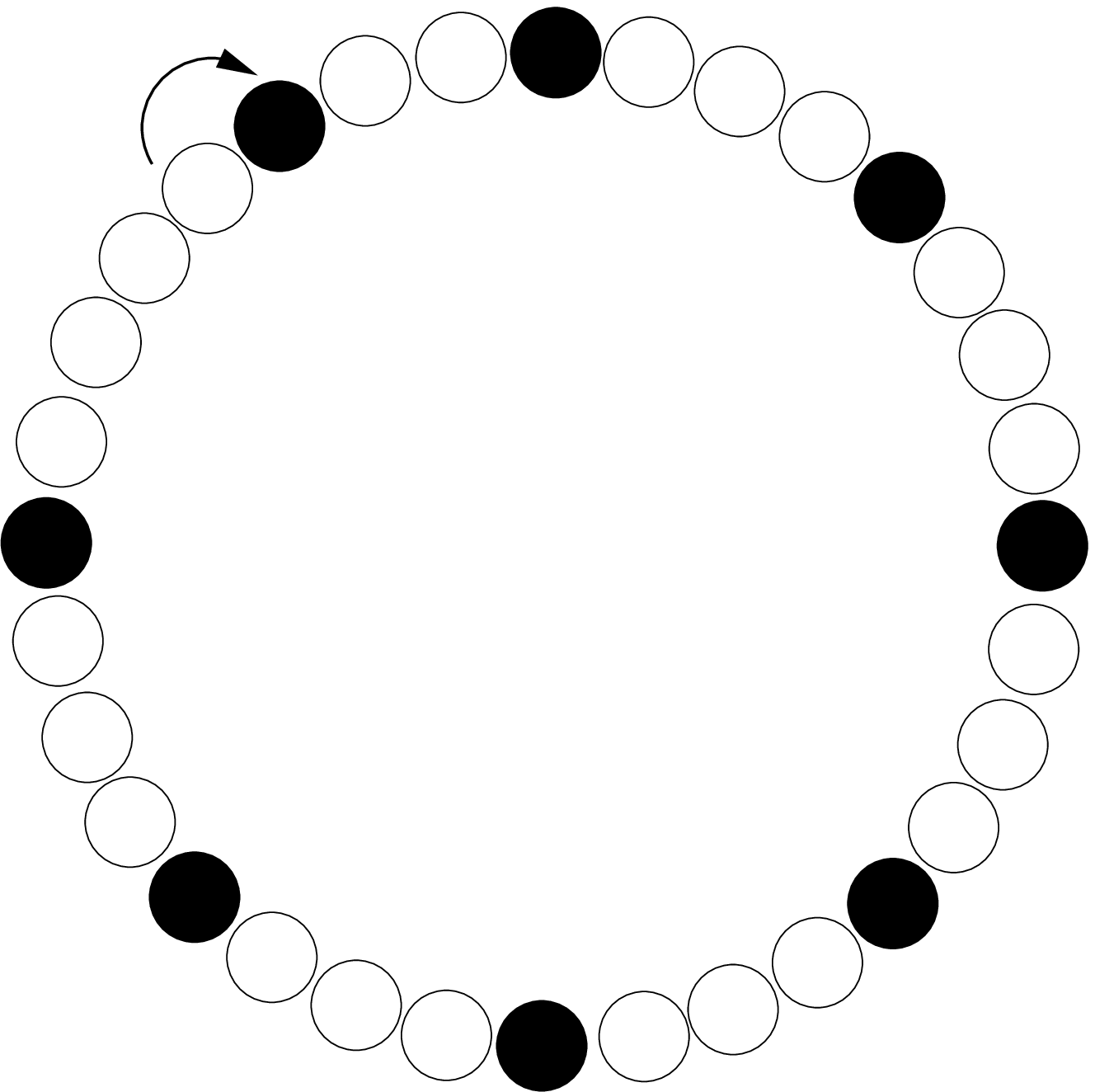}
\label{inhomo-6}
\ee 
This configuration will have a lower weight as compared to the fully symmetric
one. In the average, including also this one, the universe will
appear more or less as follows:
\be
\epsfxsize=4cm
\epsfbox{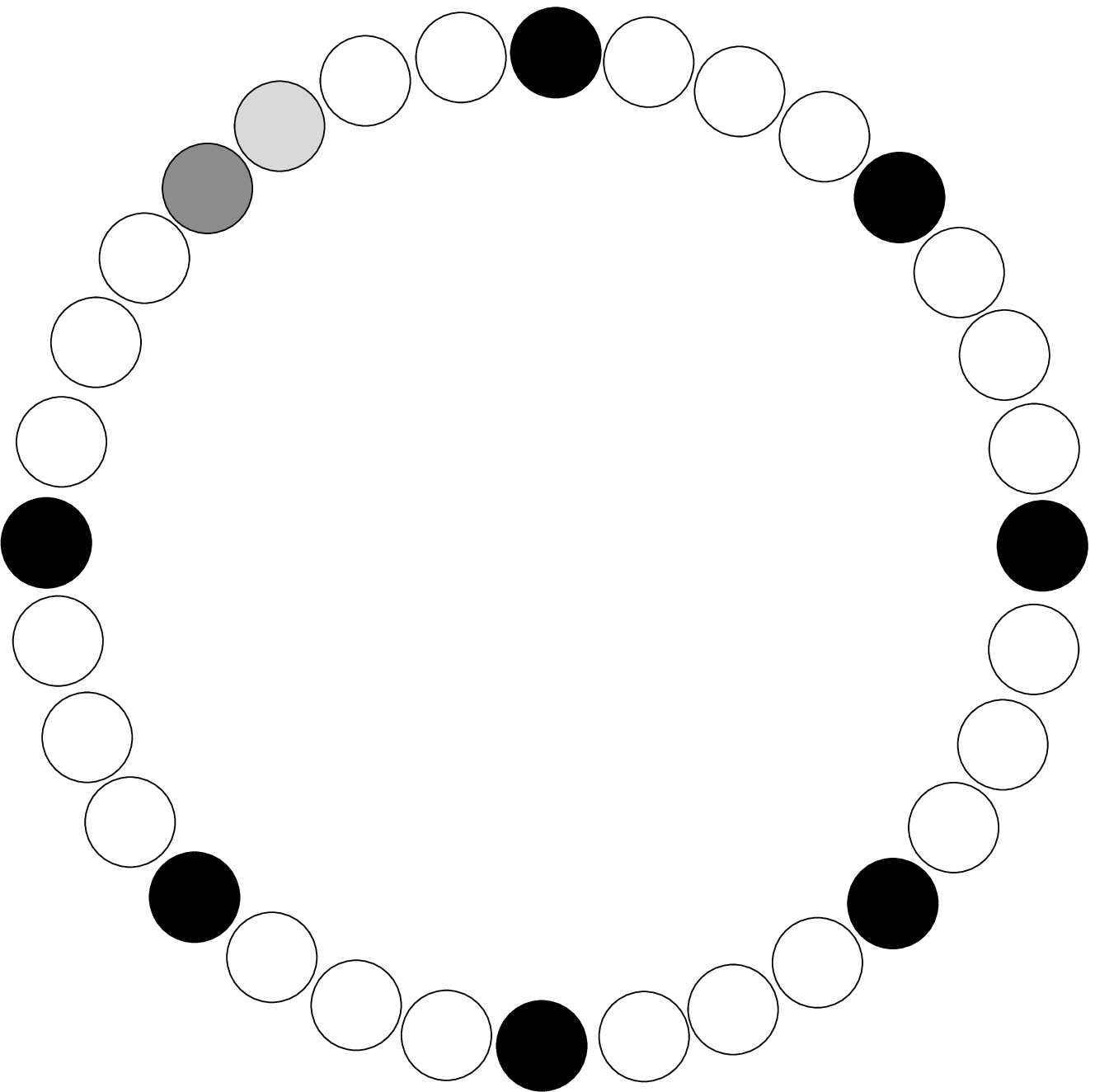}
\label{inhomo-7}
\ee 
where we have distinguished with a different tone of grey
the two resulting adjacent occupied cells, as a result of the
different occupation weight.
For the same reason as before, we don't have to consider summing 
over all the rotated configurations, in which the inhomogeneity
appears shifted by 4 cells, because all these are indeed the very same 
configuration as \ref{inhomo-6}. This is therefore the way inhomogeneities
build up in our space, in which the ``pure'' spheric geometry is
only the dominant aspect. We will discuss in section~\ref{UncP} how heavy
is the contribution of non-maximal configurations, and therefore what is
the order of inhomogeneity they introduce in the space.

\subsection{``Wave packets''}
\label{wavep}

Let's suppose there is a set of configurations of space that differ 
for the position of one energy cell, in such a way that 
the unit-energy cell is ``confined'' to a take a place in 
a subregion of the whole space. Namely, we have a sub-volume $\tilde{V}$ 
of the space with unit energy, or energy density $1 / \tilde{V}$. 
For $N$ large enough as compared to $\tilde{V}$,  
we must expect that all these configurations have almost the same weight.
Let's suppose for simplicity
that the subregion of space extends only in one direction, so that we work
with a one-dimensional problem: $\tilde{V} = r$. 
The ``average energy'' of this 
region of length $n \sim r $,  averaged over this subset
of configurations, is:
\be
E \, = \, {1 \over  n} \, = \, {1 \over  r} \, .
\label{Er}
\ee      
This is somehow a familiar expression: if we call this subregion   
a ``wave-packet'' everybody will recognize that this is nothing else than
the minimal energy according to the Heisenberg's Uncertainty Principle.
In terms of colours, each cell of space is ``black'' or ``white'', 
but in the average the region is ``grey'', 
the lighter grey the more is the ``packet'' spread out
in space (or ``time'', a concept to which we will come soon).
If we interpret this as the mass of a particle present
in a certain region of space, we can say that the particle is more heavy
the more it is ``concentrated'', ``localized'' in space. Light particles are 
``smeared-out mass-1 particles''.

\subsection{The ``time'' ordering}
\label{timev}

We have seen that, at any ``energy'' $N$, the dominant configurations
are $p$-spheres, spaces in which we can identify a radius $r \sim N$, which
turns out to scale linearly with the total energy.
We can therefore introduce an ordering in the whole phase space, that
we call a ``time-ordering'', through the identification 
of the time coordinate with $N : \, t \leftrightarrow N$. We consider
then the set $\Phi (N) \equiv \{ \Psi (N) \}$ 
of all configurations
at any dimensionality $p$ and volume $V \gg N$ ($V \to \infty$ at fixed $N$),
and call ``history of the Universe'' the ``path'' $N \to \Phi (N)$. 
Notice that $\Phi(N)$, the ``phase space at time $N$'', includes also
tachyonic configurations. 

A property of $\Phi (N)$ is that
$\Phi(N) \supset \Phi(M)$, $\forall ~ M < N$. This is the reason why
we perceive a history basically consisting in a progress toward
increasing time. Higher times bear the memory of the past, lower times.
The opposite is not true,  because ``future'' configurations are not contained
in those at lower, i.e. earlier, times. Indeed, in order to
be able to say that an event $B$ is the follow up of $A$, $A \neq B$ 
(time flow from $A \to B$), at the time we observe $B$ we need to
also know $A$. This precisely means $A \in \Phi(N_A)$ \underline{and}
$A \in \Phi(N_B)$, which implies $\Phi(N_A) \subset \Phi(N_B)$.
Time reversal is not a symmetry of the system \footnote{Only by restricting to
some subsets of physical phenomena one can approximate the description 
with a model symmetric under reversal of the time coordinate, at the price
of neglecting what happens to the environment. As we will
see, this is done at the cost of approximating masses and
couplings to constant values, and thereby giving up with the possibility of a 
higher predictive power of the theory (see also discussion in 
section \ref{detprob}).}.

From \ref{Spm} and \ref{bound3}, namely that the maximal entropy is the one of
three spheres, and scales as $S_{(3)} \sim N^2$, we can derive that
the ratio of the weight of the configurations at time $N-1$,
normalized to the weight at time $N$, is of the order:
\be
W(N-1) \, \approx \, W(N) \, {\rm e}^{-2N} \, .
\ee
At any time, the contribution of past times is therefore negligible as compared
to the one of the configurations at the actual time.

\subsection{The observer}
\label{obsv}

An observer is a subset of space, a ``local inhomogeneity''
(if one thinks a bit about,
this is what after all a person or a device is:
a particular configuration of a portion of space-time!). 
Since we are here talking of space as a finite set of 
cells, that one intuitively is led to visualize in his mind as a 
hyper-segment, one may ask if there are privileged subsets, for instance the
cells close to the border of the segment, or at the center.     
Indeed, attributing a spheric geometry to this space means that the space, 
always ``compact'' because of the finiteness of $N$,
is provided with ``boundary conditions'' such that the borders close-up
to themselves. Let's for concreteness imagine that we are in one dimension.
We have in this case a segment. The configurations of cells
at an extreme of this segment are ``continued'' at the other extreme, 
in such a way that it does not matter whether the segment really
ends at one end, because the opposite end looks like if it was glued
in a way that we can ``continuously'' flow out from one side and enter through
the other side, like in a circle. 
Being really a sphere, or ``looking like'' a sphere 
doesn't make any difference. 

Wherever it is placed, the observer breaks the homogeneity of space.
As such, it \emph{defines} a privileged point, the
observation point. 
The observer is only sensitive to 
its own configuration. He, or it, ``learns'' about the full space only through
its own configurations. For instance, he can perceive that the configurations
of space of which he is built up change with time, and \emph{interprets}
this changes as due to the interaction with an environment.
There is no ``instantaneous'' knowledge: we know about objects placed at a 
certain distance from us only through interactions, light or gravitational
rays, that modify our configuration. But we know that, for instance,
light rays are light rays, because we compare configurations through a certain
interval of time, and we see that these change as according to
an oscillating ``wave'' that ``hits'' our cells. When we talk about energies,
we talk about frequencies. We cannot talk of periods and 
frequencies if we cannot compare configurations at different times.

\subsection{Masses}
\label{masses}

As discussed in section \ref{wavep}, the energies of the inhomogeneities , the
``energy packets'', are inversely 
proportional to their spreading in space: $E \sim 1 / r$.
Indeed, this is strictly so only in the case the configurations constituting
the wave packet exhaust the full spectrum of configurations, Namely, 
let's suppose we have a wave packet spread over 10 cells.
If we have 10 configurations contributing to the ``universe'',  in each
one of which nine cells corresponding to this set are ``empty'',
i.e. of zero energy, and one occupied, with the occupied cells 
occurring of course in a different position 
for each configuration, then we can rigorously say that the energy of the
wave packet is 1/10.
However, at any $N$ the universe consists of an infinite number of 
configurations, which contribute to ``soften'' (or strengthen)
the weight of the wave packet.
A priori, the energy of this wave packet could therefore be lower (higher)
than 1/10.
We are therefore faced with an uncertainty in the value of the mass/energy
of this packet, due to the lack of knowledge of the full spectrum
of configurations. As we already mentioned in section~\ref{eSp}, and will
discuss more in detail in section~\ref{UncP}, this uncertainty is at most
of the order of the mass/energy itself. For the moment, let's therefore accept
that such ``energy packets'' can be introduced, with a precision/stability 
of this order.
According to our definition of time, the volume of space increases with
time. Indeed, it mostly increases as the cubic power of time
(in the already explained sense that the most entropic configuration 
behaves in the average like a three-sphere), while the total energy increases
linearly with time. The energy of the universe therefore ``rarefies''
during the evolution ($\rho(E) \sim 1 / N^2  \sim 1 / {\cal T}^2$).
It is reasonable to expect
that also the distributions in some sense ``rarefy'' and
spread out in space. Namely, that also the sub-volume $\tilde{V}$ in which
the unit-energy cell is confined, and representing
an excitation of energy $1 / \tilde{V}$, spreads out as time goes by.

If the rate of increase of this volume is $d r / dt = 1$, namely, if
at any unit step of increase of time 
${\cal T} \sim N \to {\cal T} + \delta {\cal T} \sim N + 1$
we have a unit-cell increase of space: 
$r \sim n \to r + \delta r \sim n + 1$, the energy of this excitation
``spreads out'' at the same speed of expansion of the universe.
This is what we interpret as the propagation of the fundamental
excitation of a \underline{massless} 
field \footnote{Notice that, in the usual 
formulation, string theory is first defined in a non-compact space-time,
where plane waves are really ``plane'', and the energy of any such wave
is constant in time and can be arbitrarily low. However, in a 
compact space-time, also in string theory the minimal energy of the
plane waves decreases as the volume of space increases.}.

If the region where the unit-energy cell is confined
expands at a lower rate, $dr / dt < 1$, we have, within
the full space of a configuration, a reference frame
which allows us to ``localize'' the region, because we can
remark the difference between its expansion and the expansion of the full
space itself. We perceive therefore this excitation
as ``localized'' in space; its energy, its lowest energy, is always
higher than the energy of a corresponding massless excitation.
In terms of field theory, this is interpreted as the propagation
of a \underline{massive} excitation.

Real objects in general consist of a superposition of ``waves'', 
or excitations, and possess energies higher than the fundamental one.
Nevertheless, the difference between what we call massive and massless
objects lies precisely in the rate of expansion of the region of space
in which their energy is ``confined''.

The appearance of unit-energy cells at larger distance would be 
interpreted as ``disconnected'', belonging to another excitation,
another physical phenomenon; a discontinuity consisting in a ``jump'' by
one (or more) positions in this increasing one-dimensional ``chess-board''
implying a non-minimal jump in entropy. A systematic expansion of the
region at a higher speed is on the other hand what we call a 
\underline{tachyon}.
A tachyon is a (local) configuration of geometry that ``belongs
to the future''. In order for an observer
to interpret the configuration as coming from the future, 
the latter must corresponds to an energy density lower than the present one.
Indeed, also this kind of configurations contribute to the mean values
of the observables. Their contribution is however highly suppressed, as we
will see in section~\ref{UncP}.

\subsection{Mean values and observables}
\label{vev}

The mean value of any (observable) quantity ${\cal O}$
at any time ${\cal T} \sim N$ is the sum of the contributions to ${\cal O}$
over all configurations $\Psi$, weighted according to their volume of 
occupation (their geometric occupation) in the
phase space:
\be
< {\cal O} > \, \propto \, \int_{\cal T} W (\Psi)\, {\cal O}(\Psi) 
\, d \Psi \, .
\ee
We have written the symbol $\propto$ instead of $=$ because,
as it is, the sum on the r.h.s. is not normalized. 
The weights don't sum up to 1, and not even do they sum up to a finite
number: in the infinite volume limit, they all diverge. However, as we 
discussed in section~\ref{ddf}, what matters is their 
relative ratio, which is finite because the infinite volume factor
is factored out. In order to normalize mean values,
we introduce a functional that works as ``partition
function'', or ``generating function'' of the Universe:
\be
{\cal Z} \, \stackrel{\rm def}{=} \int_{\cal T} W(\psi) \, {\cal D} \Psi  
\, = \, \int_{\cal T} {\rm e}^{S(\Psi)} \, {\cal D} \Psi \, ,
\label{ZPsi}
\ee
where ${\cal D} \Psi$ precisely means the sum over all possible configurations
$\Psi$. The sum has to be intended as always performed at fine volume.
In order to define mean values and observables,
we must in fact always think in terms of finite space volume, 
a regularization condition to be eventually relaxed. 
The mean value of an observable can then be written as:
\be
< {\cal O} > \, \stackrel{\rm def}{\equiv} \, {1 \over {\cal Z}}
\int_{\cal T} W (\Psi)\, {\cal O}(\Psi) \, d \Psi \, .
\label{meanO}
\ee
Mean values therefore are not defined in an absolute way, 
but through an averaging procedure in which
the weight is normalized to the total weight of all the configurations,
at any finite space volume $V$.

From the property stated at page~\pageref{noneq} 
that at any time ${\cal T} \sim N$
there do not exist two inequivalent
configurations with the same entropy, and from the fact that less
entropic configurations possess a lower degree of symmetry, we can already
state that:
\begin{itemize}
\item 
\emph{At any time ${\cal T}$ the average appearance of the universe is that
of a space in which \underline{all} \underline{the symmetries are broken.}}  
\end{itemize}
\noindent
The amount of the breaking, depending on the weight of non-symmetric
configurations as compared to the maximally symmetric one, involves
a relation between the energy (i.e. the geometry deformation) and the time
spread/space length, of the space-time deformation, as it will be
discussed in the next section.

\section{The Uncertainty Principle}
\label{UncP}

According to \ref{meanO}, the mean values of the observables 
do not receive contribution
only from the configurations of extremal or near to extremal entropy: 
all the possible configurations at a certain time
contribute. There is therefore an uncertainty in the value of the energy 
due to the lack of exact knowledge of all the terms contributing to
the sum \ref{meanO}.

Let's consider the contribution to the ``vacuum energy'' of the
neglected configurations. 
In order to see what is the amount of approximation we are introducing
when considering just the maximal entropy configurations, we can 
proceed by considering that non-extremal configurations correspond to 
un-freezing degrees of freedom, which parametrize the deviation
from the extremal entropy, due to different dimensionalities and
combinatorics within them.
This results in a decrease of the volume occupied in the phase space.
In full generality,
we can therefore account for the contribution of the extra
degrees of freedom to the ``partition function'' \ref {ZPsi}
by summing and integrating
over an infinite series of ``extra-coordinates'', which reduce 
the maximal entropy. According to \ref{Spm}, and considering that
there are no non-equivalent configurations with the same entropy, 
we can write the full contribution as:
\be
{\cal Z} ~~ \gsim ~~ \sum_{n=1} 
\int d^n L \; 
{\rm e}^{S_0 \left[1 -  L_1 \times \ldots \times L_n \right]} \, ,
\label{Zall}
\ee
where $S_0$ is the entropy of the three-sphere.
The extra terms give here the deviation with respect to the entropy 
due to the un-freezing of infinitely many coordinates 
$i \in \{1,\ldots,n  \}$, $n \in N$, from size one: $1 \to L_i$.
Any contribution is integrated over the entire axis of possible values:
$1 \leq L_i \leq \infty$, and we sum over all possible configurations,
containing an arbitrary number of such coordinates.
Notice that, leaving open the number of these, $n \in N$, we include here also
the degrees of freedom parametrizing the geometry of the spaces described
by these coordinates.  Of course, there cannot be
weights lower than 1, as it would instead seem to happen from 
expression~\ref{Zall}: when writing expressions like the above, 
we have in mind the eventual computation of mean values, as according 
to~\ref{meanO}, and therefore always 
intend to refer to a normalized result. We speak therefore of relative weights.
Expression \ref{Zall} can be integrated and gives:
\be
{\cal Z} ~~ \gsim ~~ {\rm e}^{S_0} \sum_n {1 \over S_0^n } \, = 
\, {\rm e}^{S_0} \left( 1 + {1 \over S_0 -1} \right) \, .
\label{Zallint}
\ee  
This result tells us that the contribution of non-extremal configurations,
accounted in the second term of the sum on the r.h.s.,
is highly suppressed as compared to the one of the configuration of maximal
entropy. Indeed, as soon as we move even just one cell out of the configuration
of maximal entropy, we loose (powers of) $N$ in combinatoric factors 
contributing to the weight of the configuration.

In section~\ref{timev} we have established the correspondence 
between the ``energy'' $N$ and the ``time'' coordinate that orders the history
of our ``universe''. Since the distribution of the $N$ degrees of 
freedom basically determines the curvature of space, it is quite right 
to identify it with our concept of energy, as we intend it after the 
Einstein's General Relativity equations. 
However, this may not coincide with the
\emph{operational} way we define energy, related to the way we measure it.
Indeed, as it is, $N$ simply reflects the ``time'' coordinate, and coincides 
with the global energy of the universe, proportional to the time. 
From a practical point of view, what we
measure are curvatures, i.e. (local) modifications of the geometry, and we 
refer them to an ``energy content''. An exact measurement of energy
therefore means that we exactly measure the geometry and its 
variations/modifications within a certain interval of time. 
On the other hand, 
we have also discussed that, even at large $N$, 
not all the configurations of the universe at time $N$ admit an interpretation
in terms of geometry, as we normally intend it. The universe as we see it
is the result of a superposition in which also very singular configurations 
contribute, in general uninterpretable within the usual conceptual framework 
of particles, or wave-packets, and so on.    
When we measure an energy, or equivalently a ``geometric curvature'', 
we refer therefore
to an average and approximated concept, for which we consider only a subset of
all the configurations of the universe. Now, we have seen that the 
larger is the ``time'' $N$, the higher is the dominance of the most probable
configuration over the other ones, and therefore more picked is the average,
the ``mean value'' of geometry. The error in the evaluation of the
energy content will therefore be the more reduced, 
the larger is the time spread we consider, 
because relatively lower becomes the weight of the configurations we ignore.
From \ref{Zallint} we can have an idea of what is the order of the 
uncertainty in the evaluation of energy.
According to \ref{Zall}
and \ref{Zallint}, the mean value of the total energy, receiving contribution 
also from all the other configurations, results to be ``smeared'' by an amount:
\be
< E  > ~ \approx ~ E_{S_0} \, + \, E_{S_0} \times \, {\cal O} 
\left( 1 / S_0  \right) \, .
\label{Emean} 
\ee  
That means, inserting $S_0 \approx N^2 \equiv t^2 \sim \, E^2_{S_0}$:
\be
< E  > ~ \approx ~ E_{S_0} \, + \, \Delta E_{S_0} ~ \approx ~ 
E_{S_0} \, + \, {\cal O} \left( {1 \over t} \right) 
\, .
\label{deltaEmean} 
\ee  
Consider now a subregion of the universe, of extension 
$\Delta t$ \footnote{We didn't yet introduce units distinguishing between 
space and time. In the usual language we could consider this region as being
of ``light-extension'' $\Delta x = c \Delta t$.}. 
Whatever exists in it, namely,
whatever differentiates this region from the uniform spherical ground geometry
of the universe, must correspond to a superposition of 
configurations of non-maximal entropy. 
From our considerations of above, we can derive that it is not possible to
know the energy of this subregion with an uncertainty lower than the inverse
of its extension. In fact, let's 
see what is the amount of the contribution to this energy given by the
sea of non-maximal, even ``un-defined'' configurations. As discussed,
these include higher and lower space dimensionalities, and any other kind
of differently interpretable combinatorics. 
The mean energy will be given as in \ref{Emean}.
However, this time the maximal entropy
$\tilde{S}_0(\Delta t)$ of this subsystem will be lower than the upper bound
constituted by the maximal possible entropy of a region enclosed in a time 
$\Delta t$, namely the one of a three-sphere of radius $\Delta t$:
\be
\tilde{S}_0(\Delta t) ~ < ~ \left[ \Delta t \right]^2 \, ,
\label{Stilde}
\ee 
and the correction corresponding to the
second term in the r.h.s. of \ref{deltaEmean} will just constitute a lower 
bound to the energy uncertainty \footnote{The maximal energy can be 
$E \sim \Delta t$ even for a class of non-maximal-entropy, non-spheric 
configurations.}:
\be
\Delta E ~~ \gsim ~~ {{\Delta t} \over S_0 (\Delta t)} \, ~~
\approx \, {1 \over \Delta t } \, . 
\label{Etuncertainty}
\ee
In other words, \emph{no region of extension $\Delta t$
can be stated to possess an energy lower than $1/ \Delta t$}.
When we say that we have measured
a mass/energy of a particle, we mean that we have measured 
an average fluctuation of the configuration of the universe around the 
observer, during a certain time interval. 
This measurement is basically a process that takes place along the time 
coordinate. 
As also discussed also Ref.~\cite{spi}, during the time of the ``experiment'', 
$\Delta t$, a small ``universe'' opens up for this particle. Namely, what we 
are probing are the configurations of a space region created in a time 
$\Delta t$. According to \ref{Etuncertainty}, the particle
possesses therefore a ``ground''
indeterminacy in its energy:
\be
\Delta E \, \Delta t ~\gsim ~ 1\, . 
\label{HUP}
\ee
As a bound, this looks quite like the time-energy Heisenberg
uncertainty relation. From an historical point of view, 
we are used to see
the Heisenberg inequality as a ground relation of Quantum Mechanics, ``tuned'' 
by the value of $\hbar$. Here it appears instead as a ``macroscopic relation'',
and any relation to the true Heisenberg's uncertainty looks only formal. 
Indeed, as I did already mention, we have not yet introduced units in which
to measure, and therefore physically distinguish, space and time, and
energy from time, and therefore also momentum. Here we have for the
moment only cells and distributions of cells. However, 
one can already look through where we are getting to: it is not difficult to
recognize that the
whole contruction provides us with the basic formal structures we need
in order to describe our world. Endowing it
with a concrete physical meaning will just be a matter of appropriately
interpreting these structures.
In particular, the introduction of $\hbar$ will just be a matter of
introducing units enabling to measure energies in terms of time (see discussion
in section~\ref{stringT}).

In the case we consider the whole Universe itself,
expression \ref{Zallint}
tells us that the terms neglected in the partition function,
due to our ignorance of the ``sea'' of all the possible configurations
at any fixed time, contribute to an ``uncertainty''
in the total energy of the same order as the inverse of the age of the 
Universe:
\be
\Delta E_{\rm tot} \, \sim \, {\cal O} \left( 1 \over {\cal T}  \right)
\, .
\ee
Namely, an uncertainty of the same order as the imprecision due
to the bound on the size of the minimal energy steps at time ${\cal T}$.

$1/ S_0 \sim 1 / {\cal T}^2$ basically corresponds to 
the parameter usually called ``cosmological constant'', 
that in this scenario is not constant. 
The cosmological constant therefore 
not only is related to the size of the energy/matter density of the 
Universe (see Ref.~\cite{spi}), setting thereby the minimal measurable ``step''
of the actual Universe, related to the Uncertainty Principle \footnote{See also
Refs.~\cite{estring,lambda}.},
but also corresponds to a bound on the effective precision of calculation
of the predictions of this theoretical scenario.
Theoretical and experimental uncertainties are therefore of the same order.
There is nothing to be surprised that things are like that: this is 
the statement that the limit/bound to an experimental access to the
Universe as we know it corresponds to the limit within which such a
Universe is in itself defined. Beyond this threshold, there is a ``sea''
of configurations in which  
i) the dimensionality of space is not fixed;
ii) interactions are not defined, iii) there are tachyonic contributions,
causality does not exist etc...
beyond this threshold there is a sea of...uninterpretable 
combinatorics.

\begin{itemize}
\item \emph{It is not possible to go beyond the Uncertainty 
Principle's bound with the precision in the measurements, because this bound 
corresponds to the precision with which the quantities to be measured 
themselves are defined}.   
\end{itemize}

\section{Deterministic or probabilistic physics?}
\label{detprob}

We have seen that masses and energies are
obtained from the superposition, with different weight, of configurations
attributing unit-energy cells to different positions, that concur
to build up what we usually call a ``wave packet''. Unit energies appear
therefore ``smeared out'' over extended space/time regions.
The relation between energies and space extensions is of the type
of the Heisenberg's uncertainty.
Strictly speaking, in our case there is no uncertainty: in themselves,
all the configurations of the superposition
are something well defined and, in principle,
determinable. There is however also a true
uncertainty: in sections \ref{eSp} and \ref{masses} 
we have seen that to the appearance of the Universe, and therefore to the 
``mean value'' of observables, contribute also higher and lower than three
dimensional space configurations, as well as tachyonic ones. In 
section~\ref{UncP} we have also seen how,
at any ``time'' $N$, all ``non-maximal''
configurations sum up to contribute to the geometry of space by an
amount of the order of the Heisenberg's Uncertainty. 
This is more like what we intend
as a real uncertainty, because it involves the very possibility of
defining observables and interpret observations according to 
geometry, fields and particles.
The usual quantum mechanics relates on the other hand the concept
of uncertainty with the one of probability: the ``waves'' (the set of
simple-geometry configurations which are used as bricks for building the
physical objects) are interpreted as ``probability waves'',
the decay amplitudes are ``probability amplitudes'', which allow to
state the probability of obtaining a certain result when making
a certain experiment. In our scenario, there seems to be no room
for such a kind of ``playing dice'': everything looks well determined.
Where does this aspect come from, if any, namely
where does the ``probabilistic'' nature
of the equations of motion originates from and what is its meaning in our
framework?

\subsection{A ``\sl{Gedankenexperiment}''}
\label{gedank}

Let's consider a simple, concrete example of such a situation.
Let's consider the case of a particle (an ``electron'')
that scatters through a double slit. This is perhaps the 
example in which classical/quantum effects manifest their peculiarities
in the most emblematic way, and where at best the deterministic vs. 
probabilistic nature of time evolution can be discussed.

\begin{figure}
\centerline{
\epsfxsize=4cm
\epsfbox{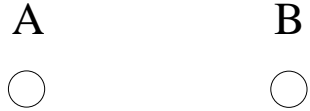}
}
\vspace{0.3cm}
\caption{A and B indicate two points of space-time, 
symmetric under reflection or 180$^0$ rotation.
They may represent the positions on a target place where light, 
or an electron beam, scattering through a double slit, can hit.}     
\label{exmp1}
\end{figure}

\begin{figure}
\centerline{
\epsfxsize=4cm
\epsfbox{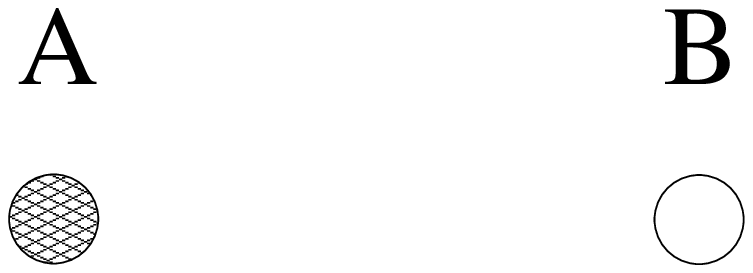}
}
\vspace{0.3cm}
\caption{Point A is marked by some property that distinguishes it from point
B.}     
\label{exmp2}
\vspace{0.5cm}
\centerline{
\epsfxsize=4cm
\epsfbox{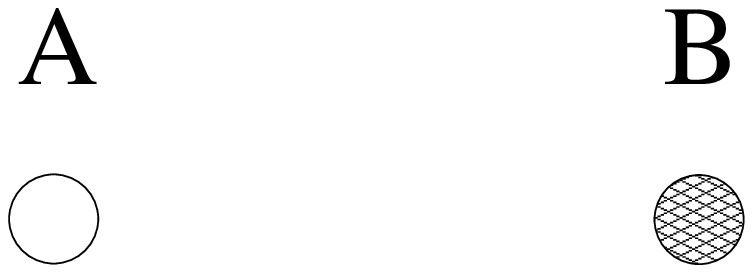}
}
\vspace{0.3cm}
\caption{The situation symmetric to figure~\ref{exmp2}.}     
\label{exmp3}
\end{figure}

As is known, it is possible to carry out 
the experiment by letting the electrons
to pass through the slit only one at once. In this case, each electron
hits the plate in an unpredictable position, but in a way that as time
goes by and more and
more electrons pass through the double slit, they build up the 
interference pattern typical of a light beam. This fact is therefore advocated 
as an example of probabilistic dynamics: we have a problem with a symmetry
(the circular and radial symmetry of the target plate, 
the symmetry between the two holes of the intermediate plate, etc...);
from an ideal point of view, in the ideal, abstract world in which formulae
and equations live, the dynamics of the single
scattering looks therefore absolutely unpredictable, although in the whole
probabilistic, statistically predictable \footnote{The 
probabilistic/statistical interpretation comes
together with a full bunch of related problems. For instance, the fact that
if a priori the probability of the points of the target plate 
corresponding to the 
interference pattern to be hit has a circular symmetry, as a matter of fact
once the first electron has hit the plate, there must be a higher probability 
to be hit for the remaining points, otherwise the interference pattern would
come out asymmetrical. These are subtleties that can be theoretically solved
for practical, experimental purposes in various ways, 
but the basic of the question remains, 
and continues to induce theorists and philosophers to come back to the 
problem and propose
new ways out (for instance K. Popper and his ``world of propensities'').}.    
Let's see how this problem looks in our theoretical framework.
Schematically, the key ingredients of the situation can be 
summarized in figure~\ref{exmp1}. 
This is an example of ``degenerate vacuum''
of the type we want to discuss. Points A and B are absolutely 
indistinguishable, and, from an ideal point of view, we can perform a 180$^0$
rotation and obtain exactly the same physical situation. 
As long as this symmetry exists, namely, as long as the \emph{whole universe},
including the observer, is symmetric under this operation, there is
no way to distinguish these two situations, the configuration and
the rotated one: they appear as only one configuration, weighting 
twice as much. Think now that A and B represent 
two radially symmetric points in the target plate of the double slit 
experiment. Let's
mark the point A as the point where the first electron 
hits. We represent the situation in which we have distinguished the properties
of point A from point B by shadowing the circle A, figure~\ref{exmp2}.
Figure~\ref{exmp3} would have been an equivalent choice. 
Indeed, since \emph{everything else} in the Universe
is symmetric under 180$^0$ rotation,
figure~\ref{exmp2} and \ref{exmp3} represent \emph{the same} vacuum,
because \emph{nothing} enables to distinguish between figure~\ref{exmp2}
and figure~\ref{exmp3}. 

\begin{figure}
\centerline{
\epsfxsize=4cm
\epsfbox{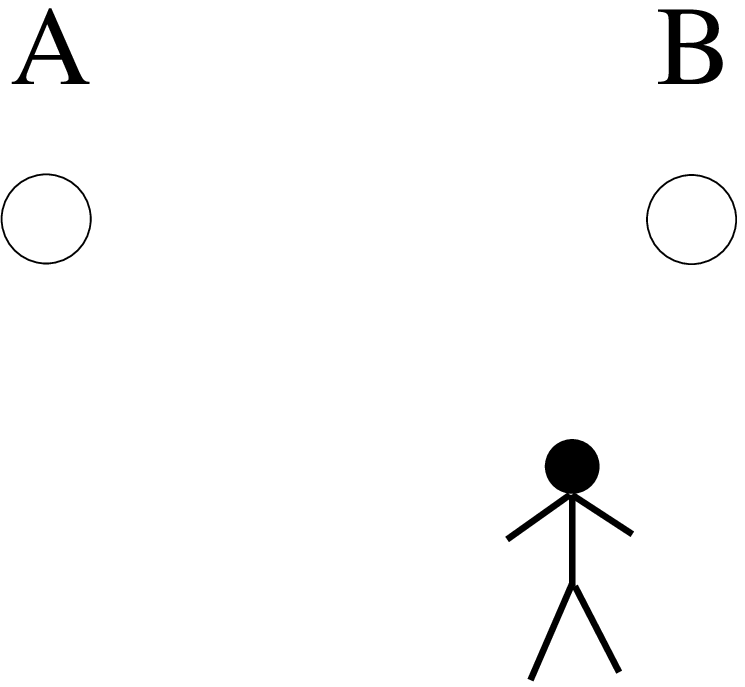}
}
\vspace{0.3cm}
\caption{The presence/existence of the observer
breaks the symmetry of the physical configuration
under 180$^0$ rotation.}     
\label{exmp4}
\centerline{
\epsfxsize=4cm
\epsfbox{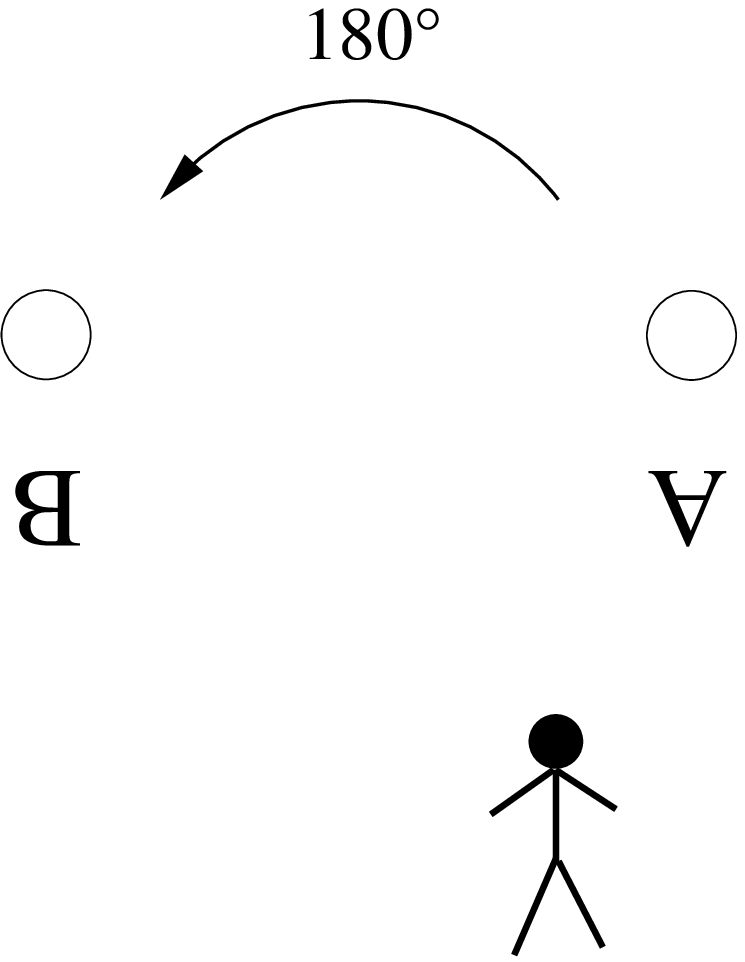}
}
\vspace{0.3cm}
\caption{The observer does not rotate. Now the rotated situation
is not equivalent to the previous one.}     
\label{exmp5}
\end{figure}

\begin{figure}
\centerline{
\epsfxsize=4cm
\epsfbox{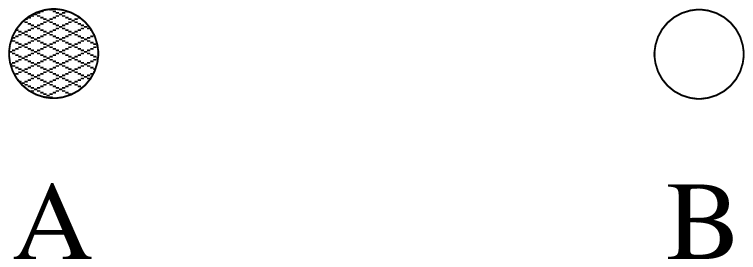}
}
\vspace{0.3cm}
\caption{The presence of an observer able
to detect a motion according to a group action is something that breaks 
the symmetry of the Universe under this group, otherwise the
action would not be detectable. Here we represent the observer as
something that distinguishes A from B.}     
\label{exmp6}
\end{figure}

\begin{figure}
\centerline{
\epsfxsize=8cm
\epsfbox{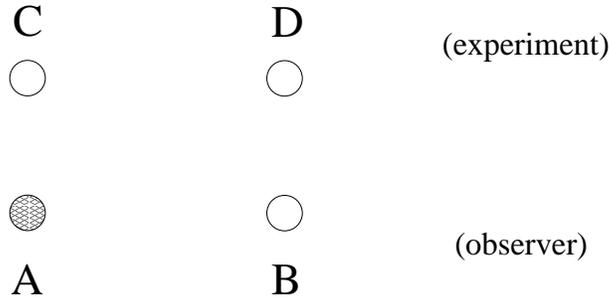}
}
\vspace{0.3cm}
\caption{In the presence of an observer, here represented by the points
A and B, even with a ``symmetric'' system, the points C, D, the Universe
is no more symmetric. Points C and D can be identified, by saying
that C is the one closer to A, D the one closer to B.}     
\label{exmp7}
\end{figure}

\begin{figure}
\centerline{
\epsfxsize=8cm
\epsfbox{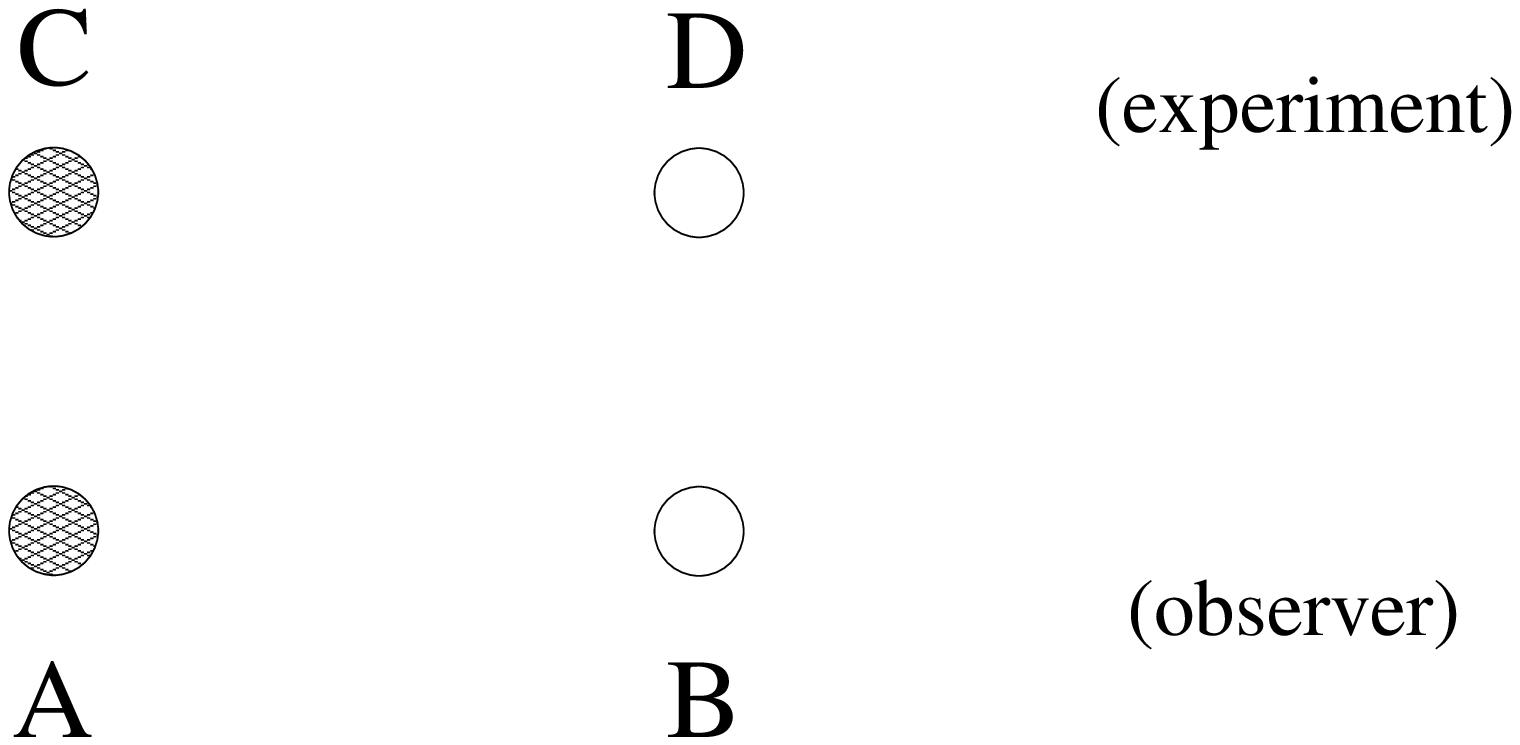}
}
\vspace{0.3cm}
\caption{The analogous of figure~\ref{exmp2} in the presence of an 
observer.}     
\label{exmp8}
\vspace{1cm}
\centerline{
\epsfxsize=8cm
\epsfbox{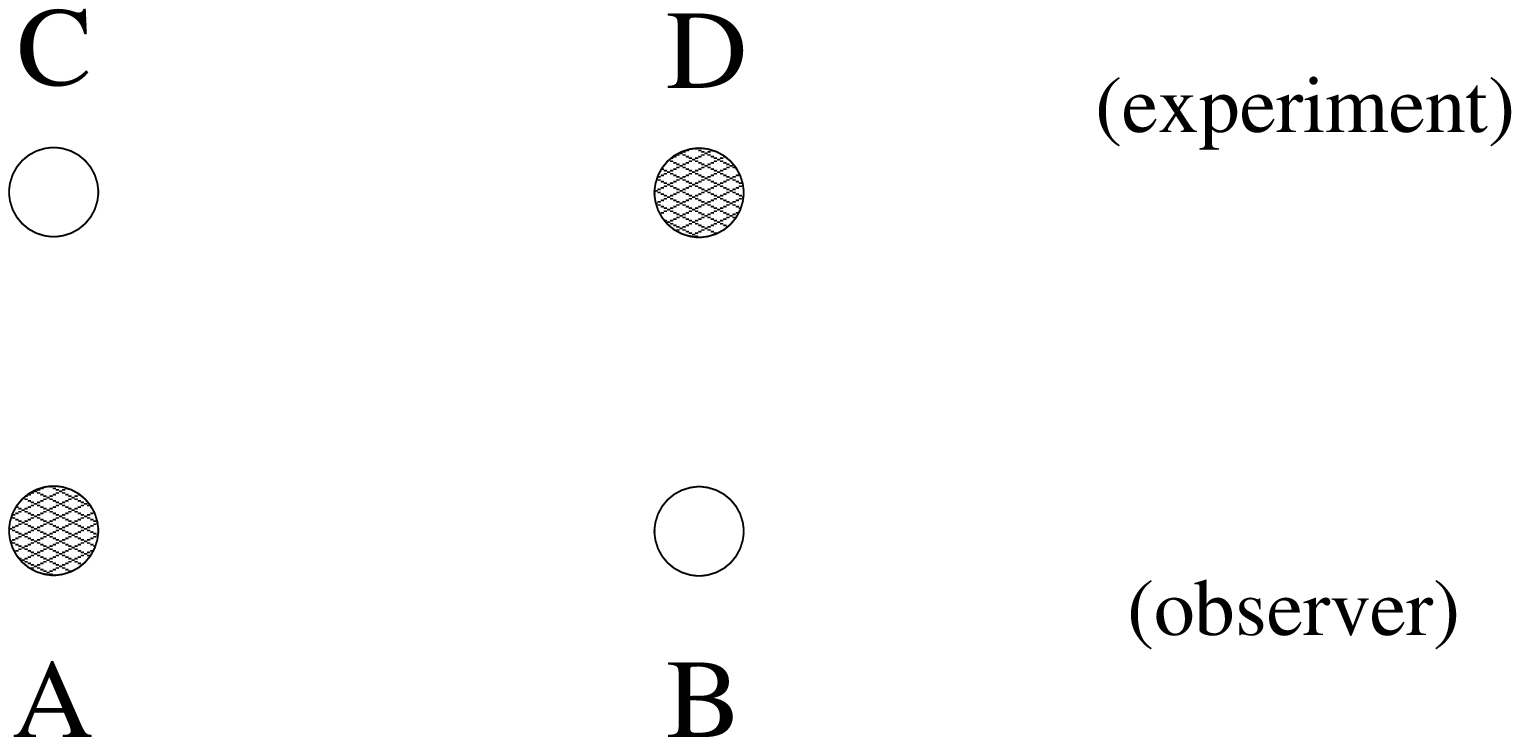}
}
\vspace{0.3cm}
\caption{The analogous of figure~\ref{exmp3} in the presence of an 
observer.}     
\label{exmp9}
\end{figure}

As we discussed in section~\ref{vev}, in our framework in the
universe all symmetries are broken. This matches with the fact that  
in \underline{any real experiment}, the environment doesn't possess
the ideal symmetry of our {\sl Gedankenexperiment}. For instance,
the target plate \emph{in} the environment, 
\emph{and the environment itself}, don't possess
a symmetry under rotation by 180$^0$: the presence of an ``observer''
allows to distinguish the two situations, as illustrated in figures~\ref{exmp4}
and \ref{exmp5}. 
There is therefore a choice which 
corresponds to the maximum of entropy.  
The \underline{real} situation can be schematically depicted 
as follows. The ``empty space'' is something like in figure~\ref{exmp6},
in which the two dots, distinguished by the shadowing, represent
the observer, i.e. not only ``the person who observes'', but more crucially 
``the object (person or device) which can distinguish between configurations''.
Now we add the experiment, figure~\ref{exmp7}.
In this case, the previous figures~\ref{exmp2} and \ref{exmp3}
correspond to figures~\ref{exmp8} and \ref{exmp9}.
It should be clear that entropy in the configuration of 
figure~\ref{exmp8} is not the same
as in the configuration of figure~\ref{exmp9}.
This means that the observer ``breaks the symmetries'' in the Universe, 
it \emph{decides} that \emph{this one}, namely figure~\ref{exmp6}, 
is the actual configuration of the Universe, i.e. the one contributing
with the highest weight to the appearance of the Universe, 
while the one obtained by exchanging A and B is not.  

The observer is itself part of the Universe, and the symmetric situation
of the ideal problem of the double slit is only an abstraction.
In our approach, it is the very presence of an observer, i.e. of an
asymmetrical configuration of space geometry, 
what removes the degeneracy of the 
physical configurations, thereby solving the paradox of equivalent 
probabilities of ordinary quantum mechanics. In this
perspective there are indeed no ``probabilities'' at all: the Universe
is the superposition of configurations in the same sense as wave packets
are superpositions of elementary (e.g. plane) waves; real waves,
not ``probability wave functions''.
This means also that mean values, given by \ref{meanO}, are 
sufficiently ``picked'', so that
the Universe doesn't look so ``fuzzy'', as it would if rather
different configurations contributed with a similar weight. 
Indeed, the fuzziness due to
a small change in the configuration, leading to a smearing out
of the energy/curvature distribution around a space region, 
corresponds to the Heisenberg's uncertainty, section~\ref{UncP}.
The two points on the target plate correspond to
a deeply distinguished asset of the energy distribution, the curvature
of space, whose distinction
is well above the Heisenberg's uncertainty.

When objects, i.e. special configurations of space
and curvature, are disentangled beyond the ``Heisenberg's scale'',
``randomness'' and
``unpredictability'' are rather a matter of the infinite number
of variables/degrees of freedom which concur to
determine a configuration, i.e., seen from a dynamical point of view,
``the path of mean configurations'', their time evolution.
In itself, this Universe is though deterministic. Or, to better say,
``determined''. ``Determined'' is a better expression, 
because the Universe at time
$N^{\prime} \sim {\cal T}^{\prime} = {\cal T} + \delta {\cal T} \sim N +1$ 
cannot be obtained by running forward the configurations at time 
$N \sim {\cal T}$.
The Universe at time ${\cal T} + \delta {\cal T}$ is not the 
``continuation'', obtained through equations of motion,
of the configuration at time ${\cal T}$; it is given
by the weighted sum of all the configurations at time
${\cal T} + \delta {\cal T}$, as the universe at time ${\cal T}$ was
given by the weighted sum of all the configurations at time ${\cal T}$. 
In the large $N$ limit,
we can speak of ``continuous time evolution'' only 
in the sense that for a small change of time, the dominant
configurations correspond to distributions
of geometries that don't differ that much from those at previous time.
With a certain approximation we can therefore speak of evolution in 
the ordinary sense of (differential, or difference) time equations. 
Strictly speaking, however, initial conditions don't determine the future.

Being able to predict the details of an event, such
as for instance the precise position each electron will hit on the plate,
and in which sequence, requires to know the function
``entropy'' for an infinite number of configurations,
corresponding to any space dimensionality at fixed ${\cal T} \approx N$,
for any time ${\cal T}$ the experiment runs on.
Clearly, no computer or human being can do that. 
If on the other hand we content ourselves with an approximate predictive 
power, we can roughly reduce physical situations to certain ideal schemes, 
such as for instance ``the symmetric double slit'' problem. 
Of course, from a theoretical point of view we lose  
the possibility of predicting the position 
the first electron will hit the target (something anyway practically 
impossible to do), but we gain, at the
price of introducing symmetries and therefore also concepts like
``probability amplitudes'', the capability of predicting with a good
degree of precision the shape an entire beam of electrons will draw on the
plate. We give up with the ``shortest scale'', and we concern ourselves
only with an ``intermediate scale'', larger than the point-like one,
shorter than the full history of the Universe itself. 
The interference pattern arises as the dominant mean configuration,
as seen through the rough lens of this ``intermediate'' scale.

\section{String Theory}
\label{stringT}

Till now, we have spoken of ``light rays'', ``gravitational fields'',
radius, curvature, in one word
we have made an extensive use of the language of geometry and field
theory, in order to make more concrete the discussion, but, despite of the 
language, we have always worked in a discrete formulation of the combinatoric
problem.
Indeed, for $N$ sufficiently large, it is not only possible but convenient to
map to a continuous description, in that this not only makes things easier
from a computational point of view, but also better corresponds
to the way the physical world shows up to us, or, more precisely, to
the interpretation we are used to give of it.

If we want to pass to a description in terms of continuous variables,
we must introduce a ``length'' to use as a measure: 
in the continuum, lengths must be measured in terms of a given unit.
Differently from the discrete formulation, in which all quantities:
the extension of space, the amount of ``energy'', the ``time'', could be
measured in terms of ``number of cells'', in the continuum
we must a priori introduce a distinguished unit of measure for any 
type of measurable quantity. 
To start with, we must introduce a unit of length, that we call
$\ell$. This not only serves as a measure, but it can be chosen to coincide with
the elementary size, the radius of the unit cell. In this way, we introduce
what we call the ``Planck length'', $\ell_{\rm Pl}$.

Energies and momenta 
are conjugate to space lengths, relation~\ref{Er}, and the natural unit in 
which they are measured is the inverse of the Planck length. This leads
to the introduction of the Planck Mass $m_{\rm Pl}$ and the unit 
of conversion between the energy/momentum and space/time
scale, the Planck constant $\hbar$ according to the relation:
\be
\left[E,P   \right] \, \sim \, {1 \over \left[ R, t  \right]} ~ \leadsto
~ m_{\rm Pl } \, \stackrel{\rm def}{\equiv} \, 
{\hbar \over \ell_{\rm Pl}} \, . 
\ee
This corresponds to the usual relation between these quantities, apart
from the fact that here doesn't appear any power of the ``speed of light''.
In fact, till now we have paired the concepts of energy/momentum/mass because
we have not yet distinguished the unit of measure of time from the one of 
space. Indeed, were all the objects either massless, or permanently at rest,
this distinction would be unnecessary. We need to disentangle time from space
in order to measure the rate of expansion of objects, ``inhomogeneities''
in the average geometry of space, as compared to the rate of the expansion
of the space itself. As discussed,
non-trivial massive objects correspond to subregions that spread out
at particular rates, giving therefore rise to a full spectrum of 
non-trivial ``speeds''. We measure these speeds in terms of $c$,
the rate of expansion of the radius of the three-sphere with respect
to $N$, intended as the time. In section~\ref{cexp} we will discuss
how this can be identified with the ``speed of light in the vacuum''.
Obviously, the formulation in terms of discrete numbers and combinatorics
corresponds to a choice of units 
for which all these ``fundamental constants'' are 1.   

From the perspective of a theory on the continuum,
in themselves these scales could be considered as free parameters.
One could think to be forced to introduce them as regulators,
but that in principle they are free to take any 
possible value. However, being $\ell$ the unit in which the length
of space is measured, by varying it one varies the ``unit of volume'', 
or equivalently ``the size of the point'', $v$. When considering the full
span of volumes $V$ we obtain a series of equivalent sets describing the same 
system, equivalent histories of the Universe. 
Running in the set 
$\{V /v  \}$ by letting both $V$ and $v$ take any possible value results
in a redundancy reproducing an infinite number of times
the same situation. Similar arguments hold for the Planck constant
$\hbar$ and the speed of expansion $c$. In particular,
fixing the speed of expansion to a constant $c$
allows to establish a bijective map between the time $t$ and
the volume $V$ of the three-spheres.  Varying the map $t \to V $
through a change of $c$ would lead to an over-counting in the
``history of the Universe''
\footnote{Also introducing a space dependence $c = c(\vec{X})$ would be 
a nonsense, because the functional \ref{ZPsi} always gives
the Universe as it appears at the point of the observer, the 
``present-time point'', say $\vec{X}_0$. 
Saying that $c(\vec{X}^{\prime}) \neq c(\vec{X}_0) $ would be like
saying that volumes appear at $\vec{X}^{\prime}$ 
differently scaled than
how they appear at $\vec{X}_0$: this is a matter of properly
reducing observables to the point of the observer, through a rescaling.}. 
In summary, we would have \underline{classes} of Universes, 
parametrized by the values of $c$ and $\ell_{\rm Pl}$. 
The real, effective phase space is therefore the coset:
\be
[{\rm eff.~phase~space}] \, = \, [{\rm phase~space}] 
\Big/ \{ c \} ,
\{ \ell_{\rm Pl} \} \, ,
\ee 
and, more in general:
\be
[{\rm eff.~phase~space}] \, = \, [{\rm phase~space}] \Big/ 
\{ {\rm symmetries} \} \, .
\ee 
When passing to the continuum, at large $N$, we must therefore look for
a mapping of the combinatoric problem to a description in terms of continuous
geometry, which i) contains as built-in the notion of minimal length,
finite speed of propagation of informations, i.e. locality of physics,
in which ii) energies are related to space extensions through relations
such as \ref{Er}, and in which iii) the ``evolution'' is labelled through
a correspondence between configurations and a parameter that we call ``time''.
Through the relation between this parameter and the curvature, or the 
``radius'' of the maximally entropic configurations, this parameter too
can be viewed as a coordinate. Measuring also time in terms of unit cells,
the same units we use to measure the space,
corresponds to fixing the ``speed of expansion'' to 1 (later on we will
see how this can be seen as the ``speed of light'').

In section~\ref{detprob} we discussed how, for
the practical purpose of reducing the combinatoric problem
of the configuration of the Universe to a ``humanly solvable'' level,
it is somehow necessary to introduce simplifications, which lead,
up to a certain extent, to some degree of indeterminacy.
String Theory arises as one such an approach to the problem: 
it is a way of representing a theory in which
the minimal length is not zero, the ``point'' has size $\ell_{\rm Pl}$,
and whose ``space-time'' indeed possesses a ``time coordinate'', which
appears built-in, although singled out from the other coordinates by
a Minkowskian choice of signature of the metric.
To be more precise, the Planck length
is the length of the ``duality invariant, unified''
string theory, otherwise also called ``M-theory''. The single
perturbative string constructions have their own ``minimal length''
$\ell_{s}$, differently related to $\ell_{\rm Pl}$, according to the size and
configuration of the slice of the string space they correspond to. 
Quantization of the string modes, in terms of raising and lowering operators,
realizes a viable implementation of the quantum, probabilistic approximation
of the description of physical phenomena, as discussed in the previous section.
In practice, it reduces to a geometric problem the ``central'' value of
observables, while implementing through quantization a way of
dealing with the deviations from the ``classical geometric solution'',
and from the definability itself of observable quantities in terms
of geometry and propagating fields.

In how many dimensions should we expect to be able to construct such a 
theory? Relativity and locality tell us that we need spinorial and vectorial
degrees of freedom. Namely, that we must build the space as a spinorial
space. If we want to describe
the corresponding degrees of freedom through a fiber of
``extra'' vectorial coordinates, we need therefore
twice as many coordinates as the dimensions of the space.
We have seen that the dominant configuration of space is three.
This is related to a specific choice of the scaling of energy $\sim N$
as compared to the scaling of space, which appeared as the natural one.
String theory is built as a geometric theory endowed with a quantization
principle, which implements the Heisenberg's Uncertainty Relations, under which
all ``non-dominant'' configurations are ``covered''.
In our set up, the space/momentum and/or time/energy relations appear
as in the Heisenberg's inequalities. We expect therefore that \emph{quantum}
string theory should be non-anomalous when built out of a number
of coordinates which allows a correspondence of these descriptions. 

Let's clarify this point.
On the ``combinatoric side'' we ``center'' the theory
around three dimensions, 
that indeed are 3+1, because to the three of space we must add the time,
or equivalently, the curvature. Correspondingly, quantum string theory 
should be something that needs 12 coordinates for its complete description. 
However, one of these
should be a ``curvature''. Projecting on a flat limit should allow a
description in terms of 11 coordinates. 
Indeed, a perturbative description would
require to single out one of these coordinates, 
to be used as the parameter, the ``coupling'' around which to expand. 
In practice, we should expect to be
able to construct a perturbative slice of such a geometric representation 
of our combinatoric problem with 10 space-time coordinates. This should be
the dimension in which the theory is perturbatively non-anomalous.
Indeed, this is the critical dimension of perturbative superstring theory.
Supersymmetry is needed in order to perturbatively introduce spinors
besides vectors \footnote{One can construct the bosonic string in a higher
number of dimensions; this however is just a rephrasing of the problem,
in which the fermionic degrees of freedom are mapped to bosonic ones.}.
However, on the combinatoric side, three space dimensions
are only the ``dominant'' choice. On the string side, this appears as the 
necessary dimensionality of the ``base'' only once this is put in relation
to the Uncertainty Relations. Namely, this procedure encodes a choice of
``starting point'' for the approximation of the configurations, three
space dimensions, plus a rule implementing the ignorance due to neglecting
the rest, the Uncertainty Relations. In this way, it is only upon 
quantization that string theory shows out an anomaly when built on other
numbers of coordinates. Alternatively, one could think to choose a different
relation between ``energy'' $N$ and radius, or time. We would then get
a different uncertainty relation, of the type:
\be
(\Delta X)^{\alpha} \Delta p \, \geq \, {1 \over 2} \hbar \, ,
\ee    
where $\alpha$ is an exponent, $\alpha \neq 1$. In this case,
quantum string theory would be non-anomalous in a different number
of dimensions.

Our argument is not a proof that the critical dimension of the superstring
is 10. It gives however the flavour of why it is so.
Indeed, there are several things that don't match, between
our combinatoric problem and superstring as it is defined in its
basic construction.
First of all, in our problem time is a coordinate that labels
classes of configurations, with
a combinatoric the more and more complex as time increases.
As it is built, superstring theory appears instead to possess a symmetry under
time reversal. Moreover, string theory
appears to possess an infinite number of degrees of freedom, and this
not only, obviously, in the non-compact case, but also when all the
coordinates are compactified. Even in this case
the string spectrum contains an infinite tower of energy/momentum states,
obtained both as Kaluza Klein momenta or windings.
This however is not completely unexpected: a perturbative construction
is built on a flattened space. What we are doing is therefore approximating
the true, curved space through a series of ``plane waves'' in a basically
non-compact space.
The matching of string time and physical time is achieved once the string
space is fully compactified and the redundancy produced by the scale invariance
removed by the introduction of masses.
In this configuration, also the symmetry under time reversal is
broken (see Ref.~\cite{spi}).

\
\\

In our set up we have geometries of coordinates, 
characterized by the (local) value 
of the curvature, and, under certain circumstances, we can recognize a
path through sets of geometries/configurations, that as discussed
we \emph{interpret} as a time evolution.
Energies and fields belong to our interpretation of this
``evolving set of geometries'': they parametrize the moving sources
of geometry and curvature, and the traditional expansion
in terms of harmonic excitations parametrizes the modes of the 
evolving shape of space. Under certain respects, these statements may
sound trivial, and, after Einstein's General Relativity,
they are, either explicitly or implicitly, familiar. 
Nevertheless I think it is good to stress this point. From this point
of view, the entire description in terms of quantum fields \emph{is}
a description of (a subset of) the evolving geometries of the Universe.    

The combinatoric description of the universe and its string representation
differ in many respects, due to the fact that the latter gives a field
theoretical interpretation and implementation of discrete configurations.
In the combinatoric problem, that, we underline, \emph{is} the fundamental
formulation of the problem, we have only ``size-one'' energies.
Lower energies result only through averaging over longer 
times (or space intervals).
On the string side we have on the other hand excitations that correspond
to different energies; after the identification of the unit length with
the Planck length, 
we can recognize that string theory smoothes down, averages
over sets of discrete configurations: in particular,
all the sub-Planckian string excitations
correspond to collections of discrete configurations.
Energies (and momenta) of the compactified string are however built not
only over a basis of Kaluza Klein modes:
\be
E_m \, \sim \, {m \over R} \, ,  
\label{KK}
\ee    
where $R$ is a compactification radius, but also as ``winding'' modes:
\be
E_n \, \sim \, {n R} \, .
\label{winm}
\ee 
These are normally introduced in toroidally compactified perturbative 
string vacua, where lengths are measured in terms of the appropriate 
string length $\ell_{s}$ (as we said, each perturbative string construction 
has its own proper length; this is an artifact of the representation of 
the whole theory through a set of perturbative slices). 
However, we can already by now say that, in the cases
of interest for us, namely for the string configurations that dominate in the 
phase space, the proper string length and the Planck length can be 
identified \footnote{This occurs because in these cases 
the coupling of the theory is one, see Ref.~\cite{spi}.}. 
Under identification of the string length with the Planck length,
$\ell_{s} = \ell_{\rm Pl}$, we can view the first Kaluza Klein excitations
of \ref{KK}, namely those with $m < R$, as the sub-Planckian ones.  
As we have seen in section~\ref{UncP},
the contribution of \emph{all} the possible configurations at any time
is ``contained'' in the uncertainty corresponding to the Heisenberg's
principle. At any time the various configurations
contribute and give origin to a spreading-out of the unit values of energies,
producing a varied spectrum, that only in the ``average'', and up to
the Heisenberg's bound, corresponds to a three-sphere. Saying 
``in the average'' precisely means that a three-sphere is rigorously only the
very maximal entropy configuration, while the non-maximal ones contribute
by spreading out mean values according to the Heisenberg's relations
\footnote{Indeed, 
when in \cite{spi} I say that in the dominant string configurations
Lorentz and rotation invariance is broken
by a shift in space time, so that the breaking is of the order
of the inverse of a proper length, i.e. of the order of masses, or matter
densities, themselves, I am precisely saying that
the spherical symmetry is broken by deviations of the order
of the inverse of time or length, i.e. of the order of under-Planckian
masses.}.
Accounting in its spectrum for mass excitations lower than the Planck scale, 
and coming with an endowed ``quantization principle'' through
(anti)-commutators etc..., string theory collects therefore in one 
``averaged'' description the effect of a superposition of configurations
of maximal and ``close-to-maximal entropy''. 
As discussed in \cite{spi},
string vacua describe massive excitations through a superposition
of plane waves living on spaces with shorter radius than the whole 
three-dimensional space. They are therefore ``localised'' objects;
indeed, as we have discussed in section~\ref{wavep}, 
these ``wave packets''
are short-extension inhomogeneities of the geometry. We will come back
to this point, with a discussion of both the string and the combinatoric
point of view, in sections~\ref{shifts}, \ref{whytime}.

\subsection{T-duality}
\label{Tduality}

The winding modes \ref{winm} can be viewed as the Kaluza Klein modes 
built over the so-called ``T-dual'' radius, $\tilde{R} = 1/R$, somehow
enabling to introduce in the game also lengths ``shorter'' than the minimal 
one, $\ell_{s} = \ell_{\rm Pl}$.
The existence of a minimal length is assured in string theory by
the existence of a symmetry, called T-duality, of the compactified string,
that basically maps energy excitations built as Kaluza Klein momenta over
length scales below the string length to dual
excitations built as windings, in practice Kaluza Klein momenta over the 
inverse scale. So, once reached the string length, the system ``bounces''
back above this scale. We already pointed out that, 
in order to concretely ``solve'' the combinatoric problem of the
physical world, it is useful to
think in terms of symmetries, and eventually discuss
how and how much these symmetries are broken or up to what extent preserved.
This is somehow unavoidable. 
The toroidally compactified string, with its symmetry under T-duality,
is an example of such a theoretical ``simplification'' of the 
true physical situation, 
which allows to reduce the amount of degrees of freedom.
Seen in this way, T-duality appears as a convenient description; 
\emph{there is nothing particularly fundamental in it} from a physical
point of view: fundamental is the regularization of the phase space
through the introduction of a minimal length, a property of which string 
theory provides an implementation.
T-duality is eventually softly broken in the very dominant vacuum
\cite{spi}. Nevertheless,
it turns out to be convenient to parametrize physics
in terms of strings and T-duality, namely by first introducing a
strong simplification via string theory on a compact space, in which
T-duality is a way of introducing a regulator; the real world is
then better approximated by softly breaking this symmetry.

Thinking in terms of T-duality and string theory allows to
work around a vacuum that we can keep under control. 
Through this approach, we have access to 
some relevant properties of the dominant configurations of the Universe, 
which can in this way be viewed as a soft perturbation of a ``simple'' vacuum. 
For instance, in this way we 
know what are the terms necessary to keep it non-anomalous, and therefore
what is the configuration that does not generate an uncontrolled, 
possibly infinite number of terms, something that would correspond to
generate ``new dimensions'' and lead to a configuration more entropic than 
expected.
E.g. we know that we must work in 10 dimensions and not in 8 or 14 etc...
But for the real physical world there is no ``softly broken T-duality''.

\section{Macroscopical and microscopical entropy}
\label{mimacro}

In its basic construction, superstring theory appears embedded in a 
non-compact space. In particular, the time coordinate appears
to be allowed to span over the infinite real axis without constraints.
On the other hand, we already remarked that any perturbative construction
implicitly corresponds to some decompactification limit.
As we discussed in sections~\ref{ddf} and~\ref{vev}, the number of ``cells''
of the target space of our combinatoric problem, what we called the
``volume'' $V$, is arbitrarily large, something we indicate by saying that
we take the limit $V \to \infty$. This however does not mean that what
we call ``space'' in the ordinary sense is at any (finite) time infinitely 
extended. Indeed, cells without an assigned geometry have no intrinsic meaning
as a space in our ordinary sense: it is only after we have assigned a geometric
interpretation that we give them also such a meaning.
In particular, we have seen that the configuration of the universe 
at any finite time in the average classically corresponds to a three-sphere 
of radius $N = (c) {\cal T}$, the
light-equivalent of the age of the universe. At any finite time we see
\emph{all} the volume $V$, but configurations that depart from the average one
of radius $c {\cal T}$ appear as perturbations of the mean geometry, 
which are not all interpretable in geometric terms. As we have discussed,
they all fall under the ``cover'' of the Uncertainty Principle, and are
related to what we interpret as the quantum nature of physical phenomena.
In other words, this is like to say that at any time ${\cal T}$ we 
indeed \emph{do see} the entire space, but this can be reduced to the ordinary 
geometric interpretation of space only up to a distance $c {\cal T}$, i.e.
as long as classical light rays have carried to us the information about it.
\begin{itemize}
\item
\emph{Only up to $R = c {\cal T}$ such a ``whole'' can be 
interpreted according to our ``classical'' concept of space, 
which cannot be disentangled from time, 
because we ``see'' space only through light rays coming to us}.
\end{itemize}

\noindent
Our perception of space is produced by light rays that come to us
travelling in space-time.
\begin{itemize}
\item
\emph{The space ``outside'' the horizon is certainly infinitely 
extended, and somehow we see it, but it contributes to our perception and 
measurements only for an ``uncertainty'' of mean values, accounted for by the
Heisenberg's uncertainties}.
\end{itemize}
\noindent 
From a classical point of view, at any finite time
${\cal T}$ space exists only up to $R = c {\cal T}$.
From this point of view, namely from the point of view of writing a theory
in terms of differential geometry, at finite time the space is always compact,
of finite volume.

In order to
represent a mapping of our combinatoric problem at any finite time,
we must therefore consider also string theory on a compact space.
This implies that it must be considered in
a \underline{non-perturbative} regime, where in particular, owing
to compactness of the space, supersymmetry is broken.
In order to understand then what kind of ``universe'' comes out
of all the possible configurations, or ``string vacua'', at any fixed volume,
we must find out those that correspond to the maximal entropy in
the phase space.
String configurations on a compact space are obtained by compactifying
the string target space on certain spaces, which may be continuous and
differentiable, or even singular. In any case, a compactification
leads to a reduction of the symmetry of the initial theory. The more
singular is the space on which the string is compactified, the higher
is the amount of symmetry reduction. The spectrum of ``massless states'' is
related to the surviving symmetries of the target space.
By considering not just the massless states, but the full string spectrum,
it is easy to see that
the more the target space is ``singularised'', the more ``diversified'' is the 
spectrum one obtains, as a consequence of the symmetry reduction, and
the higher is also the number of possibilities such a configuration has
of being realized in the phase space.

A way of constructing configurations is to divide the string space through the
action of a symmetry group.
The method of obtaining configurations by taking a quotient
is typical of string constructions, not only in the case of orbifolds, where
one mods out by a discrete symmetry; also in the case of Calabi Yau 
constructions or generic compactifications, the ``massless'' degrees of freedom
are related to the surviving (continuous) deformations of the compact space.
In that case, we can at least formally say that one mods out by a 
continuous symmetry.

\begin{figure}
\centerline{
\epsfxsize=8cm
\epsfbox{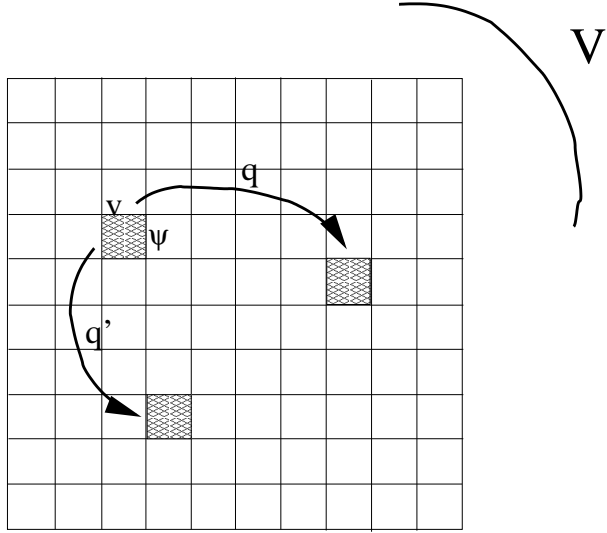}
}
\vspace{0.3cm}
\caption{The configuration $\Psi$, occupying a volume ${\rm v}$ in the initial
phase space of total volume $V$, 
is mapped to equivalent sub-volumes by the action
of the elements $q$ of a symmetry group $Q$, $q, q^{\prime} \in Q$.}     
\label{grid-1}
\end{figure}

Let's consider the action of a group $Q$, and
consider the configuration $\Psi$ obtained by
dividing the string space by $Q$. 
The volume $V$ of the phase space of the initial configuration
gets divided into cells of volume:
\be
{\rm v} \, = \, {V \over Q} \, .
\ee 
The configuration $\psi$ occupies one of these cells. Let's say it is supported
on ${\rm v}$. Elements $q, q^{\prime} \in Q$ map 
the cell/sub-volume ${\rm v}$ to equivalent cells/sub-volumes,
as illustrated in figure~\ref{grid-1}.
There are therefore $Q = V / {\rm v} $
ways of realizing this configuration. The occupation in the whole
phase space is therefore enhanced by a factor $Q$ as compared
to the initial one. By reducing the symmetry, we have enhanced the
possibilities of realizing a configuration in equivalent ways, in the same
sense as, in the two-cells example of section~\ref{setup}, 
by assigning different colours,
black and white, we have the possibility of realizing the configuration
``one-white/one-black'' in more ways than ``white-white'' or ``black-black''.

\subsection{Microscopical entropy}
\label{micre} 

The entropy considered in section \ref{setup},
given as the logarithm of the occupation in the phase space
of all the combinatorics at given volume, can be considered the
``macroscopical entropy'' of the space of all possible configurations
of degrees of freedom in a certain volume. This is directly related
to the weight of the contribution of a certain configuration to 
the way the Universe appears.
The higher is entropy, the more a configuration contributes, as it should.
On the other hand,
one can consider each configuration as a ``universe'' in itself.
There is a ``string partition function'', built over a
bunch of states, the states of the string spectrum.
These states contribute to the string partition function
with weights that can be related to their
probability in the specific string vacuum they belong to.
It is therefore possible to speak of a ``microscopical entropy'',
the entropy of the specific ``gas of states'' which contribute
to create the ``geometry'' of a string vacuum. In turn, each string vacuum
can be viewed as belonging to a ``gas of spaces'', where the
\underline{macroscopical} entropy is defined. It turns out that
the two entropies are ``dual'' to each other: 
the microscopical entropy decreases
as the distribution of states of a single vacuum is more concentrated;
this means that the symmetry of this vacuum is highly reduced,
and in turn the occupation of the whole configuration
in the full phase space is enhanced. 
Let's consider the effect of a symmetry reduction on the occupation
volume, the ``weight'' $W$ of a configuration. Let's suppose we start with
a symmetry $Q$. Let's now enhance/reduce, at fixed volume, the symmetry of
the configuration to $Q^{\prime}$, such that the volume $|| Q^{\prime}  ||$
of the group $Q^{\prime}$ is $\alpha$ times the volume of the group $Q$:
$|| Q^{\prime}  || = \alpha || Q ||$, something that we shortly indicate as
$\alpha: Q \to \alpha Q$, using the same notation for the volume factor 
$\alpha$ and the operator $\alpha$ that performs the transformation from 
one group to the other one. The weight of the configuration in
the full phase space behaves as a homogeneous function of the symmetry
factor, of degree 1:
\be
Q \, \to \, \alpha Q \, , ~~ \Rightarrow \, W_{\rm macro} (Q) \, \to \, 
W_{\rm macro} (\alpha Q) \,
= \, \alpha W_{\rm macro} (Q) \, .
\ee
On the other hand, at the \underline{microscopical} level, 
the weight is a homogeneous function of degree -1:
\be
W_{\rm micro} (Q) \, \to \, W_{\rm micro} (\alpha Q) \, = \, \alpha^{-1}
W_{\rm micro} (Q) \, .
\ee
This, for any fixed volume $M$. We can therefore write:
\ba
W_{\rm macro} & = & A(M) \, Q \, ; \nn \\
&& \\
W_{\rm micro} & = & B(M) \, Q^{-1} \, . \nn
\ea
Writing $W_{\rm macro} = \exp S_{\rm macro}$, 
$W_{\rm micro} = \exp S_{\rm micro}$, we obtain that:
\be
S_{\rm macro} \, = \, - S_{\rm micro} \, + \, F(M) \, ,
\label{SMm}
\ee
where $F(M) = \ln \left( A/B (M) \right)$.
The entropy considered in Ref.~\cite{spi} is precisely
the microscopical entropy, $S_{\rm micro}$.
Indeed, while in the macroscopical picture we work ``at fixed energy $N$'',
in the microscopical one we classify the ordering through ``space''
volumes $M \geq N$. In order to derive the dominant configuration
of the universe,
in the macroscopical picture we look for the highest entropy at fixed 
energy $N$, in the microscopical one at the minimal entropy at fixed volume.
The two approaches are dual to each other. The ``macro'' picture is the
picture of energies, the ``micro'' one is the picture of space volumes,
namely of the size of the space of the space-coordinates.
Indeed, the two ``coordinates'' are conjugates each other; this somehow 
reflects the Uncertainty Relation(s):
\be
\Delta E,\, p ~~ \gsim ~~ {1 \over \Delta t, \, X  }  \, .
\ee
These relations are ``symmetric'' in energies (i.e. attributes)
and space-time.
In one picture, we fix the attributes and have an uncertainty in the size
of space-time. In the other one, we fix the size of space(-time) and have
an uncertainty in the energies.

If we work at fixed $M$ (``fixed volume'' in the language of \cite{spi})
we can neglect the additive constant $F(M)$, that in this case results
in a multiplicative rescaling of all the weights, irrelevant to the
purpose of computing mean values of observables. We recover therefore
the relation of Ref.~\cite{spi}, expressing the weight of a configuration
at a given volume as:
\be
W(\Psi)_V \, = \, {\rm e}^{- S \, ( \equiv S_{\rm micro}) (\Psi)} \, .
\ee
Notice that $A(1) = B(1)$, implying that $F(N =1)$ vanishes: the
macroscopical and microscopical weight meet at $N = 1$, where there is only
one possible configuration, in both the macroscopical and the microscopical
sense (this is however a limit case well beyond the regime of validity 
of the string approximation).

As it is written, $S_{\rm micro}$ appears to be 
the mapping of the weight of a configuration to the ``tangent space''.
As discussed in Ref.~\cite{spi}, this logarithmic representation of space is
the environment in which perturbative string is normally constructed,
and operations such as the orbifold projections naturally defined.
This is the reason why, as discussed in \cite{spi}, $Z_2$ orbifold shifts,
introducing $1/2$ projections
on the tangent space, and therefore on the occupation volumes in the 
logarithmic representations,
lead to square-root scalings of physical masses. 
Logarithmic masses are related to the microscopical entropy, as well
as physical masses are related to the occupation volume in the ``pulled-back'',
real space.

\subsection{Mean values and observables in the string picture}
\label{vevst}

Since expression \ref{meanO} is normalized, any difference
in the normalization of the macroscopical and microscopical definitions
of entropy, eq.~\ref{SMm}, becomes irrelevant.
Expression \ref{meanO} is therefore valid both at the discrete level, 
for the combinatoric problem of the full phase space, 
and at the ``microscopic level''
of those particular subsystems, ``collections of macroscopic configurations'',
which are the string vacua. In the first case,
the weight $W(\Psi)$ of a configuration is given by:
\be
W (\Psi) \, \stackrel{\rm def}{=} \, \exp S_{\rm macro} (\Psi) \, ,
\label{WexpS}
\ee
where $S_{\rm macro}$ 
is the entropy of the configuration in the full phase
space. In the second case,
\be
W (\Psi) \, \propto \, \exp - S_{\rm micro} (\Psi) \, ,
\label{Wexp-S}
\ee
where $S_{\rm micro}$ is the entropy of the states within
each string vacuum, and the symbol ``$\propto$'' can be substituted 
by ``$=$'', as long as the final scope is that of computing mean values. 
Analogously, the partition function ${\cal Z}$ admits
the two corresponding interpretations.
Notice that on the string side the partition function ${\cal Z}$
can be normalized only when the space is compact, and therefore curved,
with a non-vanishing curvature due to the fact that we work at finite volume.
Any perturbative construction is built on a \emph{non-compact} space,
and somehow corresponds to the limit $N \to \infty$.
In the orbifold language, the condition of working in a curved, compact
space is attained at the maximal twisting \cite{spi}. 
Nevertheless, one can built vacua with a non-vanishing partition
function by explicitly breaking supersymmetry~\footnote{There are also
constructions in which the partition function vanishes also when 
supersymmetry is explicitly broken.}.

According to \ref{WexpS}, \ref{meanO} receives contributions mostly from
the configurations with the highest macro (resp. lowest micro) entropy.
One may ask what happens if ${\cal O}$ diverges on some non-extremal 
entropy configuration. In this case, the main 
contribution to the mean value of the observed quantity could come from 
seemingly negligible vacua. This however 
is a wrong way of approaching the problem,
a way of thinking that makes only sense in the traditional approach of
observable quantities defined in terms of operators acting on
an effective action, written as an explicit function of
terms determining the dynamics of the system.
In this approach, physical quantities such as masses, energies, couplings
have a value, a ``weight'' depending on their occupation in the phase
space: particles that interact more are heavier etc..., as discussed in
\cite{spi}. Owing to their very basic definition, observables
cannot ``blow up'' on rare configurations: they acquire a non-negligible
contribution from their being in correspondence with 
(relatively) often realized processes (see discussion in Ref.~\cite{spi}).

\subsection{Entropy and curvature}
\label{Srho}

In Ref.~\cite{spi} we discussed how, 
upon sufficient ``singularization'', the string vacuum
becomes necessarily curved, with an entropy equivalent to the one
of a three-dimensional black hole: $S \sim {\cal T}^2$, where ${\cal T}$
is the age, or equivalently the radius, of the Universe. 
On the other hand, we also observed that, in principle, string theory on
a compact space is always embedded in a curved space. However,
a perturbative construction is forcedly build around a de-compactification
limit of some coordinates, contributing to the definition of what
is considered the ``coupling'' of the theory. In string perturbation
it is therefore possible to work with supersymmetric vacua and flat spaces.
The string representation of the geometric/combinatoric problem
we are discussing in this work is somehow a mapping,
in which the degrees of freedom describing the geometries of space 
and their evolution are represented
in terms of coordinates of a fibered space, ``folded out'' in a flat
representation, that corresponds to a logarithmic, ``tangent space''
description.   
String theory ``lives'' on a fibered space that parametrizes
in terms of fields and waves the combinatoric problem, which is 
a problem in itself entirely related to the physical space-time.
In other words, with string theory we build-up an extra space, a fiber
over the physical space, in order to parametrize in a continuous way
the combinatorics of the geometry of the space.
The ``macroscopic'' curved geometry results from a process of
singularization of this ``folded out'' space, obtained e.g. through
progressive orbifolding. 
Let's consider for instance the probabilities of the states as they
appear in the string (one loop) partition function, once the space
is sufficiently curved to make the latter non-vanishing and therefore
the probabilities normalizable:
\be
P \, \cong \, {\int d \tau \, {\rm e}^{-E^2 \tau } \over {\cal Z} } 
\, \to ~ \approx {1 \over E^2} \, {\cal Z}^{-1} \, . 
\label{Plog}
\ee
Compare this expression with the usual definition of probability for a 
statistical system at equilibrium:
\be
P \, = \, {{\rm e}^{-\beta E} \over {\cal Z}} \, .
\label{Pexp}
\ee
A comparison of these expressions clearly
suggests that the string (perturbative)
partition function is defined on the logarithm, on the tangent space.

Entropy reduction is obtained through compactification of the string
target space on more and more curved geometries. Spaces with higher
curvature have in fact a smaller symmetry group. In Ref.~\cite{spi}
we have described this process in terms of orbifold projections.
Before attaining the maximal twisting of string vacua, 
we analyse entropy only in perturbative slices of 
the string theory (at maximal twisting, the vacuum is necessarily
non-perturbative because no decompactification limit can be taken).
On these slices, we have a partial view of the story, and we can work
on a compact subset of the full string space. There, we can apply projections
that reduce entropy. This corresponds to focusing on a subset
of ``coordinates'', of which we can fix the (sub)volume, 
ignoring the fact that other coordinates are non-compact.     
We control therefore the reduction of entropy separately on the various
slices that patch together to build up the string vacuum.

In Ref.~\cite{spi} we have discussed how orbifolds are the privileged
constructions for following the process of entropy reduction in string 
theory. In order to understand how the reduction works,
we can figure out the situation by representing the initial physical system
as a gas of identical particles. 
We quote here the discussion presented in Ref.~\cite{spi}.

\begin{figure}
\centerline{
\epsfxsize=6cm
\epsfbox{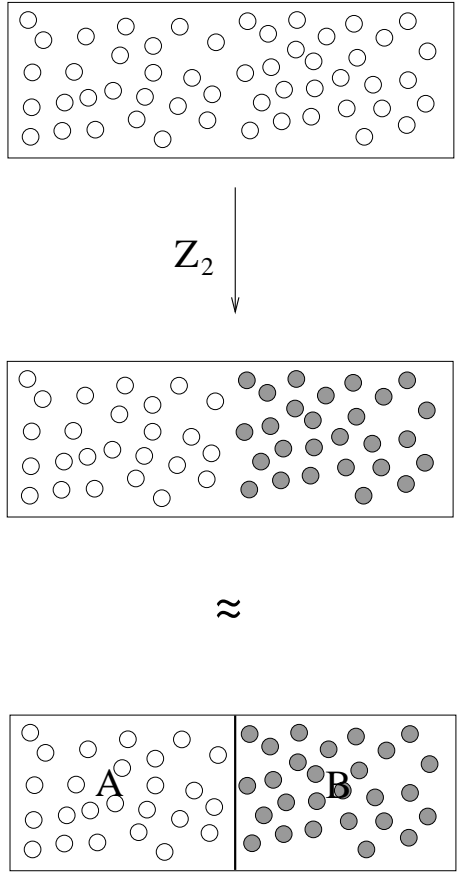}
}
\vspace{1cm}
\caption{the effect of an orbifold projection is that of reducing the 
symmetry of the system, separating the spectrum in two sectors ``confined'' 
to different parts of the phase space.}
\label{s-orb1}
\vspace{.8cm}
\end{figure}

\begin{figure}
\centerline{
\epsfxsize=13cm
\epsfbox{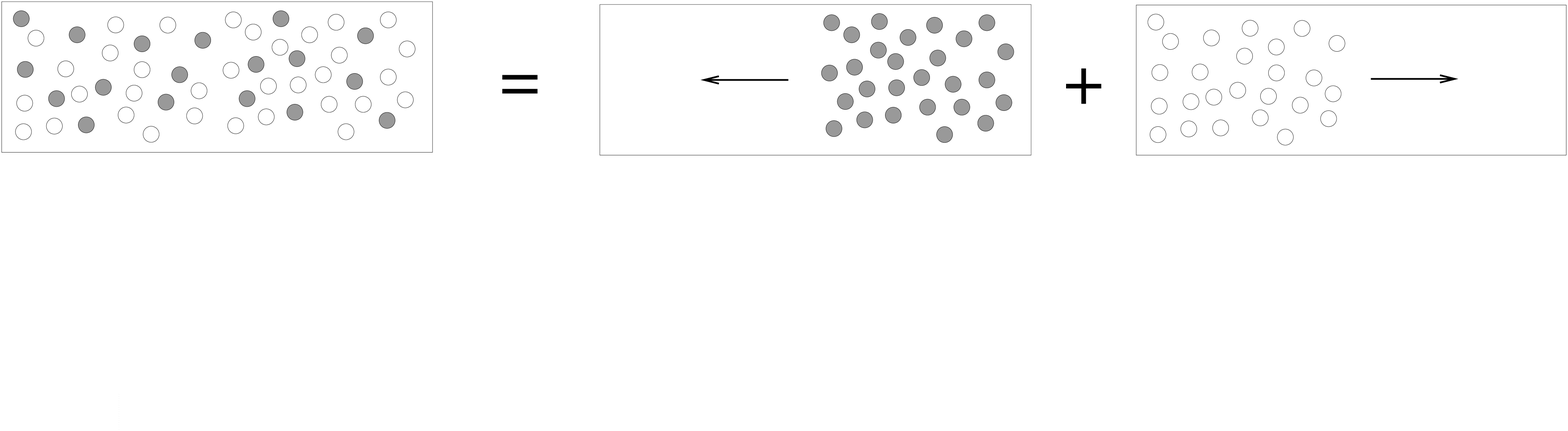}
}
\vspace{0.1cm}
\caption{The entropy decrease due to an orbifold projection can be understood
if we represent the system as the sum of two parts each one
consisting of a gas of particles. The orbifolding process is the reverse
of the expansion of each of the two gases from half of the phase space until
they occupy the entire phase space.}
\label{s-orb2}
\vspace{.8cm}
\end{figure}

A string vacuum (and in particular an orbifold) can be thought 
as a gas, in which different particles act as sources for the 
singularities of the space (from a gravitational point of view, particles
\emph{are} singularities of space-time, being sources of gravitational field, 
singular points of the curvature). Let's consider the case of a $Z_2$
orbifold. The orbifold projection
introduces a ``distinction mark'' on half of the states, it labels them
in another way. The situation is depicted in figure~\ref{s-orb1}.
\noindent
The resulting orbifold follows now the laws
of the composite systems: the probability of the various configurations of
the new string ``vacuum'' are given by the product of the probability of
the unprojected part (``A'' in figure~\ref{s-orb1}) times the
probability of the twisted part (``B'' in figure~\ref{s-orb1}):
\be
P_{A + B} ~ = ~ P_A \times P_B \, .
\label{PA+B}
\ee  
Let's assume we know the entropy of the string vacuum before
the orbifold projection.
We want to see how this quantity changes when we apply the 
$Z_2$ projection. 
Intuitively, it is clear that, since a projection reduces the amount of 
symmetry, it reduces also the volume of the phase space at 
disposal for the degrees of freedom. 
As a consequence of the increased concentration of the
probability distribution, we expect that also entropy should be reduced.
In order to see more precisely how things work, consider that
a $Z_2$ projection divides the initial system in two parts, as is clear from
the partition function:
\be
{\cal Z}  ~ \stackrel{Z_2}{\longrightarrow} ~ 
{1 \over 2} \left( {\cal Z} \ar{0}{0}
\, + \, {\cal Z} \ar{0}{1} \right) \, 
+ \, {1 \over 2} \left( {\cal Z} \ar{1}{0}
\, + \, {\cal Z} \ar{1}{1} \right) \, .
\label{Zz2}
\ee
This corresponds to the process illustrated in figure~\ref{s-orb1}.
In order to understand in which direction the variation of entropy goes,
we can view the process of separation
of the phase-space into two sectors as the opposite of
the adiabatic expansion illustrated in figure~\ref{s-orb2}.
\noindent
In this example, the system is constituted by the sum of two subsystems of
particles, that for convenience we have labelled with a different colour.
Intrinsically, they are on the other hand indistinguishable;
when they are together,
such as on the l.h.s. of the figure, all particles are on the same
footing, having at disposal the full phase space. Our system is the sum
of the two systems on the r.h.s., and entropy is the sum of the
two corresponding entropies. It is then clear that, owing to the process of
``separation'' induced by the orbifold operation, entropy decreases.

We can compute the amount of this reduction. Although represented
as a ``gas'', the system does not follow the laws of gas thermodynamics:
the particles don't have a ``temperature'' related to a ``kinetic energy''.
In order to understand what happens to entropy, we must only think in terms
of phase space and probability distributions.
We can consider the whole system as ideally divided into two
subsystems, A and B.
The probability of each half system is 1/2 of the total probability:
\be
P_A \, = \, P_B \, = {1 \over 2} \, ,
\label{pA1/2}
\ee 
When this distinction is just virtual, as it is before the orbifold
projection, the total probability of the system is $1 = 1/2 + 1/2$.
As illustrated in figure~\ref{s-orb1},
the $Z_2$ projection acts like the insertion of a wall between
A and B, that prevents a mixing of the two parts. In this case,
the probability of the configuration ``A$\,$B'' is the product 
of the two probabilities, because the phase space too has become a
product:
\be
P_{AB} \, = \, P_A \times P_B \, = \, {1 \over 2} \times {1 \over 2} \, = \,
{1 \over 4} \, .
\label{pAB1/4}
\ee
Entropy of the initial system $S_{A+B}$ is given by the sum $S_A \, + \, S_B$:
\be
S_{A+B} ~ = ~ 2 \, S_A ~ = ~   \ln 2 \, . 
\label{Sa+b}
\ee
After the projection, we have instead:
\be
S_{AB} ~ = ~ - P_A P_B \ln  P_A P_B ~ = ~ {1 \over 2} \ln 2 \, . 
\label{Sab}
\ee
Entropy is therefore reduced by half.

This example suggests that it should be possible to
translate the problem of entropy minimization
in the string configurations into geometric terms.
The single particles of the gas can in fact be viewed as sources of singularity
for the geometry. The less entropic configurations are also the more
singular ones from a geometric point of view.
Let's introduce the ``normalized curvature'' 
$\rho (R)$:
\be
\rho(R)_{\vec{x}} \, \stackrel{\rm def}{=} 
\, {R (\vec{x}) \over \int dV \, R (\vec{x}) }
\, . 
\label{Rnorm}
\ee
Like a probability, this ``density of curvature'' sums up to 1 over the
entire space. Here a comment is in order.
We are used to think that, being
energy/masses sources for the curvature, they should scale
proportionally to the curvature, as according to the Einstein's equations.
On the other hand, from \ref{Plog} we see that
the probability of the elementary states on which the string representation
is built scales as the inverse squared of the energy, and not of a length: 
$P \sim 1 /E^2$. This may seem contradictory with the identification
of expression~\ref{Rnorm} with a probability density. However,
perturbatively the string fibered space representing
the combinatoric problem of the ``geometries'' of space and its evolution
is built over a set of plane wave states. Any ``point-like'' source
of singularity, the ``particle'', doesn't appear as an elementary state
of the string construction: any picked energy distribution
must be built as a superposition of plane waves; the more is this wave
packed picked around a point, the higher is the spreading in the momenta,
according to the Heisenberg's relation of space-momentum uncertainty,
$\Delta X \Delta P \geq \hbar / 2$. 
The ``point-like'' sources are build as superpositions of plane
waves. This implies that a high $\rho (R)_{\vec x}$ is spread over
also low values of $E^2$. 
Curvature and energy are therefore somehow reciprocal: 
${\rm Energy}^{-2} \leftrightarrow \approx \rho(R)$.
A high curvature density means
that the superposition of waves contains excitations that range
from low to high energy.
Owing to the reciprocal relation of space and energy spreading
(the Heisenberg's uncertainty relation), also point-like and wave-like
representations are somehow ``reciprocal'': for what
matters a representation in terms of these elementary states,
energy is here equivalent to a length, rather than to its inverse.
Expression \ref{Rnorm} works therefore indeed as the equivalent of a 
probability for the string representation of the combinatoric problem:
it is defined as the probability on a geometric representation of the string 
space not in terms of plane waves, but of superpositions of them,
building up ``point-like'' sources of space curvature.  
This allows us to introduce a new definition of entropy:
\be
S \, = \, - < \rho (R) \ln \rho (R) > \, .
\label{srlnr}
\ee
This expression is equivalent to the usual statistical definition:
\be
S \, = \, - < P \ln P > \, ,
\ee
but it better fits the representation of the string space as a geometric space.
In the points in which the curvature vanishes, expression \ref{srlnr}
does not diverge, because they are to be considered as attained 
through analytic continuation
from the neighbouring regular points with non-vanishing curvature.  
Working with a ``regularized'' space, with arbitrary
but finite volume $V$, and therefore a non-identically-vanishing
curvature, allows us to compare entropies of any different 
configurations of the phase space.

\
\\
{\sl example: curvature density and entropy for orbifolds}

\noindent
Orbifolds are constructions in which the curvature can be considered
entirely concentrated on the fixed points, while the rest of the space is
flat. There are formulae which allow to compute the (total) curvature
of an orbifold. Were this a homogeneous space with constant curvature, we 
could then normalize the curvature by dividing the value of the curvature by
the volume of the compact space. We would have therefore:
\be
\rho(R) \, = \, {R \over V R} \, = \, {1 \over V} \, .
\label{Rconst}
\ee
Orbifolds can instead be viewed as limits of a regular 
configurations, corresponding to a space in which each fixed point 
$i = 1, \ldots, n$ is blown up to a small sphere of volume
$v_i$. If the full orbifold space has volume $V$, and for simplicity we 
use the same notation in order to indicate volume and space, we have
that the full space/volume can be written as:
\be
V \, = \, \left\{ v_1, \ldots \, v_n  \right\} \cup \, 
\left\{ V - \left\{ v_1, \ldots, v_n  \right\} \right\} \, . 
\ee
Entropy in $\left\{ V - \left\{ v_1, \ldots, v_n  \right\} \right\}$ 
can be computed by analytic continuation, and therefore turns out
to vanish. In order to see
what happens on the $n$ fixed points, consider that:
\be
\sum_{i=1}^n v_i R_i \, = \, \int R \, dV  \, .
\label{nvR}
\ee
For simplicity, we can choose $v_i = v_j = v$. We have therefore:
\be
R_i \, = \, { \int R \, d V  \over n v } \, ,
\label{Ri}
\ee 
from which we obtain:
\be
\rho_i \, = \, {1 \over nv} \, .
\ee
The curvature density at the fixed points should correspond to
a limit regularized by subtracting infinities: 
\be
\rho_i \, = \, \lim_{v \to 0} 1 / v \, = \, 1 / n \, .
\label{limRv}
\ee
This expression corresponds to considering $v$ as the unit of measure
of volumes.
With this ``trick'' 
we are in the position to see how
an orbifold possesses a lower entropy than a smooth manifold with constant
curvature equal to the average orbifold curvature.
In the orbifold case, entropy is:
\be
S_{\rm orb} \, = \, - \int_{ \{ n \}} \rho(R) \ln \rho(R) \, =
\, - \sum_{i = 1}^n {1 \over n} \ln {1 \over n} \, = \, - \ln {1 \over n}
\, . 
\label{Sorb}
\ee
In the smooth case, we have instead:
\be
S_{\rm smooth} \, = \, - \int_V \rho(R) \ln \rho(R) \, = \, 
- \int_V {1 \over V} \ln {1 \over V} \, = \,
-  \ln {1 \over V} \, .
\label{Ssmooth}
\ee
Since $\{ v_1, \ldots, v_n \} \subset V$,
$nv < V$ and $1/V < 1/n \, (= 1 /nv) < 1$, 
we have $S_{\rm smooth} > S_{\rm orb}$.

The unit volume $v$ can be considered as the volume of
the ``point''. Indeed, we can 
``measure'' the overall volume in units of the volume $v$ of the 
``elementary cell''. We already pointed out that the introduction
of such an elementary unit is a necessary step for the whole
set up to make sense. This allows us to compare any volume to the volume
of a point. When doing the comparison, 
it is intended that all lengths are measured in units of 
the elementary length $\ell$, so that a curvature, or a segment,
are measured in units of $\ell$, a $d$-dimensional volume is measured
in units of $v \sim \ell^d$. Appropriate  powers of $\ell$ also adjust
densities to adimensional quantities to be inserted in logarithms
or exponentials.

\
\\

In Ref.~\cite{spi} we have seen that through orbifoldization we can obtain
a vacuum which is quite close to the dominant configuration of space-time.
We have on the other hand seen that the geometry of this space is 
basically the one of a three-dimensional black hole. In section~\ref{eSp}
we have computed the scaling of entropy in the dominant configurations:
$S \sim N^2 \leftrightarrow {\cal T}^2$, a result derived in 
Ref.~\cite{spi}(chapter 4) directly from the string partition function.
The orbifold action on the string space was a combination of freely and 
non-freely acting orbifolds. The non-trivial action on the
extended coordinates was freely acting.
In a freely acting orbifold the curvature is not ``concentrated'' on 
fixed points: these are ``shifted'', in the sense that the associated massless
states are lifted in mass as a consequence of the space translation associated
to the twist. We can consider that the curvature is ``smeared'' over 
``extended points''. Indeed, over the entire space, although reduced in volume.
In the case of a $Z_2$ orbifold, on half the initial volume.
In practice, entropy behaves as in \ref{Ssmooth}, with an appropriately
rescaled volume, e.g. $V \to V/2$. As a result, we obtain:
\be
S_{Z_2 - {\rm freely}} \, = \, \ln {V \over 2} \, ,
\label{Sfree}
\ee 
instead of $\ln V$ as in \ref{Ssmooth}.
Expression~\ref{srlnr} seems therefore to lead to a logarithmic scaling 
of entropy as a function of the coordinates, and 
this seems in contradiction with the main power-law scaling
of the black hole.
Indeed, the relation between the spheric curvature that results at the
maximal twisting, and the expression of entropy as a logarithm
of the curvature, is established
``on the tangent space'', where the internal string space ``folds out''
to a product of coordinates. The contributions of twisted and extended
coordinates sum up, so that the internal non-freely acting orbifold actions,
contributing for terms of the type \ref{Sorb}, contribute to
entropy for an irrelevant additive constant.
The logarithmic dependence
on the volume of the extended space corresponds instead to the Jacobian
of the transformation between the logarithmic representation of the 
coordinates, where one ``builds up'' the string curvature,
and the physical representation of space-time. In other words, an
entropy given as a logarithmic function of the volume is due to the 
``logarithmization'' of the space of degrees of freedom, implicit
in any perturbative string representation:
\be
\partial  r^2 \, \stackrel{\log}{\longrightarrow} \, \partial (\log r)^2 
\, = \,
{\log r^2 \over r^2 } \, \partial r^2 \, .  
\ee

\subsection{Reducing entropy}
\label{redS}

According to \ref{WexpS} and \ref{meanO}, the dominant \underline{string}
configurations are those of \underline{minimal} entropy.
A way of reducing entropy consists in freezing degrees of freedom. 
Indeed, this is the most effective way.
Let's see what is the variation in entropy between configurations 
differing by the amount of frozen coordinates, and what is the one due
to a change of volume of extended coordinates.
For simplicity consider a configuration of $n$ coordinates,
in which $p$ coordinates
are extended up to $L$, and $n-p$ are frozen to size 1 in Planck units.
The total volume is:
\be
V \, = \, L^p \, ,
\ee 
while the curvature, the average curvature, is:
\be
R \, = \, L^{p / n} \, .
\ee 
The integral of the curvature over the entire volume is therefore:
\be
\int R \, dV \, = \, L^p \times L ^{p / n} \, ,
\ee
and the ``normalized curvature'' is:
\be
\rho(R) \, = \, {1 \over L^p} \, .
\ee
The entropy of this configuration is then:
\be
S \, = \, {1 \over L^p} \ln L^p \, .
\ee
The relative variations with respect to $p$ and $L$ are:
\ba
{1 \over S} {\partial S \over \partial p} & = & 
{1 \over p} \left(1 + {1 \over \ln L^p}    \right) 
\, = \, {1 \over p} - \ln L \, ; \label{dSdp} \\
{1 \over S} {\partial S \over \partial L} & = & 
- {p \over L^p} \left( 1 + {1 \over \ln L^p }  \right) \, . \label{dSdL}
\ea
Notice the opposite sign of these variations, in agreement with the fact that
an increase of $p$ leads to a spreading out of the curvature, and therefore
an increase of entropy, while an increase of $L$ differentiates more
the extended from the reduced coordinates, producing a higher differentiation
in the geometric shape of the space, and a reduction of entropy.
Clearly, for $p$ sufficiently small and $L$ large enough,
\be
\left\vert {1 \over S} {\partial S \over \partial L}  \right\vert \Big/ 
\left\vert {1 \over S} {\partial S \over \partial p} \right\vert 
\, \approx \,  {1 \over L^p} \, \ll \, 1  \, .
\label{Sl/Sp}
\ee
Therefore, the natural direction of the process of entropy reduction
is toward a maximization of the number of ``frozen'' coordinates
(in string-orbifold language = maximal twisting).
However, as we discuss in section~\ref{entrV},
there is a ``bound'' on the maximal ``twisting''.

\subsection{String in anomalous dimensions and entropy}
\label{Sand}

We have seen that, 
by reducing the volume and the dimension of space, entropy is reduced.
Thinking in this way, one should conclude that, 
for the string in dimension $< 10$, we have a reduced
entropy as a consequence of a reduced number of degrees of freedom.
This is not true: the string is dimension $d < 10$ is anomalous: it is 
consistent only if seen as the compactification of the critical string.
Indeed, the anomaly generates an infinite number of (counter)-terms,
which correspond to switching-on new coordinates. Alternatively, we can think
of parametrizing these degrees of freedom through fields (Liouville fields); 
this is equivalent. How many extra coordinates do we have? As many as to
reach the critical dimension.
In the case of string in dimension higher than 10, the anomaly generates
infinitely many terms, and there is no ``fixed point'', dimension at which
the game stabilizes: we generate an infinite number of dimensions.

\subsection{Shifts in string theory and volume reduction in phase space}
\label{shifts}

In Ref.~\cite{spi} we have explained the sequence of masses of the
elementary particles through the relation to their occupation in  
the phase space. The more a particle interacts, the more are its decay 
channels, the heavier it is.
In particular, in order to estimate the relative occupations in the phase
space, we proceeded by counting the number of ``symmetry reductions''
the string vacuum went through in order to give rise to a certain kind
of particle. 
Let's see how this works in the case of $Z_2$ orbifold projections.
Let's suppose we start with a certain string vacuum, in which
there is a set of particles with a mass produced by a $Z_2$ orbifold
shift on the momenta, such that $m \to m + 1$. We have:
\be
M \, \sim \, 0 \to {1 \over R} \, . 
\ee
Let's act now on this set with a further $Z_2$ orbifold projection, that
reduces the symmetry by lifting half of the previous
particles with a new shift on the momenta, along another direction,
compactified on a circle with equal radius as the first one:
\be
M \to M^{\prime} \, \sim \, {1 \over R} \, + \, {1 \over R^{\prime}}\, ,
~~~~~~ R = R^{\prime}. 
\ee
We end up with a vacuum in which half of the particles
have mass $1 \over R$, and half have mass ${1 \over R} + {1 \over R}$.
As discussed in \cite{spi}, these operations are carried out in a
logarithmic representation of the physical vacuum, so that,
in the ``true'' space, the ratio of masses is not $M^{\prime}/ M = 1/2$, 
but $M^{\prime} = \sqrt{M}$ in Planck units.  

Apart from the scaling law, linear or exponential depending on whether we
work on the true space or, like in perturbation theory, 
``on the tangent space'', we see here how a $Z_2$ orbifold operation
can reduce the symmetry of the vacuum and at the same time pop out
a new kind of particles, corresponding to the lifting of half of the 
previous ones. The ratio of masses between those originating as the reduction
of the pre-existing ones,
and the new ones with a shifted mass, corresponds to the ratio of the
``volumes'' of the symmetry groups of the vacuum, inversely related to
the ratio of the volumes occupied in the phase space by the 
two configurations. Namely, with the new projection, in the phase space
there are now, among others, both the ``initial'' configuration $\Psi$
and the configuration $\Psi^{\prime} = Z_2 ( \Psi)$. In $\Psi$ there are
only particles of mass $M$; in $\Psi^{\prime}$ one-half of the particles 
originate as the reduction of the old ones, one half have new attributes.
The configuration $\Psi$ has twice as much symmetry volume 
as $\Psi^{\prime}$. In the full phase space it occurs therefore half of
the times of $\Psi^{\prime}$.
This means that the particles with mass $M$
occupy a fraction of phase space $V_M \propto 1 (\Psi) + 2 \times 
{1 \over 2}(\Psi)$, where ${1 \over 2}(\Psi)$ is the contribution of the 
particles of mass $M$ belonging to $\Psi^{\prime}$.
The new particles occupy instead a fraction
$V_{M^{\prime}} \propto 2 \times 
{1 \over 2}(\Psi) = {1 \over 2} V_M $, and consequently have mass:
$M^{\prime} = M / 2$ (in the tangent space, 
or $M^{\prime} = \sqrt{M}$ in the exponential representation.
Remember that masses are lower than 1, so that $\sqrt{M} > M$).
In terms of unit energy cells and proper volumes of particles,
this means that, in the same interval of time, the new particles
occur in the space a number of times
$n^{\prime} = \sqrt{n}$ if $n$ is the number of times of the ``old'' ones.
In practice, this means that they occupy a volume $r^{\prime} = \sqrt{r}$,
and their ``mass'' is $M^{\prime} = {1 \over r^{\prime}} = {1 \over \sqrt{r}} 
= \sqrt{M} $.

\subsection{Why must elementary masses depend on time?}
\label{whytime}

We want to see here, from the perspective of entropy and weights, why
the masses of the elementary particles should scale with (powers of) time,
instead of remaining constant. Of course, this question can be
rephrased as: why the dominant configurations produce this running
behaviour of masses? In this framework, all configurations in principle
do exist and contribute; the point is to see which ones contribute the more.
We can illustrate the situation in the following picture:
\be
\epsfxsize=8cm
\epsfbox{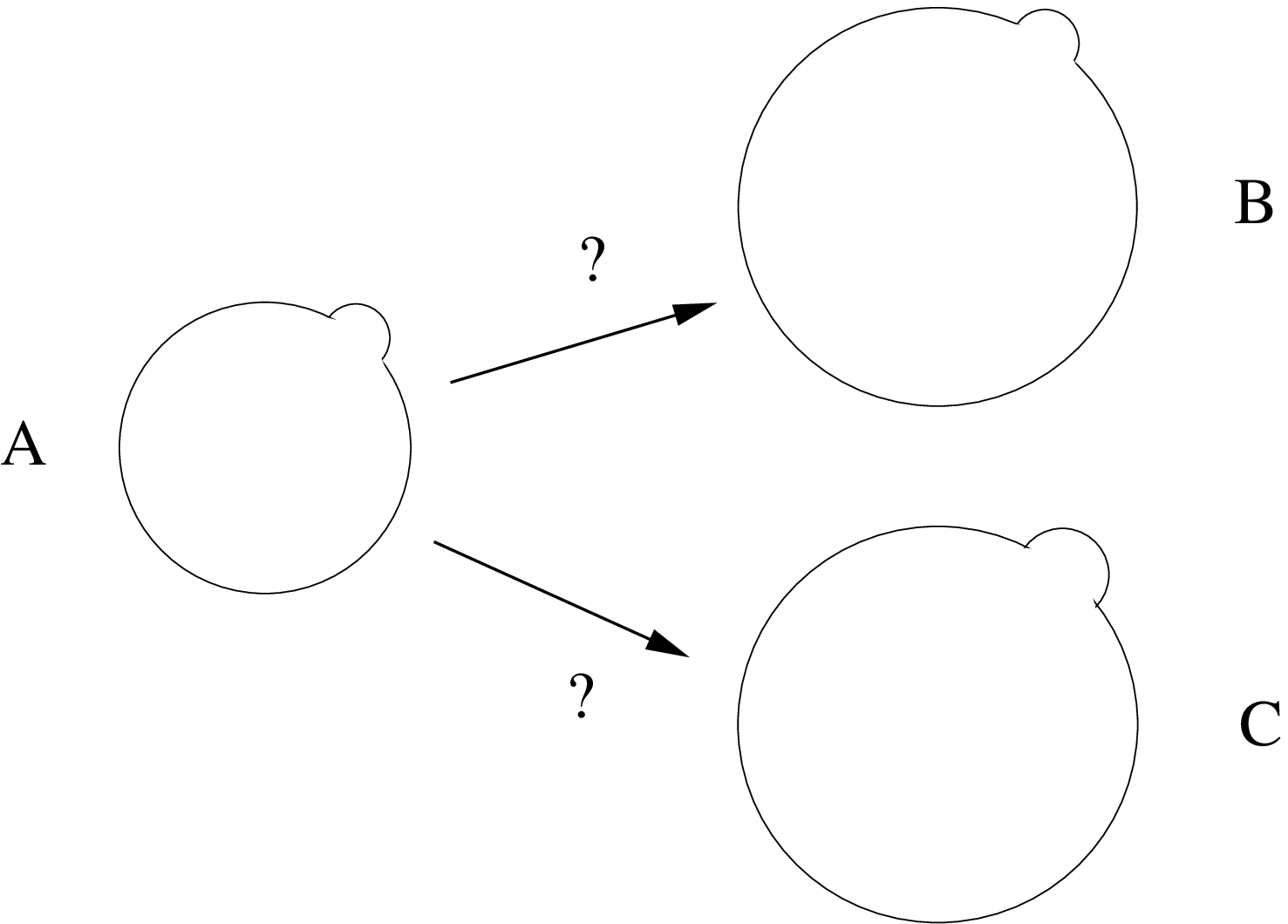}
\label{mass-scaling}
\ee
We represent a particle with a certain mass as a deformation of the
spheric geometry of the universe, the hump popping out from
the spheric surface. In picture A) we have the universe
at a certain time. In B) we represent the universe at the next time, therefore
as a sphere of larger volume, but with the deformation corresponding to the
massive particle of exactly the same size as in A) (constant mass).
In C) we represent instead the situation of the universe and particle
with a mass rescaled according to a functional relation to the radius of the
universe (time-dependent mass). If we look at the problem of entropy
from the microscopical point of view, configuration A) represents the
configuration of minimal entropy at time ${\cal T}$. In order
to keep the massive deformation at fixed size also at time ${\cal T} 
+ \delta {\cal T}$, the ``wave packet'' must be formed out of more harmonics
than in A), because, as compared to the ``period'', the radius of the universe,
B) represents a more picked configuration than A). 
In C), owing to the fact that we are keeping the same functional
relation as in A), the massive packet is built out of the same
harmonics as in A), simply based on a period rescaled to the actual time.
The configuration B) is therefore more spread out in the states of the
string spectrum than the configuration C), and is therefore 
microscopically more entropic. This means, macroscopically less entropic,
and therefore un-favoured with respect to C).
The most realized behaviour is therefore the one that keeps a functional
dependence of masses on the period, i.e. the radius, of space-time.
That this relation is an exponential one, instead of a linear one
(masses scale as powers of the age of the Universe, 
$m \sim {\cal T}^{- \alpha}$, not as fractions) depends on the fact
that under an increase $N \to N + \delta N$ the combinatorics
leads to an exponential increase of the weights: $W \sim \exp N^2$.
As discussed in Ref.~\cite{spi},
linear increments are typical of the ``tangent space'', the logarithm
of the weight, where we deal with entropy rather than directly with 
weights.

\newpage

\section{The dimensionality of space-time}
\label{dimspt}

\subsection{Entropy and volume in various dimensions}
\label{entrV}

We already mentioned that the dimensionality of space-time is a matter
of ``statistics'' of the phase space, the result of the fact that,
at any volume, the maximal entropy is attained in 
four space-time dimensions. Let's see here the problem of
the dimensionality of space-time from the string point of view,
where the Universe is dominated by the minimal entropy configurations.
In Ref.~\cite{spi} the dimension of space-time was derived to be 4 in the 
framework of orbifold constructions, where the maximal twisting of coordinates
left room to 3+1 un-twisted ones, to be identified with the 
space-time. Here we use some string tools investigated in \cite{spi}
in order to perform an analysis of entropy in various dimensions.

As discussed in \cite{spi}, the ``vacuum energy'' computed in 
string theory is normalized in such a way to account for the $D-2$
dimensional energy density of the Universe. Namely, in four
dimensions, the ``vacuum energy''. i.e. the integral of the partition 
function, gives ``1'' as ${\cal T}^2 \times$ the energy density, $\rho(E)$, 
so that $\rho(E) \approx 1 / {\cal T}^2$, with the consequence
that the total energy scales as the age of the Universe:
$E_{\rm tot} \approx {\cal T}$, enabling to view the entire Universe as a 
black hole (see Ref.~\cite{spi}). This string result derives from a 
re-interpretation of string amplitudes, which accounts for the lack
of translational invariance in this new theoretical framework, characterised
by a mismatch in the normalization of mean values
by a volume factor  (see also \cite{lambda}).

Proceeding in an analogous way, but for dimensions $D > 4$, the same
computation would produce a total energy scaling as:
\be
E_{\rm tot}^{(D)} \, \approx \, {\cal T}^{D-3} \, .
\ee
Of course, as also in \cite{spi}, all quantities are here formally
adimensional, being intended that they are measured in reduced Planck units.
Physical dimensions are adjusted by appropriate powers of the Planck mass, 
Planck scale, speed of light and Planck constant.
In dimensions $D < 4$, the scaling behaviour is different.
In these cases, obtained by further compactification of the string coordinates
below $D = 4$, there are no further consequences on the normalization
of the string coupling, and we obtain:
\be
E_{\rm tot}^{(D = 2,3)} \, \approx \, {\cal T} \, .
\ee
The ``time'' coordinate, the ``age of the Universe'', continues on the
other hand to correspond to the inverse of the temperature. 
Considering that $V^D \approx {\cal T}^{(D-1)}$, we obtain
the following scalings: 
\ba
{\cal S}^{(D)} & \approx & V^{D - 2 \over D-1} \, ~~~~~~D \geq 4 \, , \nn \\
{\cal S}^{(D)} & \approx & V \, ~~~~~~~~~~ \, \, D = 3 \, , \label{Scaling} \\
{\cal S}^{(D)} & \approx & V^2 \, ~~~~~~~~~ \, D = 2 \, . \nn
\ea
These expressions indicate that, for $D \geq 4$, the corresponding string vacua
describe higher-dimensional black holes (see for 
instance~\cite{Rabinowitz:2001ag} for the entropy of a black hole in 
$n$ dimensions).

As we discussed, on the string side
working at ``finite $N$'' must be replaced by
working at fixed volume. From the point of view
of the combinatoric approach, where we can compare segments of a certain
length with any power, namely where we can say unambiguously if
$m^p $ is greater, equal or lower than $n^q$, fixing the volume means fixing
the total number of unit cells. Whether this is expressed as the cubic 
power of the radius, $\sim R^3$, or as another power, as in
the higher dimensional spheres we are here considering, it does not matter.
On the string side, this approach is translated by expressing all
quantities as adimensional, rescaled by appropriate powers of the Planck 
length, as it was done in \cite{spi} and \cite{estring}, where we gave
all masses as appropriate roots of the age of the Universe.  

The increasing of the ratio entropy/volume in $D \geq 4$ is not such a 
surprise: it follows our intuition that, as the combinatorics of space 
increases as a consequence of the increased number of degrees of freedom,
also entropy increases. Noteworthy is that below the $D = 4$ threshold,
the ratio entropy/volume starts increasing again. 
The minimal entropy/volume is obtained for $D = 4$, which is therefore
the dimensionality ``selected'' at any fixed volume. 
Different space-time dimensionalities have a weight difference of the order:
\be
W(D = n) \big/ W(D=4) \, \sim \, {\rm e}^{- {\cal T}^{n-4}} \, .
\ee
Other space-time dimensions have therefore a negligible
weight in the phase space, as compared to $D = 4$.

\subsection{``vectorial'' and ``spinorial'' metric}
\label{vecsp}

String theory implements coordinates and degrees of freedom in a framework
of space-time-dependent fields.
Somehow by definition, in the dynamical representation of the combinatoric 
problem on the continuum, centered on the configurations of minimal 
(resp. maximal) entropy, the expansion
of space-time can be viewed as driven by the propagation of 
massless fields, which in some sense ``represent'' it. 
The space-time coordinates must
therefore correspond to massless fields; there must be a 
non-degenerate mapping between space-time degrees of freedom and field
degrees of freedom. If, by absurd,
there are more massless degrees of freedom than space-time coordinates,
it means that we have wrongly computed the dimensionality of space-time:
the number of ``coordinates'' maximally extended, i.e. with the
minimal curvature, is higher, and we get a contradiction. In the 
\underline{minimal entropy} string configurations, 
we must therefore have a correspondence:
\be
\left\{ X_0 = t, \vec{X}    \right\} \, \leftrightarrow \, 
\left\{ \phi_i (t, \vec{X})  \right\} \, , ~~~~~~~
\det \left[ \partial \phi_i(t, \vec{X}) \big/ \partial (t, \vec{X}) \right]
 \, \neq \, 0 \, ,
\label{detphi}
\ee  
for a set of fields $\phi_i$, $i = 1, \ldots, 4$. As discussed in \cite{spi},
at the minimum of entropy, rotational invariance of space is broken.
There are therefore
only ``diagonal'' degrees of freedom. Namely, in these configurations
the space built on the coordinates $x_1, \ldots, x_3$ 
doesn't possess a rotational 
symmetry: $ x_i \, \to \, x^{\prime}_i = A_{ij} x_j$. 
After having gauged away the redundant degrees of freedom, for instance
in the light-cone gauge, where only transverse degrees of freedom 
appear, the graviton field has only the two ``diagonal'' entries:
\be
g_{\mu \nu} \, = \, \{ g_{11}, g_{22}  \} \, . 
\ee 
This makes up two field coordinates. We need two more to fill up
the space-time dimension and ensure that the map \ref{detphi} is 
non-degenerate. Being the speed
of expansion of the Universe ``fixed'' to a finite constant $c$, 
the space-time is necessarily a relativistic space,
whose representations are built on spinors.
Vectorial representations can be built starting from spinorial ones, but
not the other way around.
The spin connection contains therefore a mixture of vectorial and purely 
spinorial components.

Consider now the role played by 
the field $g_{\mu \nu}$. It ``rotates'' two vectors, by contracting
their indices into a scalar, according to:
\be
V^{\mu }, \, V^{\nu} ~ \to ~ V^{\mu} g_{\mu \nu} V^{\nu} \, .
\label{vvg}
\ee
We expect that a ``purely spinorial'' spin connection in a similar way
rotates, and contracts, spinor indices.
\be
\psi^{\alpha}, \, \psi^{\beta} ~ \to ~ \psi^{\alpha} \tilde{A}_{\alpha \beta} 
\psi^{\beta} \, .
\label{psipsiA}
\ee 
Owing to the breaking of rotational symmetry, the only bi-spinors
present in the minimal string vacua
are those that pairwise build up diagonally vector coordinates. 
If we indicate the spinors associated to each bosonic coordinate as
$\phi^{\mu}_1$, $\phi^{\mu}_2$, this means that there are no
mixed states of the type:
\be
\phi^{\mu}_{\alpha} \otimes \phi^{\nu}_{\beta} \, , ~~~~~~~
\mu \neq \nu \, , ~~~~ \alpha, \beta \in \{1,2 \} \, ,
\ee
but only diagonal ones:
\be
x_{\mu} \, = \,
 \phi^{\mu}_{1} \otimes \phi^{\mu}_{2} \, .
\ee
We expect therefore the ``spinorial'' part of the spin connection to be
in bijection with a vectorial representation consisting of just two
transverse field degrees of freedom. This is actually the way the 
electromagnetic vector-potential field in these vacua works.
The field $A_{\mu}$ is a vectorial field, not a spinorial one.
On the other hand, the vector index $\mu$ must be somehow thought as 
a ``bi-spinor'':
\be
A_1 ~ \sim ~ A_{{1 \over 2} {1 \over 2}} \, .
\label{1-22}
\ee
Indeed, $A_{\mu}$, normally introduced through a gauge mechanism
applied to a scalar quantity built on a bi-spinor, somehow provides with field 
degrees of freedom a ``metric'' which contracts spinor indices to
a scalar:
\be
\overline{\psi} \partial {\! \! \! /} \psi ~ 
\stackrel{\rm gauge}{\longrightarrow} ~ 
\overline{\psi} A {\! \! \! /} \psi ~ = ~
\overline{\psi}^{\alpha} \gamma^{\mu}_{\alpha \beta} A_{\mu} \psi^{\beta} \, .
\label{ppgA}
\ee 
The gamma matrices, precisely introduced by Dirac 
in order to deal with ``square-roots'' of vectorial relations,
play the role of converter from bi-spinorial to vectorial indices.
The field $\gamma^{\mu}_{\alpha \beta} \equiv
{A {\! \! \! /}}_{\alpha \beta}$ corresponds therefore to
the ``spinorial spin connection'' $\tilde{A}_{\alpha \beta}$ introduced above.
This field provides the two missing degrees of freedom required in order
to complete the non-degenerate ``representation'' of space-time \ref{detphi}.

Being a representation of space-time
means that graviton and photon propagate at the speed 
of space-time itself. As such, they  correspond to massless fields. 
As we will discuss in section~\ref{boundary} and illustrate in 
figure~\ref{boundary2}, their propagation
occurs at the speed of expansion of the horizon,
which corresponds to $c = {\pi \over 2} R$, where $R$ is the radius of the
three dimensional sphere. 
$c$ is precisely what we call the ``speed of light''.
We stress that this relation, as all the above argument,
makes only sense in the minimal entropy configurations, in which
the dimension of space-time is four, and the geometry is mostly the 
one of a three-sphere. On the other hand,
only upon maximal shifting, besides maximal twisting, we have four true,
non-anomalous space-time dimensions (graviton + photon transverse modes
that ``stir'' the space-time). Otherwise, when there are more massless
fields, they effectively stir more space dimensions. In this case, although 
apparently we have only four space-time coordinates, in practice, 
from the point of view of the phase space, we have a higher dimensional sphere.

The four modes of graviton and photon are the only massless fields
of the minimal configurations. 
Other ``coordinates'' span subspaces of
higher average curvature, they are ``shorter''. As discussed in \cite{spi},
they describe massive, shorter range fields, and ``expand'' at a
lower rate. They can under appropriate circumstances be described
in field theory, where they belong to a ``fiber'', based on the space-time.

\subsection{Time evolution in string theory}
\label{timevS}

As it is defined, the
``time'' coordinate of the target space of string theory 
doesn't automatically correspond to the one we gave in section~\ref{timev}.
In general, the physics arising in string vacua is symmetric under 
time reversal. This symmetry must be explicitly broken:
only in this case an increase in the value of the time coordinates 
leads to a configuration with higher entropy, and this coordinate can be
given the same interpretation as the time coordinate we have introduced
in this work. As discussed in \cite{spi}, this condition is always
verified in the string configurations of minimal entropy. 
From the very philosophy of the 
combinatorics-to-string map, i.e.  centering on mean values and
accounting for deviations through
the Heisenberg's uncertainty relations, we see that
the string ``time'' coordinate is therefore a kind of ``average parameter''.
The set of minimal vacua
involved in the ``string history'' are those in which the maximal 
possible amount of coordinates are fixed at the minimal length, 
and the other ones span
a sub-manifold, the space-time, of progressing volumes $V$. 
We can build a ``history'' of the Universe as a path, a ``progress'', 
through volumes $V$ of space-time along the configurations of minimal entropy:
as illustrated in figure~\ref{expansion2}, we can view the average
configuration around the minimum of entropy
at volume $V^{\prime} = V + \delta V$ as the evolution of the
corresponding mean configuration at volume $V$, because the change in entropy
in passing from the one to the other is minimal. Passing from the one
to the other realizes the smoothest step, the one closest to a
representation in terms of a continuous flowing as due to differential
equations. Using the language of quantum field theory, we can say that
the configuration at volume $V$ ``projects'' 
with the highest amplitude onto the configuration which
has the minimal entropy at the new volume. 
Consistency of the string vacuum (absence of anomalies etc...) ensures then 
that a configuration of minimal entropy will naturally flow, through time 
evolution according to the built-in time dependence of the fields of its 
spectrum, from a minimal entropy vacuum to the next minimal entropy vacuum,
thereby realizing a consistent correspondence with the time evolution 
introduced in section~\ref{timev}.   
In this progress, any volume change cannot be a simple, 
trivial rescaling, re-absorbable in a redefinition of the scale. 
The phase space, the ``physics'', at volume
$V^{\prime} \neq V$ is really different from the one at volume $V$.
Along this progression, frozen (i.e. ``twisted'') coordinates in the minimal 
entropy configuration will remain stuck at the minimal value.

String theory provides us with a representation of
the geometric problem of minimizing entropy along this path, where
the physical configurations are described in terms
of expanding and propagating fields. These are 
quantum relativistic fields,
as implied by having fixed the speed of light and working at finite volume,
as discussed in \cite{spi}. Anomaly cancellation
ensures that what we are doing is consistent, namely that if we start in 4
dimensions, the theory does really expand only along four dimensions, there
are no additional extended/infinite coordinates, as they would be effectively
produced by the infinite degrees of freedom (i.e. counter-terms) 
of an anomalous theory. This means that, if within non-anomalous string theory
we derive that the minimal geometry at a certain volume is produced by
a certain amount of fields and particles, this is the ``right'' content, or
``spectrum'' of the theory, which remains of this type also after
we have moved to other points in the history of the minimal entropy
configurations.

\subsection{Closed geometry, horizon and boundary}
\label{boundary}

\begin{figure}
\centerline{
\epsfxsize=8cm
\epsfbox{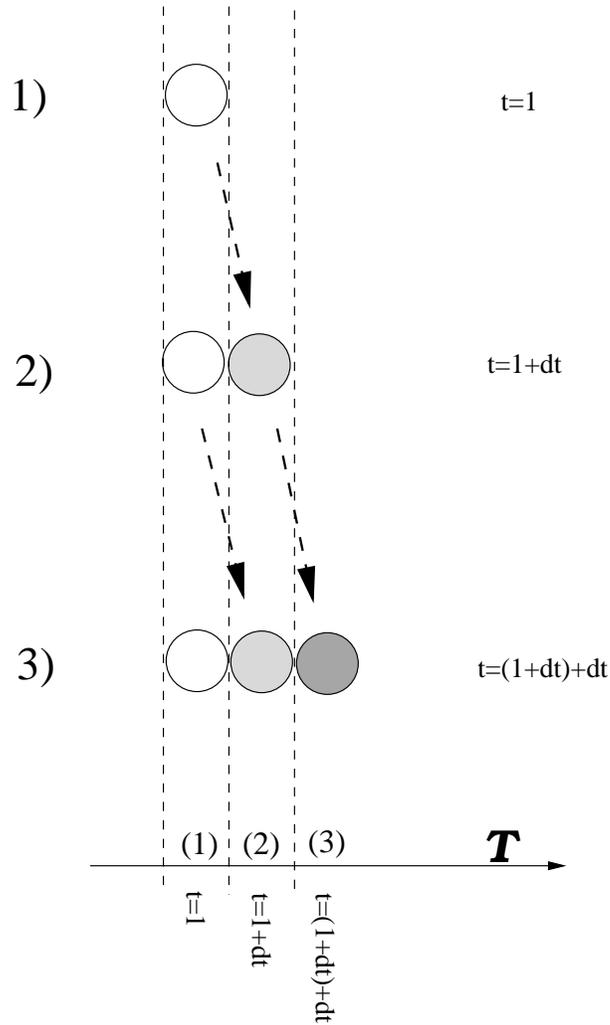}
}
\vspace{0.3cm}
\caption{The progress of the universe through increasing times $\sim$ number
of elementary cells. As the statistics grows, the configuration
gets more complex and differentiated. Configuration 3) does not ``include''
configuration 2), which in turn does not include number 1): with volume
increases also the radius, and the curvature of space decreases. 
The curvature and therefore the energy levels are different at 
different times. Configuration 1) ``flows'' to 2), and 2) to 3) only in the
sense that 2) is more similar to 3) than 1) or a configuration with
two equal balls are, and therefore this is the most realized ``evolutionary
path''.}     
\label{expansion2}
\end{figure}

\begin{figure}
\centerline{
\epsfxsize=8cm
\epsfbox{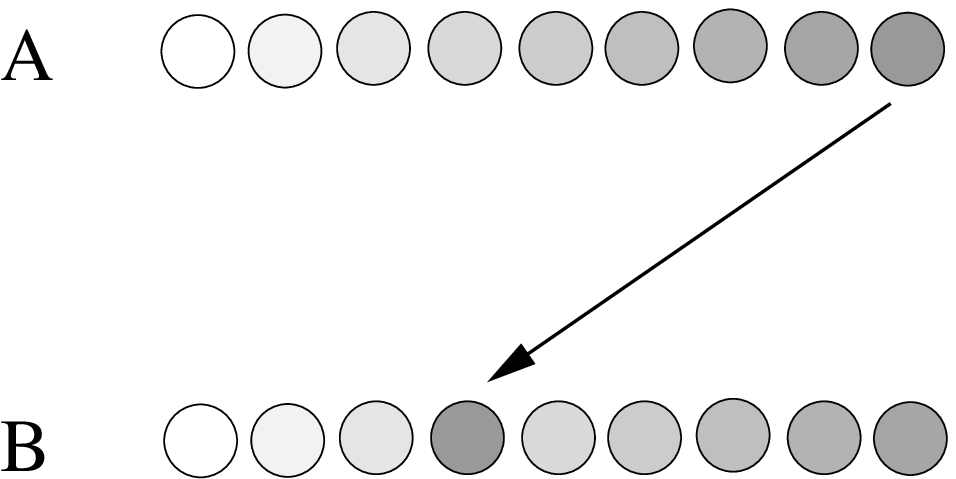}
}
\vspace{0.3cm}
\caption{Configuration B) is obtained by superposing to
the dominant configuration less entropic configurations than in A),
and therefore has a lower weight, the more and more negligible 
the higher is the ``jump'' between B) and A).}     
\label{discont-1}
\end{figure}

\begin{figure}
\centerline{
\epsfxsize=8cm
\epsfbox{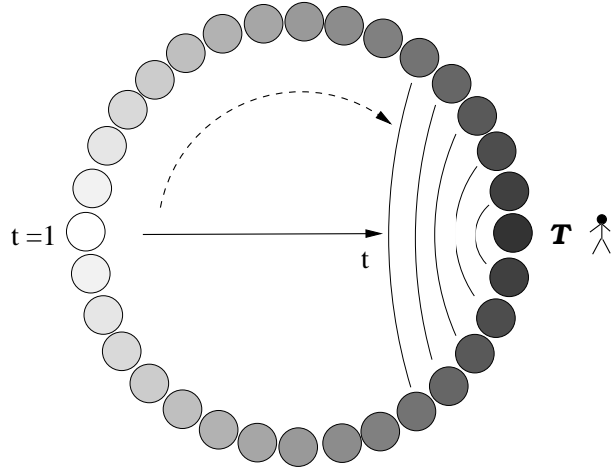}
}
\vspace{0.3cm}
\caption{The Universe at time ${\cal T}$ appears to the observer as
a set of surrounding shells made out of elementary cells.
These shells, here represented only through sections constituted
by two linked antipodal cells, go from the one closest to the
observer (the darkest one), 
which is also the closets in time, up to the farthest, the horizon,
that corresponds to the ``big-bang'' cell. The curvature of space
is on the other hand everywhere the one of the age corresponding to the
observer, ${\cal R} \sim 1 / {\cal T}^2$. Notice that the two-sphere 
corresponding to the horizon appears to ``conserve'' the total energy flux, 
$1$, of the initial unit cell.}     
\label{expansion3}
\end{figure}

We can now see how our Universe builds up.
At $t = 1$ there is only one
possible configuration, that we illustrate in figure \ref{expansion2} with
a ball representing the unit cell.
Already at the next step, $N = 2$, we have many more (infinitely many)
possibilities, corresponding to any possible space ``dimension'',
being $N^p$ no more trivial, and the combinatorics increases very rapidly. 
For what discussed in section \ref{eSp}
we can concentrate the analysis to three dimensions,
which gives a contribution of the same amount as the sum of
all the neglected dimensionalities and configurations (Uncertainty Principle).
The dominant configuration
is the one that gives a ``homogeneous'' distribution of the occupied cells 
(what at large $N$ becomes a ``spheric'' geometry) . 
But already at $N = 2$ also non-extremal
entropy configurations start to exist and give non-trivial contributions.
The result is that, on the top of a homogeneous distributions, in the Universe
start to show up inhomogeneities, of the order of the Uncertainty Relations.
We can represent this process by distinguishing the regions of the space
as balls with a different colour, figure~\ref{expansion2}.    
As time goes by (i.e. $N$ increases), we get new possibilities 
of differentiation from the basic homogeneity, in a sort of ``progressive
differentiation through steps of small perturbations''.
We indicate this with an increasingly darker coloration of the balls.
In principle, one could ask if there could be ``discontinuities''
in this progress, namely, whether there could be steps in which a 
darker ball falls between lighter balls, as illustrated 
in figure~\ref{discont-1}. 
Of course, nothing prevents
such configurations from appearing, and indeed they \emph{are present}.
However, dominance of the most entropic configurations, with deviations
milder and milder as the asset departs from the most entropic 
configuration, tells us that these discontinuities give a minor contribution,
so that, in the average, the evolution takes place in the continuous
way we illustrate in figure~\ref{expansion3}.

The progress through times/values of the average curvature and its
inhomogeneities looks therefore like the propagation of a perturbation,
and is translated, in the continuous language of string theory, 
into a ``propagation'' of the information through an expanding universe.
As time goes by, the average curvature of space decreases,
and this takes place everywhere, throughout all the space. 
The ``perturbations'' too spread out: as $1 / N^2$ for those
that we interpret as ``massless fields'', or with a lower rate for
the ``massive'' ones. To an observer,
the local physics appears to be influenced, ``produced'' if one prefers,
by the sum of the information that propagated up to him from everywhere else
in the space. We already mentioned that a detector is a particular 
configuration of space, and an experiment is the detection of variations of 
this configuration during a certain interval of time.
By letting an experiment, or a detection, to take place
in a certain interval of time, the observer can resolve for different
``smearing rates'' of the inhomogeneities, and consequently organise
the interpretation of
what happens in terms of massless and/or massive objects. 
Of course, this is only an approximation, because any possible
configuration contributes, however to an uncertainty expressed by
the Uncertainty Principle \footnote{If one wants, here is another way to 
see that the functional \ref{ZPsi}, \ref{meanO} must be equivalent to the 
Feynman's path integral, as shown Ref.~in~\cite{spi}.}. The ``experimental
observation'' of the universe surrounding an observer indeed carries
the information about the geometry of the full universe,
the progress toward an increasing ``local'' differentiation from the
average geometry being interpretable as a continuous, ``jump-less''
evolution that propagates the perturbation of the spheric geometry.
String theory provides us with a theoretical framework enabling us to
organise the interpretation of this information
in terms of particles, fields, interactions etc...
What \underline{we observe} is indeed \underline{always just our 
local physics}, but we disentangle the jungle of data by organising
them as a superposition of informations coming from different places,
and therefore ``originating'' at different times in the past.

Notice that, although
we see regions of the Universe corresponding to past times, 
we don't see ourselves in our past times: neighbouring shells
are not derived the one from the other through time evolution.

\emph{The universe at (any) present time ${\cal T}$ is given by the
(weighted) superposition of the configurations of the phase space 
\underline{at present time}}. 

In this sense, the actual configuration 
depends only on the present phase space, not on the past:
the evolution belongs to our interpretation.

For the observer the universe turns out to mainly consist
of a progression from the farthest configuration, the ``initial'' one, 
to the nearest, representing the physics at present time.
At any time, the dominant configuration is however not derived
by a process of evolution of the one at previous time, but through a
process of entropy maximization at present time. 
\begin{figure}
\centerline{
\epsfxsize=8cm
\epsfbox{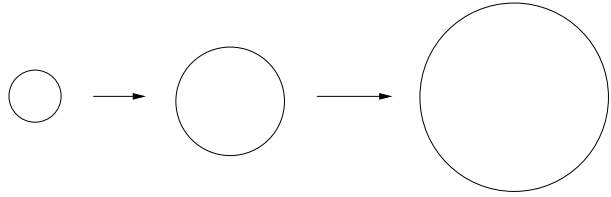}
}
\vspace{0.3cm}
\caption{During the time evolution, the three-dimensional space
build up of an increasing number of elementary cells 
expands with the geometry of a three-sphere of growing radius.}     
\label{expansion1}
\end{figure}
To an observer the space appears built
as an ``onion'': the observer is surrounded by shells, 
two-spheres corresponding to farther and older phases of the universe, 
up to the horizon, that correspond to the ``big bang'',
as illustrated in figure~\ref{expansion3}.
However, in itself there is no real ``big-bang point'', located somewhere
in this space: the argument leading to this description of the observed 
Universe is not related to a particular choice of the point of observation.
This does not mean that the Universe looks absolutely identical
for any choice of this point inside the whole space; simply, it means that
from any point the Universe appears built as an ``onion'', with informations
coming from everywhere around, going backwards in time, and space, up to
the horizon, which is an ``apparent horizon'', with no real physical location, 
but always at distance $N \sim {\cal T}$ from the observer.

Wherever the location of the observer in the space is,
what he will see is only the ``tangent'' space around him,
experienced through the modifications it produces in himself.
In this way, he will have access to the knowledge of the
average energy density of the universe, and indeed an indirect experience
of the whole universe: the ``local physics'' is the one
specific of the actual time, and as such ``knows'' about the full
extension of the space. Owing to the special ``boundary conditions''
that ``sew'' the borders of space into a sphere,
the latter will look in the average homogeneous all around in every direction:
the observer will always have the impression of being located ``at the center''
of the Universe.
Through the modification of his configuration, the ``local physics'',
(i.e. the set of all what happens to him, light rays hitting from the various
directions, gravitational fields etc..., that he will \emph{interpret}
as coming from all over the space around), he will then ``measure''
the energy density through all the space, concluding that it is 
$\rho(E) \sim 1 / R^2$. 
On the other hand, knowing that such an 
energy-density scaling law is the one of a sphere, he will deduce
that the horizon surface at a distance $R$ from himself, and with
area $\sim R^2$, has boundary conditions such that the space 
closes-up to a sphere.
The observer will then \emph{interpret}
the set of cells spread out along the horizon,
a surface with area $\sim R^2$ and
energy density $\sim 1 / R^2$, as corresponding to a point, a 
unique cell of unit volume and unit energy, ``smeared'' over
something that appears like a two-sphere.
He will therefore refer to this as to the ``big-bang'' point, 
the initial configuration, energy one at volume one,
and he will say that what he sees by looking at the horizon, is indeed
the beginning of everything. We repeat however that
this point, or surface, namely, the horizon, does not really exist as a
special point located somewhere in space: the interpretation would be the same
for any observer, located at whatever point in this space.

\subsection{Non-locality and quantum paradoxes}
\label{nonloc}

The uncertainty encoded in 
Quantum Mechanics through the Uncertainty Principle, lifting up the 
predictive power of classical mechanics to a 
probabilistic one, leads also to
non-locality, possibly violating the bound on the speed of the transfer
of information set by the speed of light $c$.

It has been a long debated question, whether this 
had to be considered
as something really built-in in natural phenomena, or simply an 
effect due to our ignorance of all the degrees of freedom involved, something
that could be explained through the introduction of ``hidden variables''
\cite{EPR}.
It seems that indeed the physical world lies on the side of true quantum 
interpretation, which, as shown by Bell \cite{bell}, is
not reproducible with hidden variables. 

At the quantum level, the bound on the travel speed of information, $c$,
can be violated by non-locality of wave-functions.  
How can we understand all that in the light of our framework? 

In our framework, the uncertainty of the Uncertainty Principle, 
at the base of quantum mechanics, is due to the fact that what we observe
and measure is the sum of an infinite number of configurations, among which
also tachyonic ones contribute. This is 
in agreement with the fact that non-locality
implies somehow a propagation at speed higher than $c$, thereby
violating causality. And indeed,
the ``dynamics'' described in the previous section, resulting
from the fact that at any time the universe is
the sum of all configurations at the actual time, averaged to the one of 
maximal entropy, implies in some sense an istantaneous transfer of 
information, although the classical geometric deformations propagate
at maximal speed $c$.
Related to this are also other properties, that in our scenario we see
as quantum effects on the large, cosmic scale. For instance, the masses
themselves. In our framework, masses depend on the space-time size
of the universe. This means that, 
at any time, an electron, or any other particle, ``knows'' how large
is the universe up to the horizon. How would such a boundary 
information determine the properties of each particle, 
even those locally produced in laboratory, if any information could only be
transferred at maximal speed $c$? This non-local, basically instantaneous
knowledge implied in the combinatoric scenario is reproduced in string theory,
a relativistic theory, essentially in two ways: 

1) through an explicit 
quantization, in order to reproduce the ordinary quantum effects, which
include non-local, tachyonic-like effects such as experiment correlations
violating Bell's inequalities and so on, and 

2) through the very basic properties of the string construction, in which,
as we discussed in Ref.~\cite{spi},
massive particles live partly in the extended space, identified with the
ordinary space-time, the space along w´hich they propagate,
and partly have a foot in the internal string space.

\begin{figure}
\centerline{
\epsfxsize=8cm
\epsfbox{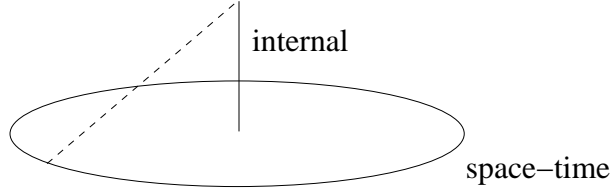}
}
\vspace{0.3cm}
\caption{A pictorial illustration of how can massive particle/fields have
knowledge of the full space-time. Here the extended but compact space-time
is represented as a disc on the horizontal plane. The internal dimension 
is represented by the vertical segment.
A field/particle confined to live tangent to the space-time
cannot ``see'' the horizon, but just the neighbours of the point where it is
sitting. A particle/field
with a foot on the internal space can lift-up its point of view
and see the horizon, having from above a global view of the disc on the 
plane.}     
\label{int-ext}
\end{figure}

Roughly speaking,
beeing extended also in the internal space provides massive states an
extra dimension from which they can ``look'' at the entire space-time, anyway
a compact space, of which they can ``see'' the boundary, because 
the internal coordinate is non-local with respect to the external ones
(see figure~\ref{int-ext}).
This is not true for the massless fields (photon and graviton), that live
entirely in the extended space-time. They therefore feel only the local
physics.

\subsubsection{Holography and orbifold}
\label{holorb}

\begin{figure}
\centerline{
\epsfxsize=6cm
\epsfbox{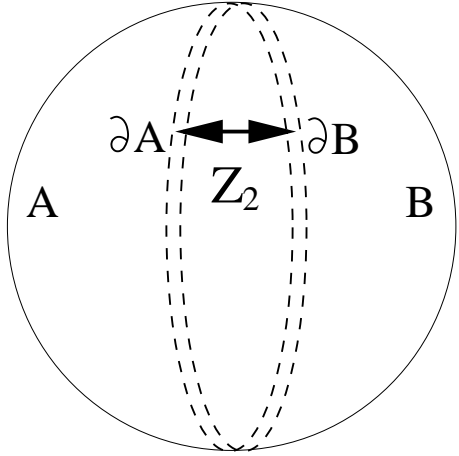}
}
\vspace{0.3cm}
\caption{Space-time is a three-sphere, here represented in lower dimension
by a two-sphere. Although without boundary, it can be thought
as the union of two hemispheres with neighbouring circles (two-spheres in
four space-time dimensions) as boundaries. The two boundaries
of the hemispheres are identified by a $Z_2$ reflection symmetry.}     
\label{boundary1}
\end{figure}

As discussed in \cite{spi},
the average dominant metric of the Universe,
i.e. a three-dimensional sphere, solves the FRW equations 
for the cosmological evolution of the metric. We have
on the other hand seen that the entropy of this universe scales like the
entropy of a black hole, namely with the volume of a two-sphere
with the same radius. This surface appears as the 
boundary of the space, that in turn appears to the observer as a ball.
Although a three-sphere doesn't have a boundary, we have nevertheless the 
impression that the Universe does have a boundary, 
constituted by the horizon of observation.
This ``holographic'' behaviour may seem contradictory. 
However, a sphere can be viewed as cut in two ``hyperdiscs'' by a hypersphere
(for instance, a two-sphere is cut by a circle in two discs).
If we pair two neighbouring hyperspheres, one serving as boundary 
for one of these two hyperdiscs, the other to the other one, and we
identify these two boundaries, together with the two bulks we build 
up a sphere. A sphere can therefore be viewed as the union of two hyperdiscs,
with a $Z_2$ identification of the boundaries (see figure~\ref{boundary1}). 
As such an orbifold, our Universe does have a boundary 
constituted by a two-sphere.
This pair of two-spheres holographically
contains the information about the 
``bulk'' (the union of the two ``bulks''), 
which can be thought as the interior of a
black hole with, as horizon, one of the two two-spheres. 
In this way we obtain two times more entropy than what we would expect,
corresponding to a rescaling $R \to R / \sqrt{2}$. 
However, things are re-established
to their correct normalization by the $Z_2$ orbifold operation.

It is interesting to have a look at the action of this operation on the
geometry of this space.
As we discussed in Ref.~\cite{spi}, a $Z_2$ operation on the extended
space coordinates of the string target space
is required in order to reach the string
configuration of minimal entropy. According to Ref.~\cite{spi},
in the string representation
the operation is a pair of freely acting projections, whose
effect on the string vacuum
is to break parity and time-reversal symmetries, as well as to
reduce to a minimum the gauge group, by breaking the weak symmetry
group, giving mass to bosons and matter states. The shift on the string side
is precisely what is needed in order to 1) reduce the number of massless
fields to coincide with the number of space-time dimensions (see discussion
in section \ref{vecsp}). If there are more
massless fields, as a matter of fact space is more than three-dimensional.
However entropy, which is basically
the Jacobian of the transformation from the string frame to the space-time
frame (see \cite{spi}, chapter 4), remains proportional to ${\cal T}^2$.
This means that we don't have anymore an $n$-sphere.
These configurations are non-extremal and 
therefore suppressed.
2) The further important effect of the shift is to break parity and
time reversal, with the consequence of making possible the identification
of the time coordinate of the string target space with the
time as we have defined it in this work. Once again, on the string side
this is related to a mass shift.

On the side of the geometry of combinatorics,
the $Z_2$ is an operation that simply identifies
two boundaries, something that looks completely
unrelated to what happens on the string side. 
The relation is that on the string side only by lifting massless 
degrees of freedom
we reduce to true three dimensions; only in three dimensions a two-sphere
has co-dimension one and can be a boundary for a semi-sphere, in the sense
above explained. Otherwise, in higher dimensions we can ``pass around''
this region, which is not anymore a boundary: from a physical
point of view, in this case we can ``go beyond the horizon and come back'' 
without problems. We cross therefore the border to tachyonic configurations,
and talking of time evolution becomes meaningless.
Therefore, the operation on the string side and on the combinatoric side
effectively do correspond.

Notice that all these shifts give rise to masses which introduce
modifications of the geometry of the order of the Heinsenbergs's
uncertainty relation.
This agrees with the fact that non-extremal configurations
(among which also tachyonic ones) contribute by an amount of the
order of the Heisenberg's uncertainty (see section~\ref{UncP}).

On the string side,
the orbifold operation halves the volume of the coordinates it acts on.
In order to see that it reduces the radius by a factor $\sqrt{2}$,
one can consider that, being space-time a surface spanned by light rays,
any operation acting on the time coordinate 
must also act on a space coordinate.
The volume elements are of the type $v = (ds)^2 = dt dx$. 
The factor 2 is therefore effectively ``distributed'' over two coordinates.

\subsubsection{Speed of light versus speed of expansion}
\label{cexp}

\begin{figure}
\centerline{
\epsfxsize=7cm
\epsfbox{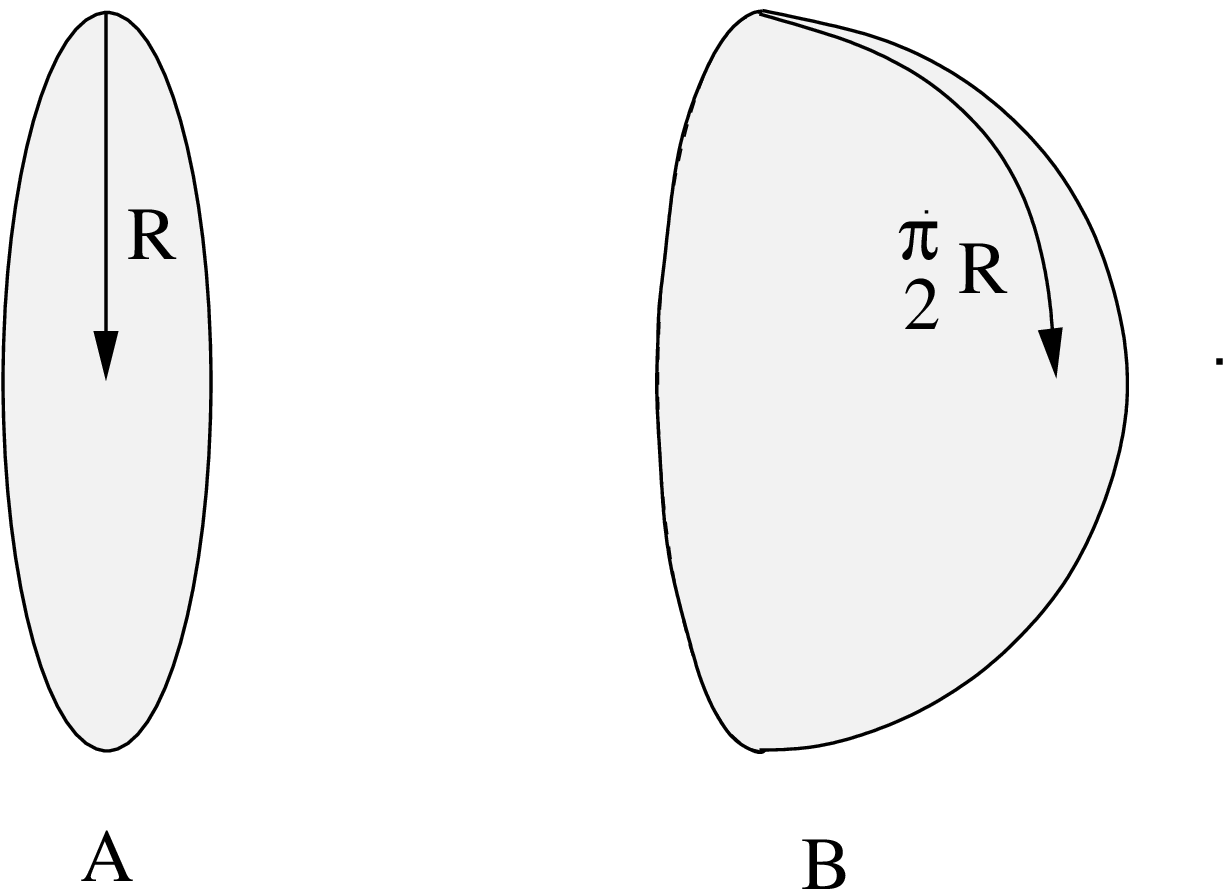}
}
\vspace{0.3cm}
\caption{A light ray seen by the observer as coming from the horizon, 
represented as the boundary of a flat space, the disc of picture A, 
follows in reality a curved path, a geodesic on the hemisphere depicted
in B. The real distance is longer: $R$ in (A) and ${\pi \over 2} R$ 
in (B).}     
\label{boundary2}
\end{figure}

As we said, to an observer the Universe appears in the average
as a ``(hyper)disk'' bounded
by a (hyper)circle, both of them with radius $R = c {\cal T}$, where ${\cal T}$
is the age of the Universe. 
However, the geometry of space-time is curved, and corresponds to the one of a
three-sphere, this too of radius $R = c {\cal T}$. 
Light paths correspond
therefore to geodesics in this curved space. A light beam that appears
to the observer to be long as much as $R = c$ times the age of the Universe
${\cal T}$, i.e. coming from the origin of time, corresponding to the horizon
as illustrate in picture A of figure~\ref{boundary2},
in reality follows a curved path, as illustrated on the right in picture B,
a path long ${\pi \over 2} c {\cal T}$. The ``true'' speed of light is 
therefore higher than the speed of expansion, or equivalently 
the ``scale factor'' $R$ of the universe is shorter, and expands 
at a lower speed, than real distances.
However, this is of no physical relevance:
it does in fact make no sense to ask what is the ``real'' speed of expansion,
the ``real'' age of the Universe, in a world in which everything looks
like if we were living in a flat space bounded by a horizon at distance
$R = c {\cal T}$. The only signal of the existence of a curvature
comes from indirect experiments, in which cosmological data are
interpreted within a theoretical framework. Namely, for what matters
parameters such as the energy density of the Universe, the matter content,
the speed of expansion of the horizon, and consequently the age of the 
Universe, everything works consistently with the identification of the
speed of light with the speed of expansion, the total energy with
the integral of the energy density
over the (hyper)disk, i.e. the ball enclosed by the horizon, and the
Universe itself with a black hole.
For what matters the cosmological evolution too, densities
and equations of motion are normalized for a space with boundary, the
``ball'' enclosed by the horizon. The two scales,
either of times or of lengths, are proportional, and
any computation of red-shifts and similar parameters is insensitive
to this detail.

\section{In few words}
\label{few}

\begin{figure}
\centerline{
\epsfxsize=8cm
\epsfbox{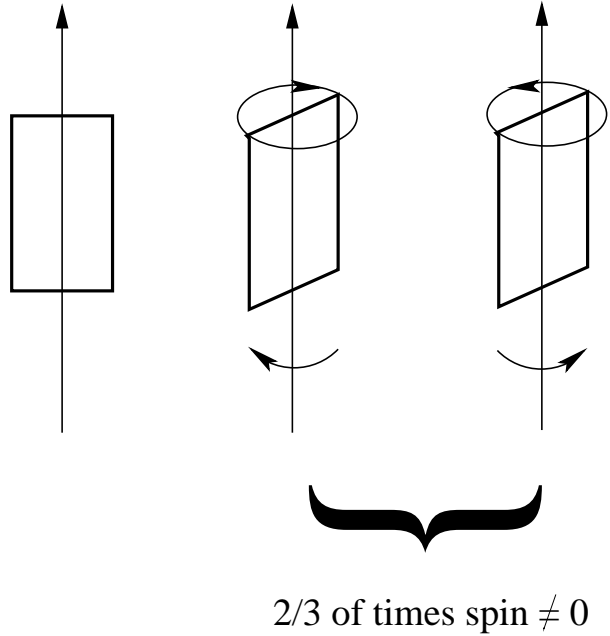}
}
\vspace{0.3cm}
\caption{A simple system consisting of a plate that can only either stay
at rest, or rotate along the two senses of the $z$ axis.}     
\label{3plates}
\end{figure}

Let's summarize in simple words the philosophy of what we are stating in
this work. The Universe appears to possess certain properties and has a 
certain aspect, because this is the most frequently realized 
configuration. In order to illustrate this point with an example, 
let's consider the elementary
case of a ``universe'' consisting just of a plate which, as possible 
configurations, can either stay at rest, or rotate along the $z$-axis,
clockwise or counterclockwise (if this is an electron, or a proton, this
schematically reproduces the configurations of spin $0$, spin $+ 1/2$, 
spin $-1/2$). The situation is illustrated in figure~\ref{3plates}.
Clearly, we have 1/3 of the probability to see the plate/particle at rest,
1/3 to see it rotating leftwards, 1/3 rotating rightwards. 
All in all, 2/3 of the times
we look at the plate, we see it in rotation, either leftwards or rightwards. 
In practice, in the average we see a rotating plate, because the situation of
``being in rotation'' has more possibilities to be realized (rotation
clockwise or rotation counterclockwise) than the configuration 
``plate at rest''. What we ``experimentally'' conclude is that the plate, with
a certain degree of approximation (very rough in this case indeed = 2/3),
has a non-vanishing angular momentum. Beyond the 2/3 degree of precision, 
the spin cannot be measured, because beyond this threshold
there is indeed no spin. This example is extremely schematic
and by no means reproduces whatever real situation. However, here we see
the basic reason why, according to our discussion, the Universe appears 
to be as it is: protons
and electrons have spin because this is the most frequently realized
configuration in the phase space, etc... The fully varied spectrum of
particles, fields, interactions, galaxy clusters etc... is the effect
of being this the configuration of minimal entropy, or equivalently the
most frequently realized configuration. 

A question arises now: is it life too part of this game? Can we see
the existence of life, and the increasing complexity of its forms, as
due to the same rationale? Does life exist because, in the appropriate
conditions, the chemical processes that we interpret
as ``life'' correspond to the most frequently realized
configuration? In principle, nothing seems to forbid it, 
and indeed at least certain aspects seem to nicely fit within this
picture (see Ref.~\cite{paleo}).  
The hardest point to be accepted is however the fact that, if the entire
story of life can be reduced to these same principles, then also 
human thoughts, actions and decisions, are ``determined'' 
(not deterministic though!), although predictable only by knowing
an infinite number of degrees of freedom. That means, in practice
impossible to predict.

\vspace{1.5cm}

\newpage

\providecommand{\href}[2]{#2}\begingroup\raggedright\endgroup

\end{document}